\documentclass[fleqn,usenatbib]{mnras}

\usepackage{newtxtext,newtxmath}
\usepackage{float}
\usepackage[T1]{fontenc}
\usepackage{ae,aecompl}
\usepackage[normalem]{ulem}
\usepackage{graphicx}	% Including figure files
\usepackage{amsmath}	% Advanced maths commands
\usepackage{svg}
\usepackage{gensymb}
\usepackage{xspace}
\usepackage[export]{adjustbox} % for aligning figures left or right
\usepackage{xcolor}
\usepackage{orcidlink}
\usepackage{soul}

\providecommand{\bjdtdb}{\ensuremath{\rm {BJD_{TDB}}}}

\providecommand{\lst}{\ensuremath{\,{\rm L_\odot}}}
\providecommand{\mj}{\ensuremath{\,{\rm M_J}}}
\providecommand{\rj}{\ensuremath{\,{\rm R_J}}}

\providecommand{\mst}{\ensuremath{\,{\rm M_\odot}}}
\providecommand{\rst}{\ensuremath{\,{\rm R_\odot}}}
\providecommand{\fave}{\langle F \rangle}
\providecommand{\fluxcgs}{10$^9$ erg s$^{-1}$ cm$^{-2}$}
\providecommand{\arcsec}{$^{\prime \prime}$}
\providecommand{\tess}{{\it TESS\,}}

\graphicspath{{./}{figures/}}

\title[Improved brown dwarf radii]{Improved radius determinations for the transiting brown dwarf population in the era of Gaia and TESS}

\author[T. W. Carmichael]{Theron W. Carmichael, $^1$\thanks{E-mail: tcarmich@ed.ac.uk}\orcidlink{0000-0001-6416-1274}
\\
$^1$ Institute for Astronomy, University of Edinburgh, Royal Observatory, Blackford Hill, Edinburgh, EH9 3HJ, UK\\
}

\date{Accepted XXX. Received YYY; in original form ZZZ}

\pubyear{2023}

\begin{document}
\label{firstpage}
\pagerange{\pageref{firstpage}--\pageref{lastpage}}
\maketitle


\begin{abstract}
\noindent I report updates to the substellar mass--radius diagram for 11 transiting brown dwarfs (BDs) and low-mass stars published before the third data release from the Gaia mission (Gaia DR3). I reanalyse these transiting BD systems whose physical parameters were published between 2008 and 2019 and find that when using the parallax measurements from Gaia DR3, 7 BDs show significant differences in their radius estimate or an improvement in the radius uncertainty. This has important implications for how these BDs are used to test substellar evolutionary models in the mass--radius diagram. The remaining 4 BDs show mass--radius estimates that are consistent with their previous pre-Gaia DR3 measurements. The 7 BDs that show significant deviation from the original mass--radius measurements are AD 3116b, CoRoT-3b, CoRoT-15b, EPIC 201702477b, Kepler-39b, KOI-205b, and KOI-415b. Of these, AD 3116b is a known member of the Praesepe cluster at an age of 600 Myr. Additionally, some of the previously smallest known transiting BDs, KOI-205b and KOI-415b, are not as small as once thought, leaving the mass--radius region for the very oldest BDs relatively sparse as a result of this work.
\end{abstract}


%%%%%%%%%%%%%%%%%%%%%%%%%%%%%%%%%%%%%%%%%%%%%%%%%%%%%%%%%%%%%%
%%%%%%%%%%%%%%%%%%%%%%%%% INTRODUCTION %%%%%%%%%%%%%%%%%%%%%%%
%%%%%%%%%%%%%%%%%%%%%%%%%%%%%%%%%%%%%%%%%%%%%%%%%%%%%%%%%%%%%%

\section{Introduction} \label{sec:intro}
The transiting brown dwarf (BD) population has grown at a remarkable rate over the past 4 years due in large part to NASA's Transiting Exoplanet Survey Satellite (\tess) mission \citep{tess}, which launched in 2018. These discoveries have been especially important because of how well the \tess mission has aided in the precise characterisation of the fundamental properties, namely the radius, of these transiting BDs. Astronomers have traditionally defined the range of masses for BDs to be between $13-80$ Jupiter masses (\mj), but more thorough studies have determined the mass range to vary between $11-16\mj$ at the lower end \citep{spiegel2011} to $75-80\mj$ \citep{baraffe03} at the higher end. These thresholds vary by several Jupiter masses because of variations in the chemical composition of the BDs and possibly because of their formation conditions. Some studies even prefer a breakdown by mass density and show a substellar boundary as low as $60\mj$ \citep{hatzes15}. Although a precise mass determination is important to confirm that an object is indeed a BD, a precise radius determination is just as crucial. One of the best ways we can precisely determine the radius of a BD is by searching for transiting BD systems. Of the 37 known transiting BD systems, we typically find them to have radii between $0.7-1.4$ Jupiter radii (\rj) \citep{grieves2021, carmichael2021}.

While we use spectra and radial velocities (RVs) of host stars to precisely measure the mass of a transiting BD, we use transits of main sequence stars by BDs provide an estimate of the radius of a BD. The precision of this estimate is directly impacted by how well we know the host star's radius as the transit depth is $\delta \propto (R_b/R_\star)^2$. Until recently, this knowledge of the host star's radius has been a frequent limiting factor on the ultimate precision and accuracy to which we could know the radii of transiting BDs.

This is where the Gaia mission has played a pivotal role in our estimation of transiting BD radii. With the Gaia mission's third data release, we now have access to precise and accurate parallaxes for all of the main sequence stars that are known to host transiting BDs. These parallaxes directly translate into stellar distances and radii (via the distance--luminosity relation $L_\star \propto d^2$), which means that these stellar radii are no longer as much of a limiting factor as before Gaia DR3. However, this means that a number of transiting BDs published previous to the launch of the Gaia mission will not have as precise or accurate stellar (and BD) radii as is currently possible. In fact, nearly a dozen different transiting BD systems published prior to 2019 have inaccurate or no parallax information for the host star of the BD. This practice of reanalysing transiting BDs has been established in the past with systems like WASP-30, where \cite{wasp30b} provided improved mass--radius measurements over the original discovery work by \cite{wasp30b_old}. A similar improvement work was done by \cite{kepler39} for the Kepler-39 and KOI-205 transiting brown dwarf systems. Other works that reanalyse previously known systems from the wider exoplanet community are \cite{schanche2020} and \cite{borsato2021}.

It is particularly important that these radius measurements are updated as they most directly enable us to make some conclusion of a transiting BD's age. This is because BDs--transiting or not--are believed to contract with age \citep{burrows01, baraffe03, ATMO2020, sonora21} and the rate of this contraction is of key interest to those who simulate the evolution of BDs. The radius of transiting BDs aids us in distinguishing young BDs from old BDs just like the mass of a BD distinguishes them from giant planets and low-mass stars.

Age is a particularly difficult parameter to measure for stars, let alone transiting BDs, but previous works have been able to determine relatively precise ages for transiting BD systems. For example, \cite{carmichael2021} used gyrochronology of a young solar analogue host star to constrain the age of a transiting BD. Others like \cite{nowak17, ad3116, david19_bd} have determined the ages of transiting BDs via associating their host stars with stellar clusters, for which we typically know the ages of very well. It is these system where the star's age is so precisely known that are the most important to have a precise and accurate radius for the transiting BD. This enables a well-informed examination of how well BD evolutionary models are able to predict the radius evolution of BDs and so allow for deeper investigations into other BD properties like chemical composition and internal structure.

Here I show updated estimates of the masses and radii of 11 different transiting brown dwarf systems that were published between 2008 and 2019. These 11 systems are AD 3116, CoRoT-3, CoRoT-15, EPIC 201702477, KELT-1, Kepler-39, KOI-189, KOI-205, KOI-415, WASP-30, and WASP-128. Of these, I provide improved mass and radius estimates for AD 3116b, CoRoT-3b, CoRoT-15b, EPIC 201702477b, Kepler-39b, KOI-205b, and KOI-415b. Section \ref{sec:observations} will present the new \tess observations used for a number of these systems and review the previously published data, and in one case, point out erroneous information. Section \ref{sec:analysis} will present the tools used to reanalyse these systems and show agreement with or improvements to previously published mass--radius determinations. Section \ref{sec:conclusion} will provide discussion on specific systems to compare them to their previously published values and will also remark on how the improved mass--radius determinations affect the substellar mass--radius diagram.

%%%%%%%%%%%%%%%%%%%%%%%%%%%%%%%%%%%%%%%%%%%%%%%%%%%%%%%%%%%%%%
%%%%%%%%%%%%%%%%%%%%%%%%% OBSERVATIONS %%%%%%%%%%%%%%%%%%%%%%%
%%%%%%%%%%%%%%%%%%%%%%%%%%%%%%%%%%%%%%%%%%%%%%%%%%%%%%%%%%%%%%

\begin{table}
\setlength{\tabcolsep}{2pt}
\centering
 \caption[]{Object names, observed \tess sectors, and observation cadence (in seconds) as of August 2022 (Sector 54). The bold numbers in the  ``Sectors Obs.'' column indicate which \tess sectors the data used in this work originate from. Though EPIC 201702477 was observed by \tess and processed through the SPOC pipeline, the transits of the brown dwarf were not captured due to the breaks in the light curve data. The transits of CoRoT-15b are completely diluted by neighboring stars, with one at $23\farcs8$ with a $\Delta G = -0.4$ (fainter) magnitudes and 2 at 30-40$\arcsec$ with a $\Delta G =+$1-2 (brighter) magnitudes. Objects with with the ``No SPOC" note were not processed through the SPOC PDCSAP pipeline or not currently made available at MAST. This means that the spacecraft motions and scattered Earth and moon light from the affected \tess sectors are not removed from the raw light curves, so I do not use them.} \label{tab:coords}
 \begin{tabular}{lcccc}
 \hline
 Object name & TIC ID & Sectors obs.& Cadence (s)\\
 \hline
 AD 3116 & 184892124 & \textbf{44}, \textbf{45}, \textbf{46} & 120 \\[1pt]
 CoRoT-3 & 392353449 &  \textbf{54} & 120\\[1pt] % 54
 CoRoT-15 & 206893389 &  6, 33 &  No detection\\[1pt]
 EPIC 201702477 & 281888055 &  45, 46 & No detection\\[1pt]
 KELT-1 & 432549364 &  \textbf{17} & 120 \\[1pt] %\ 57
 Kepler-39 & 272836943 &  14, 15, 41 & No SPOC\\[1pt] % 56
 KOI-189 & 48451130 & 14, 15, 26, {\bf 40}, {\bf 41} & 120 \\[1pt] % 54, 55
 KOI-205 & 271749758 & 14, 15, 40, 41 &  No SPOC\\[1pt] % 54, 55
 KOI-415 & 138099073 &  14, 15, 40, 41 & No SPOC \\[1pt] % 54, 55
 WASP-30 & 9725627 & \textbf{2}, \textbf{29}, \textbf{42} & 120, 600  \\[1pt] % 69
 WASP-128 & 180991313 & \textbf{10}, \textbf{36}, \textbf{37} & 600  \\[1pt] % 64
 \hline
 \end{tabular}
\end{table}

\section{Observations}\label{sec:observations}
Here I present the new \tess light curve analysis I performed for a subset of these transiting BD systems. I also briefly review the provenance of the originally published light curve and radial velocity data for each system. I use these data in most cases for the reanalysis of the systems in this work.

\begin{table}
\begin{center}
\caption[Corrected radial velocity data for KOI-189]{Corrected BJD values for KOI-189 RVs from the SOPHIE spectrograph.}\label{tab:koi189_rvs}
\begin{tabular}{c|c|c}
\hline
BJD (days) & RV ($\rm m\, s^{-1}$) & RV uncertainty ($\rm m\, s^{-1}$) \\
\hline
2455746.4495 & -78748.0 & 24.0\\
2455763.5061 & -68845.0 & 16.0\\
2455768.4028 & -71854.0 & 13.0\\
2455773.4533 & -75930.0 & 34.0\\
2455776.5059 & -78475.0 & 36.0\\
2455785.4191 & -69234.0 & 30.0\\
2455788.5163 & -67503.0 & 59.0\\
2455791.4151 & -67779.0 & 37.0\\
2455809.3730 & -79241.0 & 51.0\\
2455876.2423 & -69533.0 & 56.0\\
\hline
\end{tabular}
\end{center}
\end{table}

\subsection{\textit{TESS} data}
I use light curve data from \tess for 5 of the 11 transiting BDs that are part of this work. In some cases, the \tess light curves are in addition to the original discovery light curves. Unless explicitly stated otherwise, I use the Science Processing Operations Center \citep[SPOC;][]{jenkins_2016} \tess light curves that are available on the Mikulski Archive for Space Telescopes (MAST). These data are from the Presearch Data Conditioning Simple Aperture Photometry flux pipeline \citep[PDCSAP;][]{smith2012_pdc, stumpe2014_pdc}, which removes some systematic stellar effects but aim to retain transits, eclipses, and native stellar features such as brightness modulation. I then use the {\tt lightkurve} \citep{lightkurve} package in Python to normalise the PDCSAP light curves for transit analysis, which typically involves removing photometric modulation caused by star spots and stellar rotation. All 11 systems have been observed by \tess for at least one sector, but not every light curve produced is suitable for transit analysis. I restrict the use of \tess data for transit analysis to those systems with \tess data that has been processed through the PDCSAP pipeline. This means that only AD 3116, CoRoT-3, KELT-1, KOI-189, WASP-30, and WASP-128 use \tess data in this work. The remaining systems (CoRoT-15, Kepler-39, KOI-205, KOI-415) suffer from dilution from nearby stars or do not have light curve data processed by SPOC at the time of this work. For these systems, the algorithms in \cite{smith2012_pdc} and \cite{stumpe2014_pdc} would need to be replicated here in order to correct for spacecraft motions and scattered Earth and moon light contamination prevalent in certain \textit{TESS} sectors (most notably the \textit{Kepler} field for this work). I avoid this level of raw data reduction as it would not be at the same quality as that from SPOC and it could introduce inconsistent or inadequate treatment of these systems. The light curves from SPOC vary in cadence as they are sometimes sourced from the \tess full-frame images, which change in their exposure times between the primary and extended \tess missions.

The new \tess data used in this work are detailed in Table \ref{tab:coords} with the sectors each target was observed in along with the cadence (exposure times) for the light curve data.

\begin{table*}
\setlength{\tabcolsep}{4pt}
\centering
 \caption[]{Table of magnitudes used in SED fitting for the host stars in AD 3116, CoRoT-3, CoRoT-15, EPIC 201702477, and KELT-1. Sources: Gaia - \cite{dr3}, \tess - \cite{stassun18}, 2MASS - \cite{2MASS}, WISE - \cite{WISE}. Additional magnitudes for AD 3116 and EPIC 201702477 are sourced from \cite{ad3116} and \cite{bayliss16}, respectively.} \label{tab:magnitudes_1}
 \begin{tabular}{lccccc}
 \hline
 {}       & AD 3116 & CoRoT-3 & CoRoT-15 & EPIC 201702477 & KELT-1\\
 \hline
 \tess    & $16.057 \pm 0.008$ & $12.713 \pm 0.011$ & $14.869 \pm 0.008$ & $13.920 \pm 0.007$ & $10.224 \pm 0.006$ \\[2pt]
 $G$      & $17.416 \pm 0.020$ & $13.219 \pm 0.020$ & $15.599 \pm 0.001$ & $14.391 \pm 0.001$ & $10.591 \pm 0.001$ \\[2pt]
 $G_{BP}$ & $19.084 \pm 0.026$ & $13.623 \pm 0.020$ & $16.240 \pm 0.020$ & $14.764 \pm 0.020$ & $10.873 \pm 0.020$ \\[2pt]
 $G_{RP}$ & $16.147 \pm 0.020$ & $12.652 \pm 0.020$ & $14.823 \pm 0.020$ & $13.863 \pm 0.020$ & $10.175 \pm 0.020$ \\[2pt]
 $J$      & -                  & $11.936 \pm 0.030$ & $13.801 \pm 0.030$ & $13.268 \pm 0.030$ & $9.682 \pm 0.020$ \\[2pt]
 $H$      & -                  & $11.710 \pm 0.040$ & $13.434 \pm 0.040$ & $12.881 \pm 0.030$ & $9.534 \pm 0.030$ \\[2pt]
 $K_S$    & $13.499 \pm 0.043$                  & $11.618 \pm 0.030$ & $13.389 \pm 0.050$ & $12.766 \pm 0.030$ & $9.437 \pm 0.020$ \\[2pt]
 WISE1    & $13.330 \pm 0.030$ & $11.343 \pm 0.030$ & $13.168 \pm 0.030$ & $12.814 \pm 0.030$ & $9.414 \pm 0.030$ \\[2pt]
 WISE2    & $13.113 \pm 0.030$ & $11.394 \pm 0.030$ & $13.141 \pm 0.032$ & $12.840 \pm 0.030$ & $9.419 \pm 0.030$ \\[2pt]
 WISE3    & -                  & $11.323 \pm 0.172$ & $12.152 \pm 0.435$ & - & $9.386 \pm 0.034$ \\[2pt]
 $u$      & $22.290 \pm 0.190$ & -                  & -                  & $16.312 \pm 0.005$ & - \\[2pt]
 $g$      & $19.646 \pm 0.014$ & -                  & -                  & $14.871 \pm 0.003$ & - \\[2pt]
 $r$      & $16.675 \pm 0.005$ & -                  & -                  & $14.354 \pm 0.003$ & - \\[2pt]
 $z$      & $15.845 \pm 0.006$ & -                  & -                  & $14.137 \pm 0.004$ & - \\[2pt]
 \hline
 \end{tabular}
\end{table*}

\begin{table*}
\setlength{\tabcolsep}{4pt}
\centering
 \caption[]{Table of magnitudes used in SED fitting for the host stars in Kepler-39, KOI-189, KOI-205, KOI-415, WASP-30, and WASP-128. Sources: Gaia - \cite{dr3}, \tess - \cite{stassun18}, 2MASS - \cite{2MASS}, WISE - \cite{WISE}.}\label{tab:magnitudes_2}
 \begin{tabular}{lcccccc}
 \hline
 {} & Kepler-39 & KOI-189 & KOI-205 & KOI-415 & WASP-30 & WASP-128\\
 \hline
 \tess     & $13.826 \pm 0.001$ & $13.759 \pm 0.001$ & $13.935 \pm 0.008$ & $13.662 \pm 0.007 $ & $10.977 \pm 0.008$    & $11.867 \pm 0.006$\\[2pt]
 $G$      & $14.262 \pm 0.020$ & $14.380 \pm 0.020$ & $14.511 \pm 0.020$ & $14.114 \pm 0.020$ & $11.350 \pm 0.020$ & $12.274 \pm 0.020$\\[2pt]
 $G_{BP}$  & $14.598 \pm 0.020$ & $14.926 \pm 0.020$ & $14.999 \pm 0.020$ & $14.467 \pm 0.020$ & $11.640 \pm 0.020$ & $12.583 \pm 0.020$\\[2pt]
 $G_{RP}$  & $13.764 \pm 0.020$ & $13.703 \pm 0.020$ & $13.882 \pm 0.020$ & $13.604 \pm 0.020$ & $10.934 \pm 0.020$ & $11.814 \pm 0.020$\\[2pt]
 $J$       & $13.236 \pm 0.030$ & $12.895 \pm 0.030$ & $13.095 \pm 0.020$ & $13.049 \pm 0.020$ & $10.508 \pm 0.030$ & $11.282 \pm 0.020$\\[2pt]
 $H$       & $12.999 \pm 0.020$ & $12.377 \pm 0.020$ & $12.709 \pm 0.020$ & $12.671 \pm 0.020$ & $10.283 \pm 0.020$ & $11.011 \pm 0.020$\\[2pt]
 $K_S$     & $12.918 \pm 0.030$ & $12.288 \pm 0.030$ & $12.656 \pm 0.020$ & $12.661 \pm 0.030$ & $10.199 \pm 0.020$ & $10.942 \pm 0.020$\\[2pt]
 WISE1     & $12.858 \pm 0.030$ & $12.239 \pm 0.030$ & $12.579 \pm 0.030$ & $12.571 \pm 0.030$ & $10.162 \pm 0.030$ & $10.923 \pm 0.030$\\[2pt]
 WISE2     & $12.892 \pm 0.030$ & $12.327 \pm 0.030$ & $12.685 \pm 0.030$ & $12.595 \pm 0.030$ & $10.202 \pm 0.030$ & $10.936 \pm 0.030$\\[2pt]
 WISE3     & $12.601 \pm 0.329$ & $12.280 \pm 0.233$ & $12.396 \pm 0.294$ & $12.364 \pm 0.289$ & $10.077 \pm 0.060$ & $10.966 \pm 0.088$\\[2pt]
 \hline
 \end{tabular}
\end{table*}

\subsection{CoRoT brown dwarfs}
I use the light curve data from the CoRoT mission \citep{corot} along with the respective RV data published for the analysis of CoRoT-3 \citep{corot3b} and CoRoT-15 \citep{corot15b}. CoRoT-3 has RV data from the CORALIE, HARPS, Sandiford, SOPHIE, and TLS instruments while CoRoT-15 has RV data from HARPS and HIRES only. The \tess data for CoRoT-15 are too diluted from nearby starlight to be of use here. This is partly due to the design of the CoRoT mission as its pixel scale of $2\farcs 32$ is much better than \tess at observing the relatively crowded fields of stars that CoRoT-15 occupies. Though the \tess light curve of CoRoT-3 is also diluted, the transits are still recoverable, so I use these data in my analysis.

\subsection{\textit{Kepler} and \textit{K2} brown dwarfs}
\subsubsection{\textit{Kepler}}
Of the 4 transiting BD systems discovered by the \textit{Kepler} mission \citep{kepler}, only KOI-189 has a \tess light curve that is vetted by the SPOC. I include these data in addition to the original light curve data from \cite{diaz14} for KOI-189 in the transit analysis while I use the original \textit{Kepler} data from \cite{kepler39}, \cite{diaz13}, and \cite{moutou13} for Kepler-39, KOI-205, and KOI-415, respectively. I use the RV data from SOPHIE instrument originally published for Kepler-39, KOI-205, and KOI-415.

The RV data from the SOPHIE instrument for KOI-189 have erroneous observations times as presented in \cite{diaz14}. I show the corrected values in Table \ref{tab:koi189_rvs}. The error in the observation times (BJD) was a typographical mistake in the hundred-thousands place where the ``4" was incorrectly replaced with a ``5" (as in 2,405,746 to 2,505,746). Taken as shown in \cite{diaz14}, these RV data would be dated on May 2148, and not June 2011 as the original paper states within the text accompanying the table.

\subsubsection{\textit{K2}}
The transiting BD systems from the \textit{K2} mission \citep{k2} are AD 3116 (EPIC 211946007) and EPIC 201702477. There are data from \tess for both of these systems, but only AD 3116 shows evidence of a transit; the \tess mission does not detect transits for the EPIC 201702477 system across the 2 consecutive sectors it was observed in. This is likely because the 41-day orbital period of this BD and the breaks in the \tess data. Given this, only the \tess data for AD 3116, in addition to the light curve data from \cite{ad3116}, are used for the transit analysis while just the light curve data from \textit{K2} and ground-based facilities \citep{bayliss16} are used for EPIC 201702477. These ground-based facilities are the Cerro Tololo InterAmerican Observatory (CTIO) and the South African Astronomical Observatory (SAAO). I use the RV data from these previous works for this pair of \textit{K2} systems (\textit{Keck}/HIRES for AD 3116; HARPS and SOPHIE for EPIC 201702477).

\subsection{KELT and WASP brown dwarfs}
\subsubsection{KELT}
The single transiting BD system from the KELT survey \citep{kelt} is KELT-1 \citep{kelt1b}. I use the \tess data in addition to the ground-based data originally published for the transit analysis of this system. The ground-based light curve data are from the University of Louisville Moore Observatory (ULMO/MORC24) and the Fred Lawrence Whipple Observatory (FLWO/KeplerCam) and the RV data are from the TRES instrument at FLWO.

\subsubsection{WASP}
The WASP survey \citep{wasp} has detected 2 transiting BD systems, WASP-30 and WASP-128. I use the light curve from the TRAPPIST telescope and RV data from CORALIE for the analysis of WASP-30 \citep{wasp30b}. I note that \tess data from SPOC are available for both WASP-30 and WASP-128, but given the photometric modulation of WASP-30's light curve, the resulting normalised \tess light curve shows a slightly shallower transit compared to the TRAPPIST light curve, so this required additional use of spline detrending tools in {\tt lightkurve}. In the case of WASP-128, I use the original RV data from CORALIE and HARPS \citep{wasp128b}, but not the original light curve data, as these were not made available. Instead, I use the \tess data for the transit analysis of WASP-128.

%%%%%%%%%%%%%%%%%%%%%%%%%%%%%%%%%%%%%%%%%%%%%%%%%%%%%%%%%%%%%%
%%%%%%%%%%%%%%%%%%%%%%%%%%% ANALYSIS %%%%%%%%%%%%%%%%%%%%%%%%%
%%%%%%%%%%%%%%%%%%%%%%%%%%%%%%%%%%%%%%%%%%%%%%%%%%%%%%%%%%%%%%
\section{Analysis} \label{sec:analysis}

\subsection{Global analysis with {\tt EXOFASTv2}}\label{sec:exofast}
Here I give details on my global analysis of the stellar and BD parameters for these 11 systems using {\tt EXOFASTv2} \citep{eastman2019}. {\tt EXOFASTv2} uses the Monte Carlo-Markov Chain (MCMC) method to estimate the most likely values and uncertainties for the physical properties of the star and BD in each system. I use N=36 (N = 2$\times n_{\rm parameters}$) walkers, or chains, and run until the fit passes the default convergence criteria for {\tt EXOFASTv2}, which is typically on the order of hundreds of thousands of steps for these systems \citep[described in more detail in][]{eastman2019}.

I set either uniform $\mathcal{U}[a,b]$ or Gaussian $\mathcal{G}[a,b]$ priors on the following input parameters: parallax ($\varpi$) from Gaia DR3, spectroscopic stellar effective temperature ($T_{\rm eff}$), spectroscopic stellar metallicity ([Fe/H]), and V-band extinction ($A_V$). The two spectroscopic priors are taken from the respective results in the discovery works for each system and an upper limit is set on $A_V$ from a database\footnote{Database for extinction calculation: \url{https://irsa.ipac.caltech.edu/frontpage/}} with measurements from \cite{av_priors}. Of these priors, $\varpi$ in particular aids in establishing a precise estimate for the stellar radius, which directly affects the estimate for the BD radius. Following the method detailed in \cite{dr3_correction}, I apply the parallax zero-point correction to the Gaia DR3 $\varpi$. The systems in this paper experience a 1-2$\sigma$ change in $\varpi$ when the correction is applied\footnote{Gaia parallax zero-point correction code: \url{https://www.cosmos.esa.int/web/gaia/edr3-code}}, except for AD 3116, which only sees a $\Delta\varpi \approx 0.25\sigma$.

In addition to these priors, I input the following data with {\tt EXOFASTv2}: magnitude measurements of the host star for the SED (shown in Tables \ref{tab:magnitudes_1} and \ref{tab:magnitudes_2}), previously published RV data for each system, and new \tess light curves and previously published light curves. In some cases, the light curve data are truncated to only include the transit events with 4-5 hours of baseline before ingress and after egress. This is done to reduce the convergence time for systems with especially long temporal coverage by space telescope missions. This should not be done if a search for secondary eclipses has not been performed as secondary eclipse data provide valuable information on the nature of the companion. I let the RV offset $\gamma$ be a free parameter and use the RV jitter term, $\sigma_J$, to account for the surface activity of the star. The RV jitter term is based on a simple white noise model implemented in {\tt EXOFASTv2}. The stellar magnitudes along with the input spectroscopic data establish the boundaries on the MIST stellar isochrone models \citep{mist3, mist2, mist1} built into {\tt EXOFASTv2} and are used here to estimate the stellar parameters for each system.

\begin{figure*}
    \centering
    \includegraphics[width=0.33\textwidth, trim={0.0cm 0.0cm 0.0cm 0.0cm}]{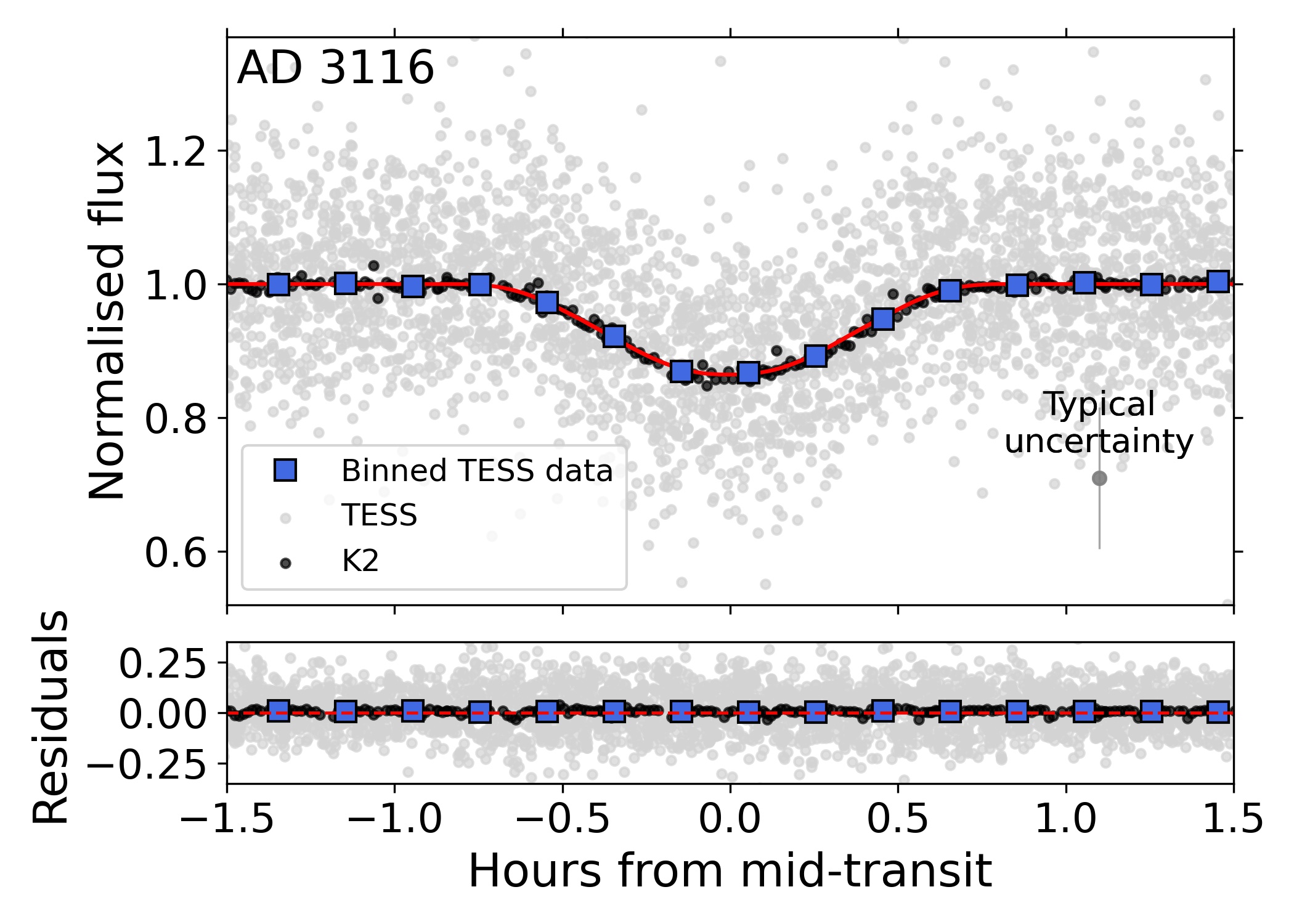}
    \includegraphics[width=0.33\textwidth, trim={0.0cm 0.0cm 0.0cm 0.0cm}]{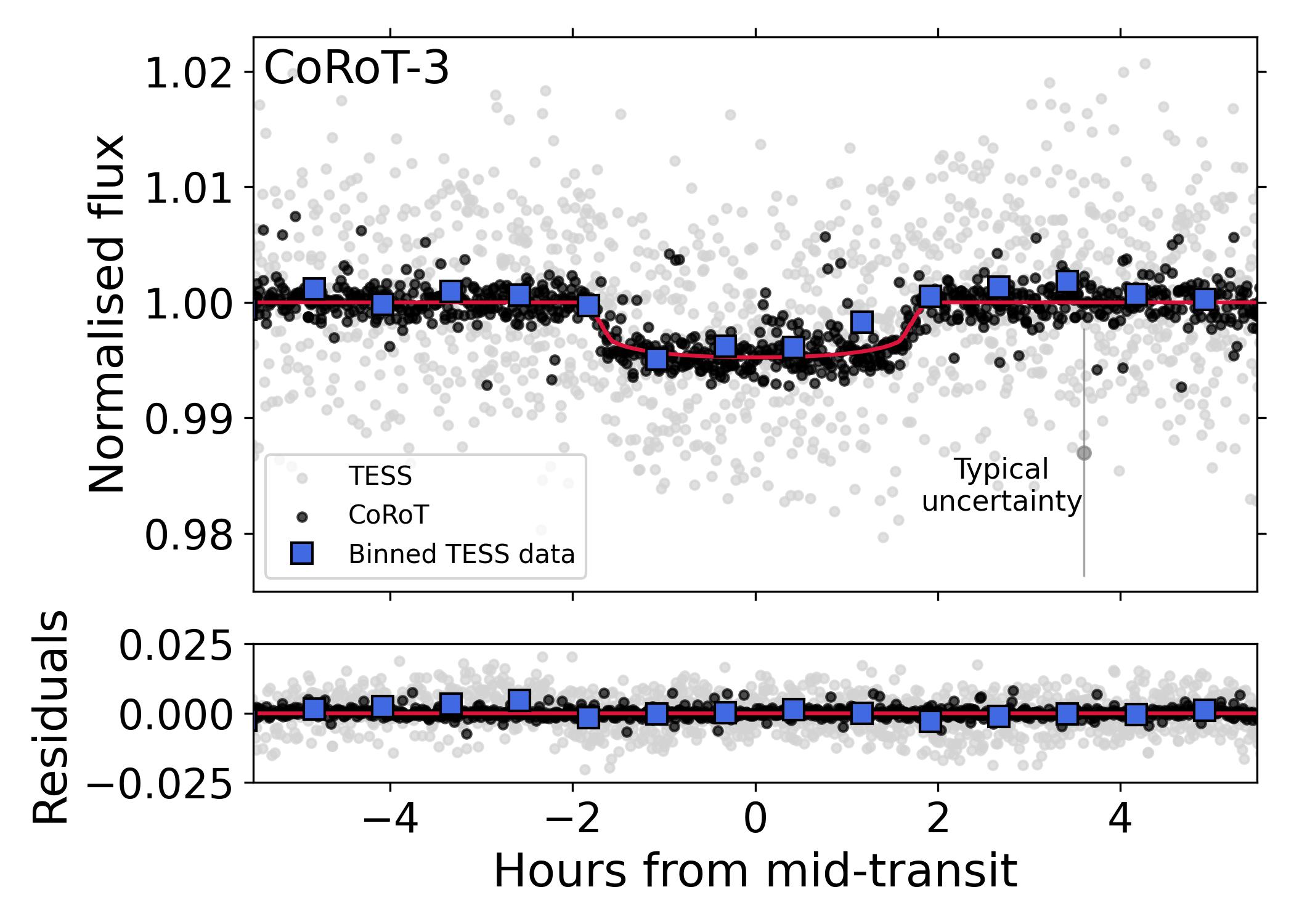}
    \includegraphics[width=0.33\textwidth, trim={0.0cm 0.0cm 0.0cm 0.0cm}]{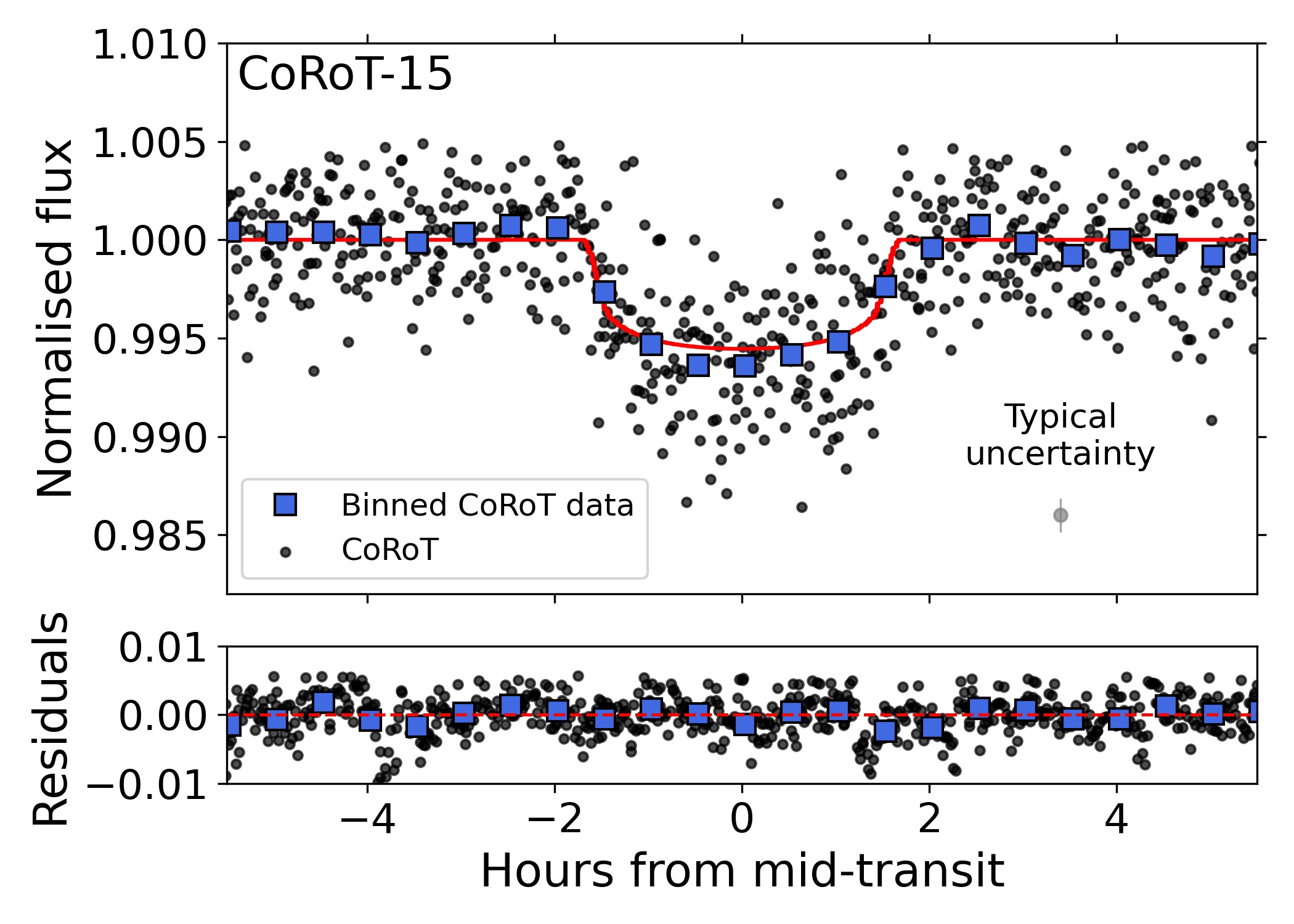}
    \includegraphics[width=0.33\textwidth, trim={0.0cm 0.0cm 0.0cm 0.0cm}]{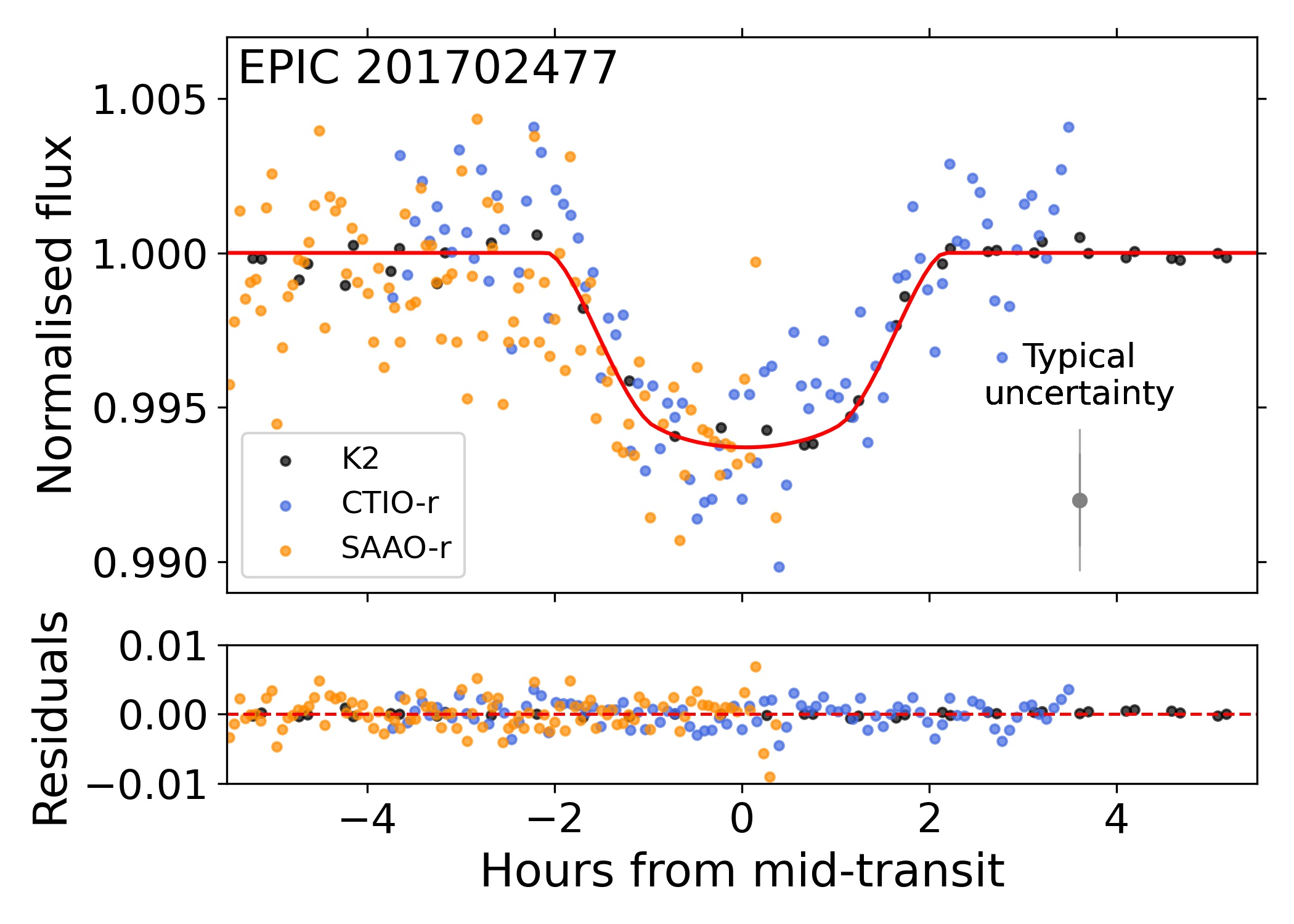}
    \includegraphics[width=0.33\textwidth, trim={0.0cm 0.0cm 0.0cm 0.0cm}]{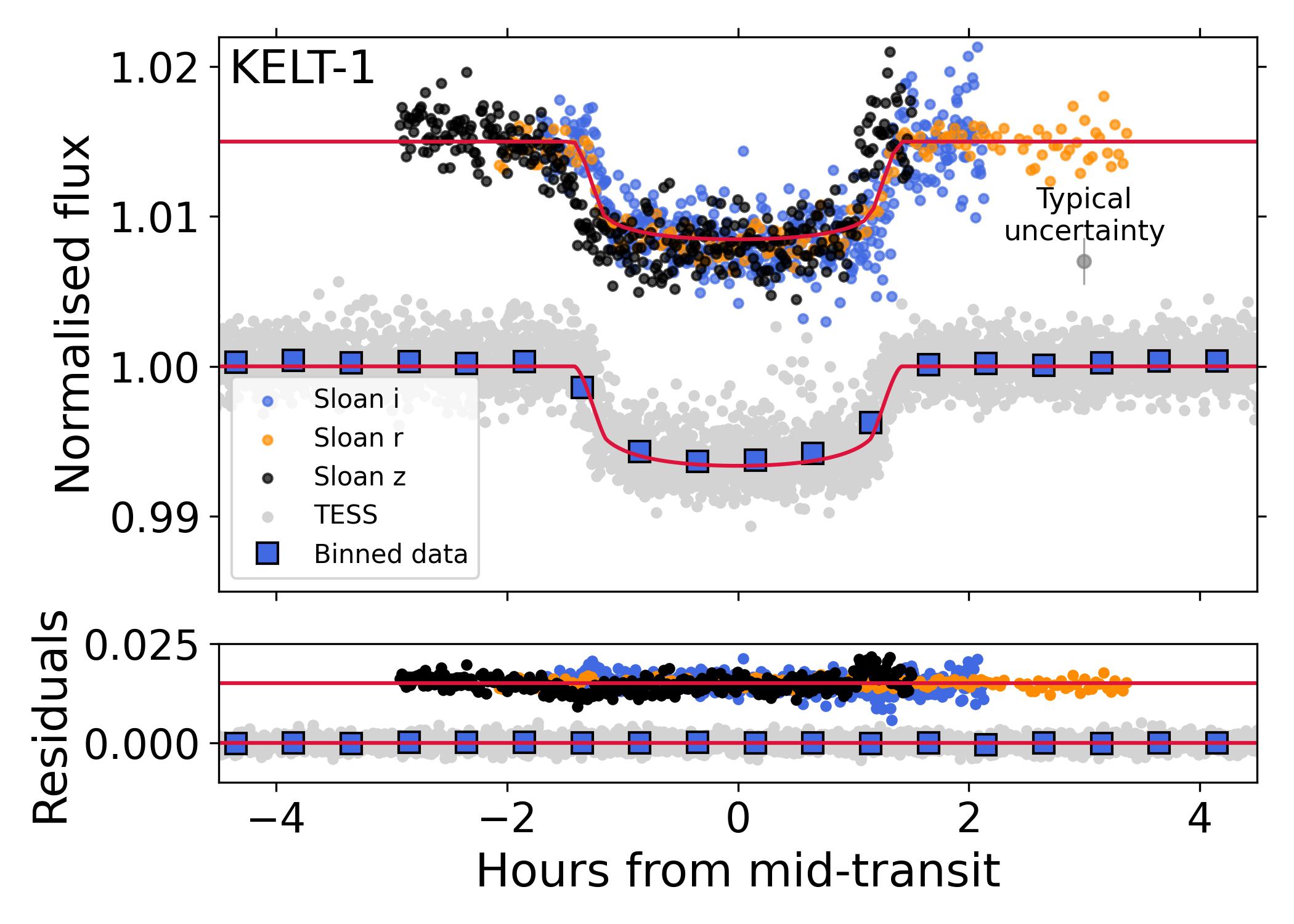}
    \includegraphics[width=0.33\textwidth, trim={0.0cm 0.0cm 0.0cm 0.0cm}]{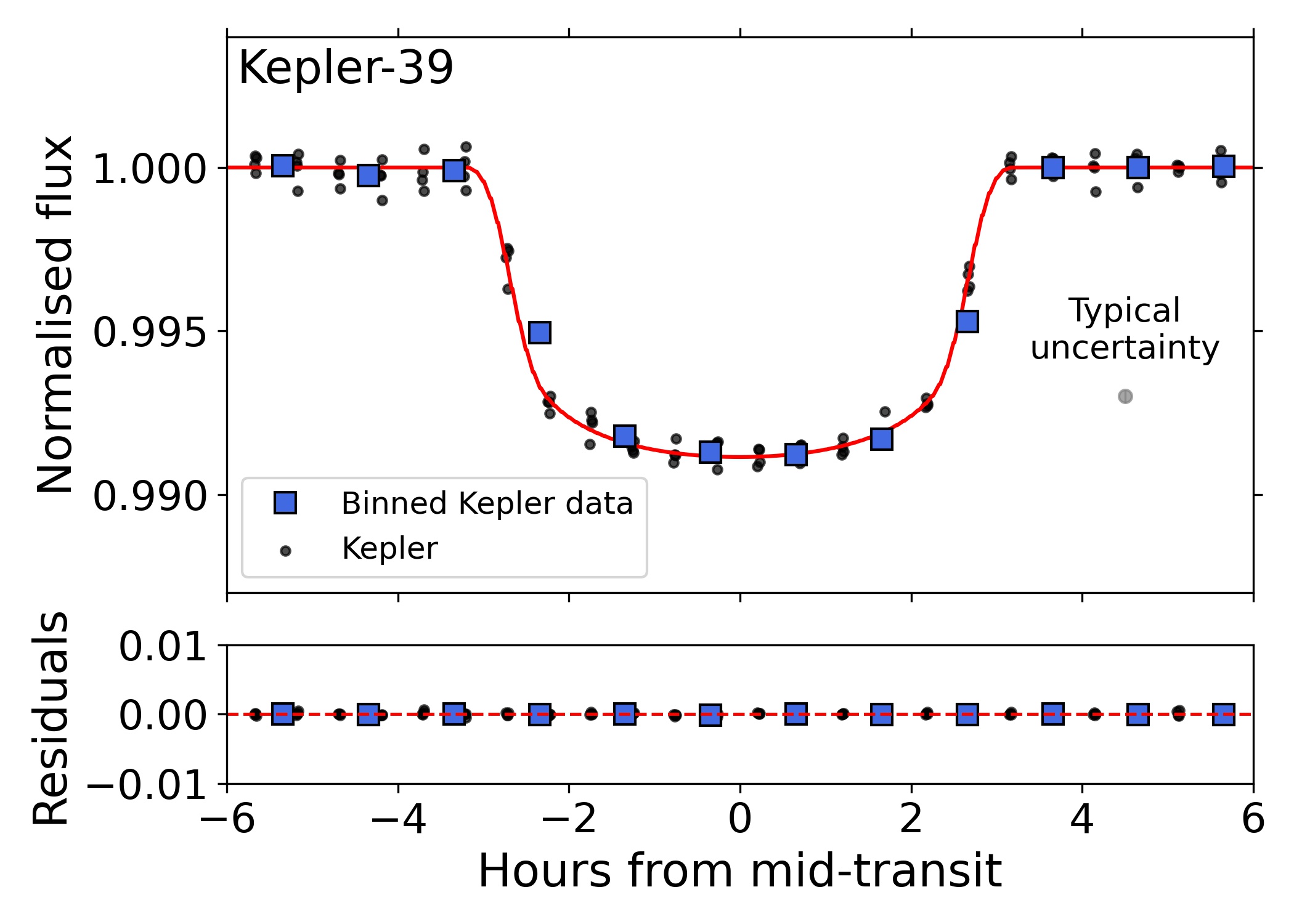}
    \includegraphics[width=0.33\textwidth, trim={0.0cm 0.0cm 0.0cm 0.0cm}]{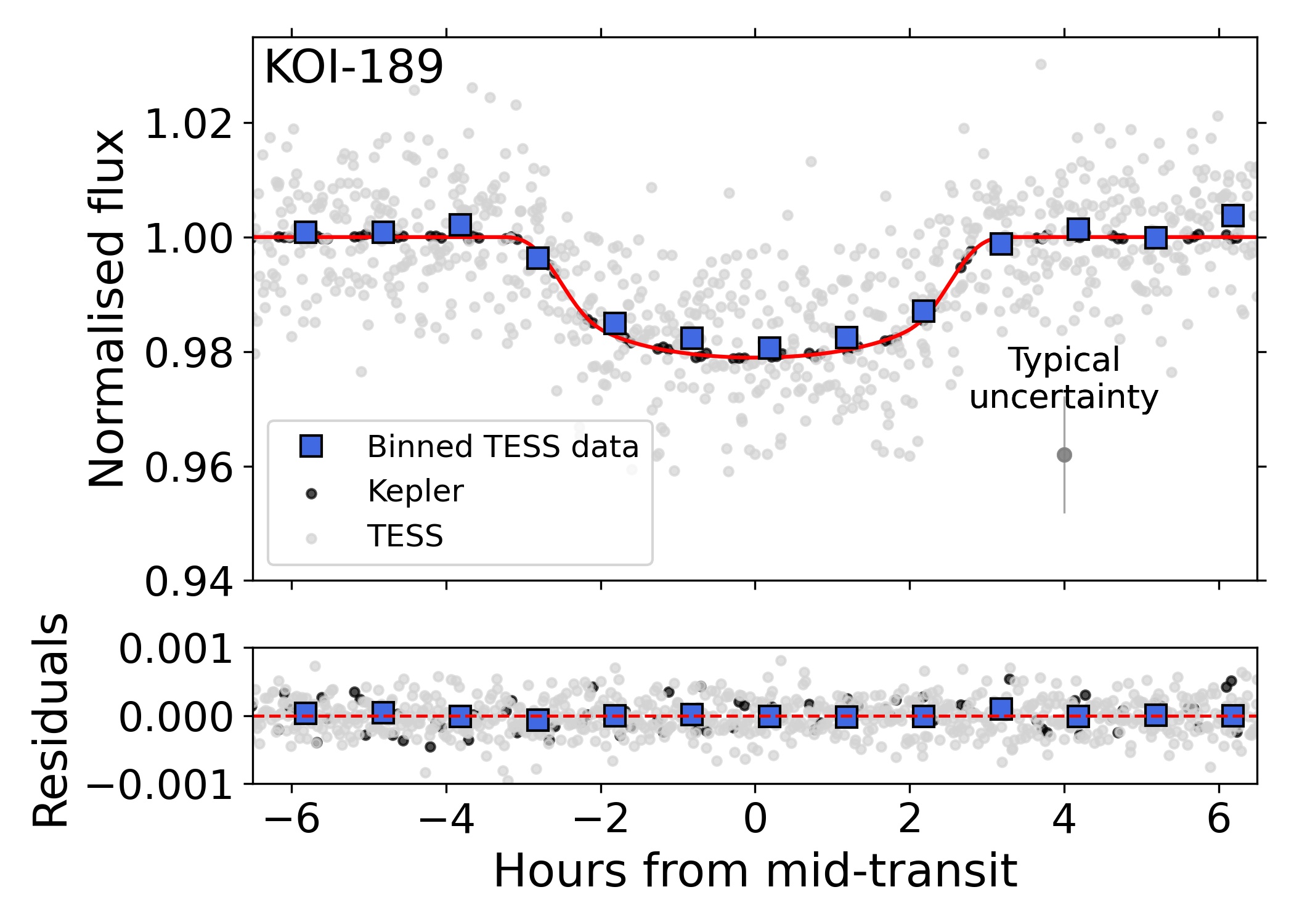}
    \includegraphics[width=0.33\textwidth, trim={0.0cm 0.0cm 0.0cm 0.0cm}]{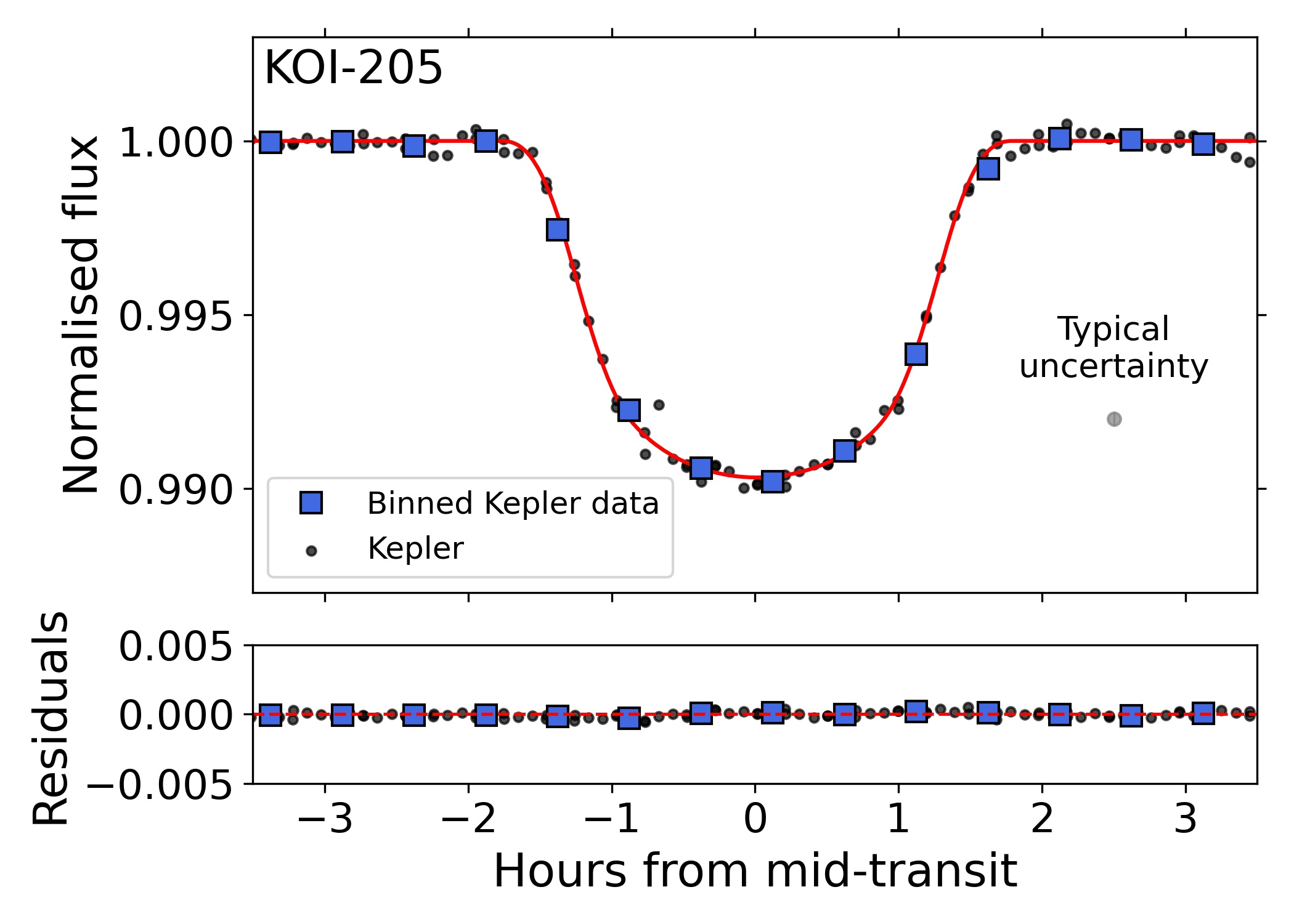}
    \includegraphics[width=0.33\textwidth, trim={0.0cm 0.0cm 0.0cm 0.0cm}]{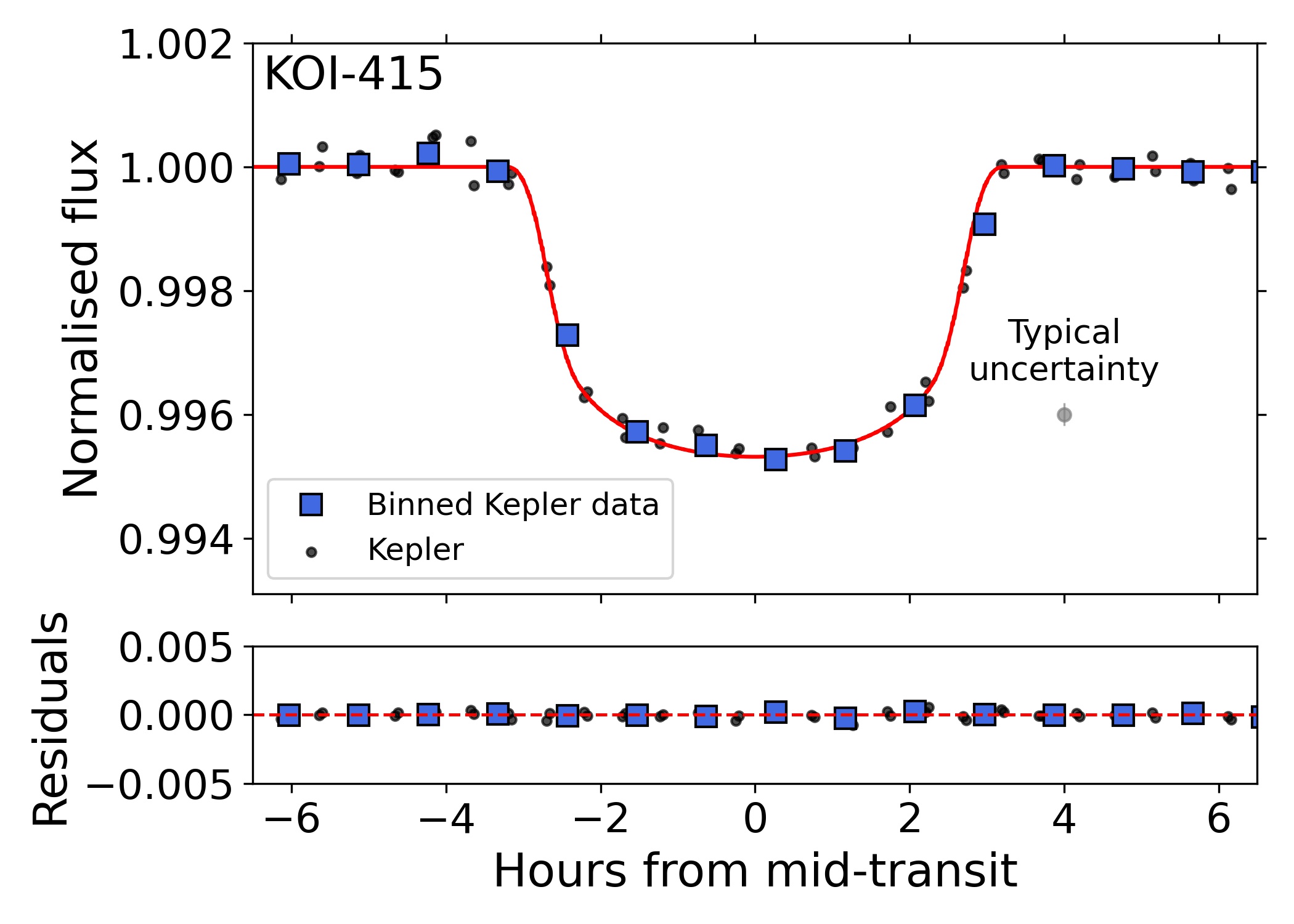}
    \includegraphics[width=0.33\textwidth, trim={0.0cm 0.0cm 0.0cm 0.0cm}]{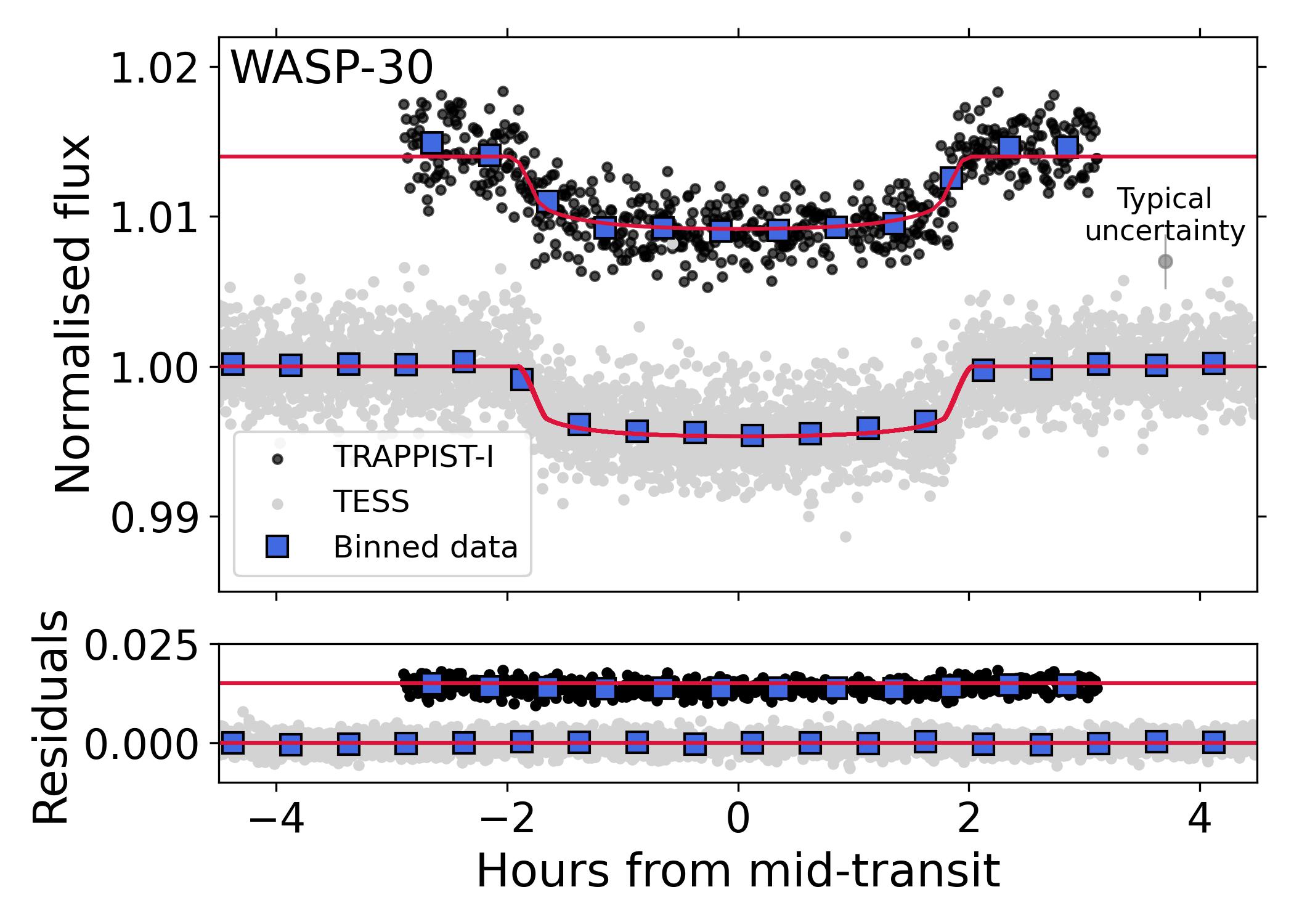}
    \includegraphics[width=0.33\textwidth, trim={0.0cm 0.0cm 0.0cm 0.0cm}]{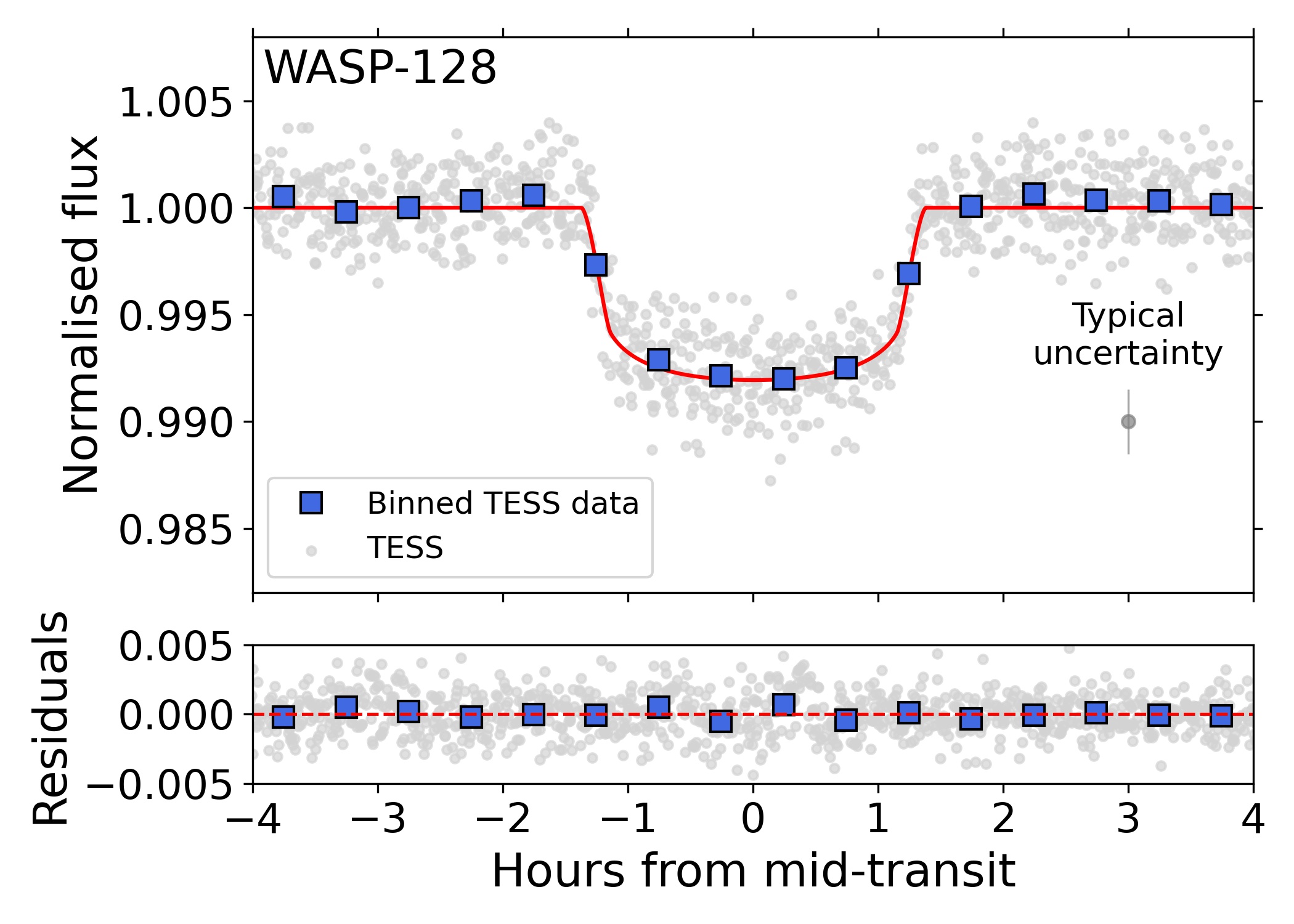}
    \caption{Transit light curves modeled in this study. The red curve indicates the best-fitting model for each transit.}
    \label{fig:transits}
\end{figure*}

\begin{figure*}
    \centering
    \includegraphics[width=0.33\textwidth, trim={0.0cm 0.0cm 0.0cm 0.0cm}]{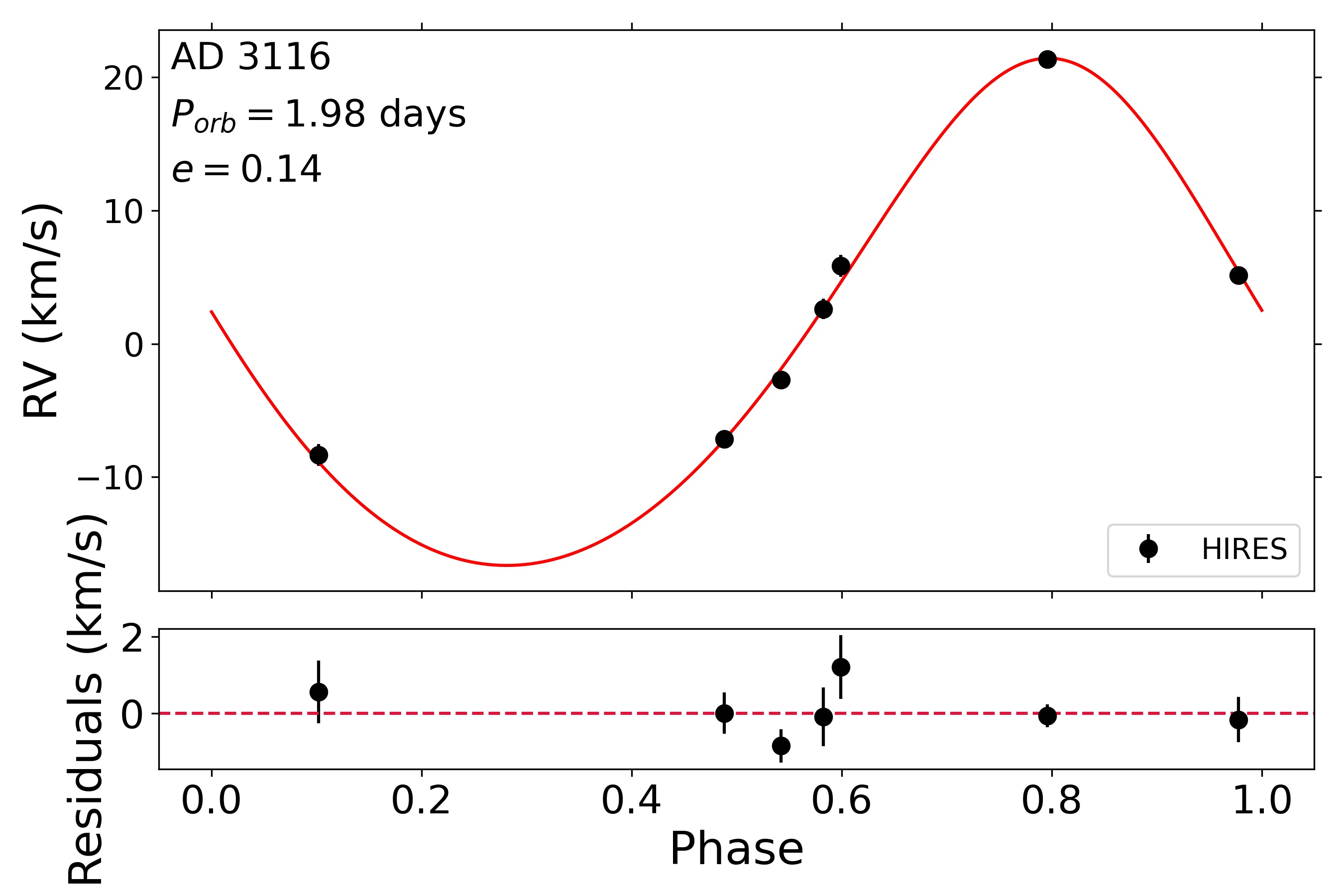}
    \includegraphics[width=0.33\textwidth, trim={0.0cm 0.0cm 0.0cm 0.0cm}]{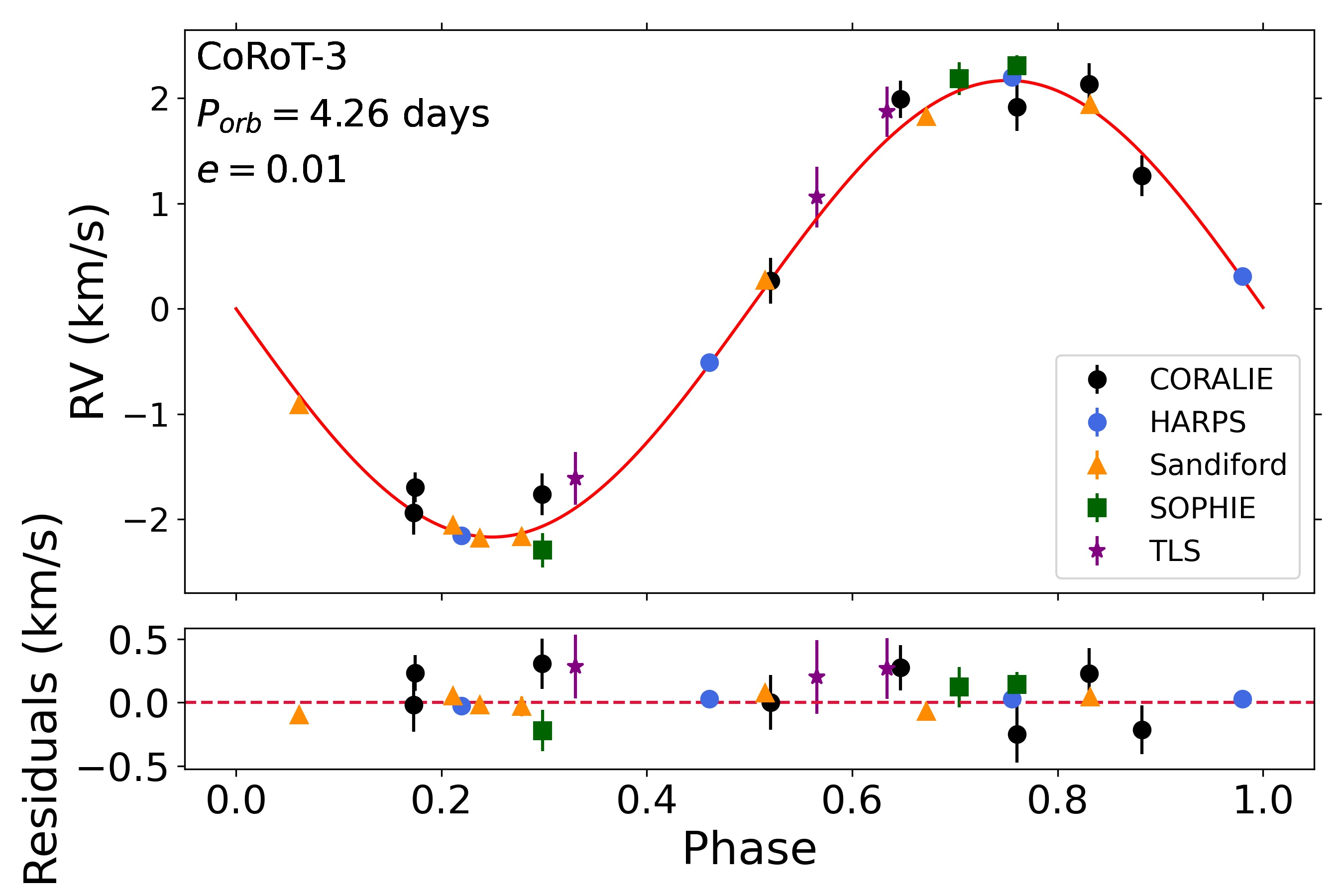}
    \includegraphics[width=0.33\textwidth, trim={0.0cm 0.0cm 0.0cm 0.0cm}]{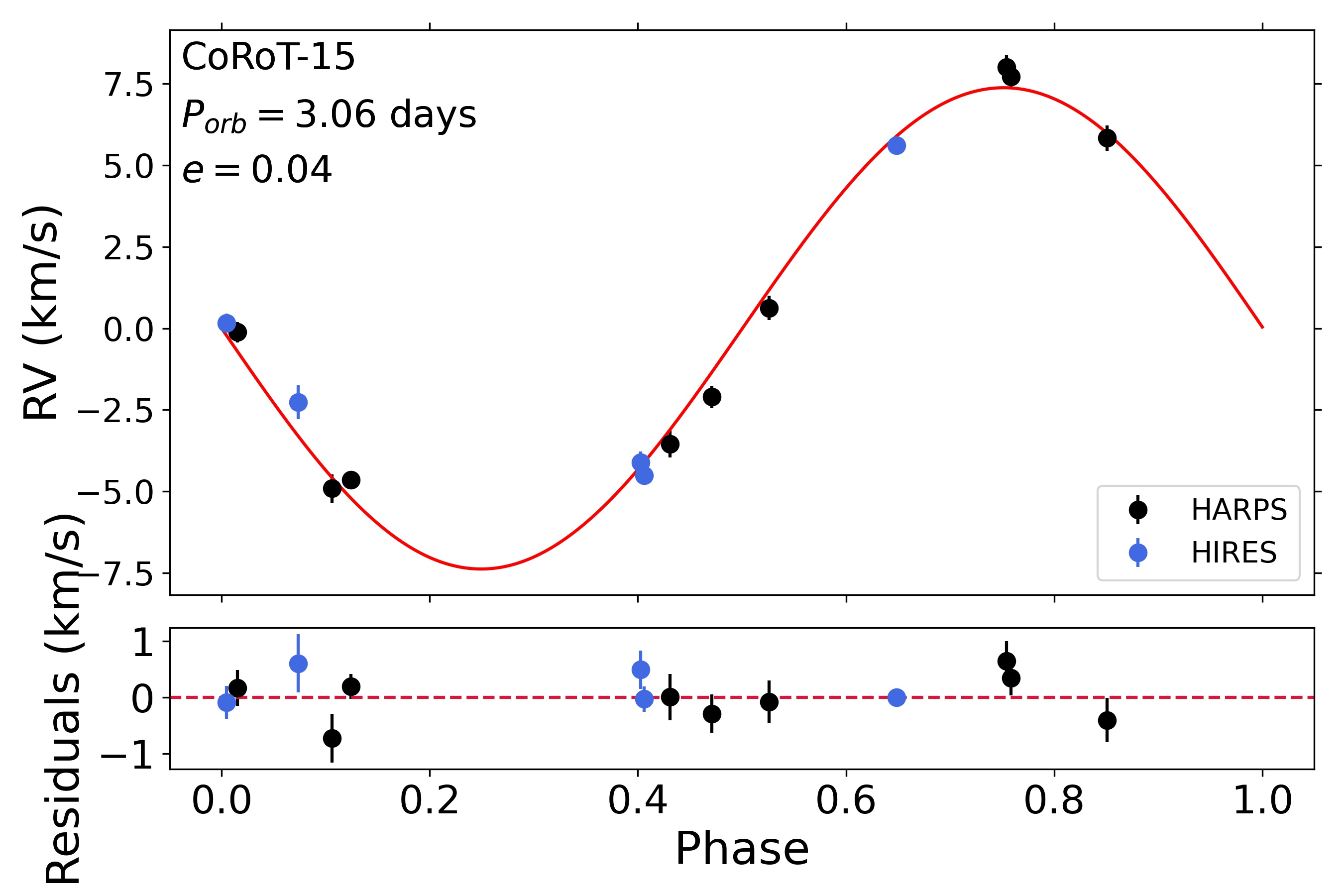}
    \includegraphics[width=0.33\textwidth, trim={0.0cm 0.0cm 0.0cm 0.0cm}]{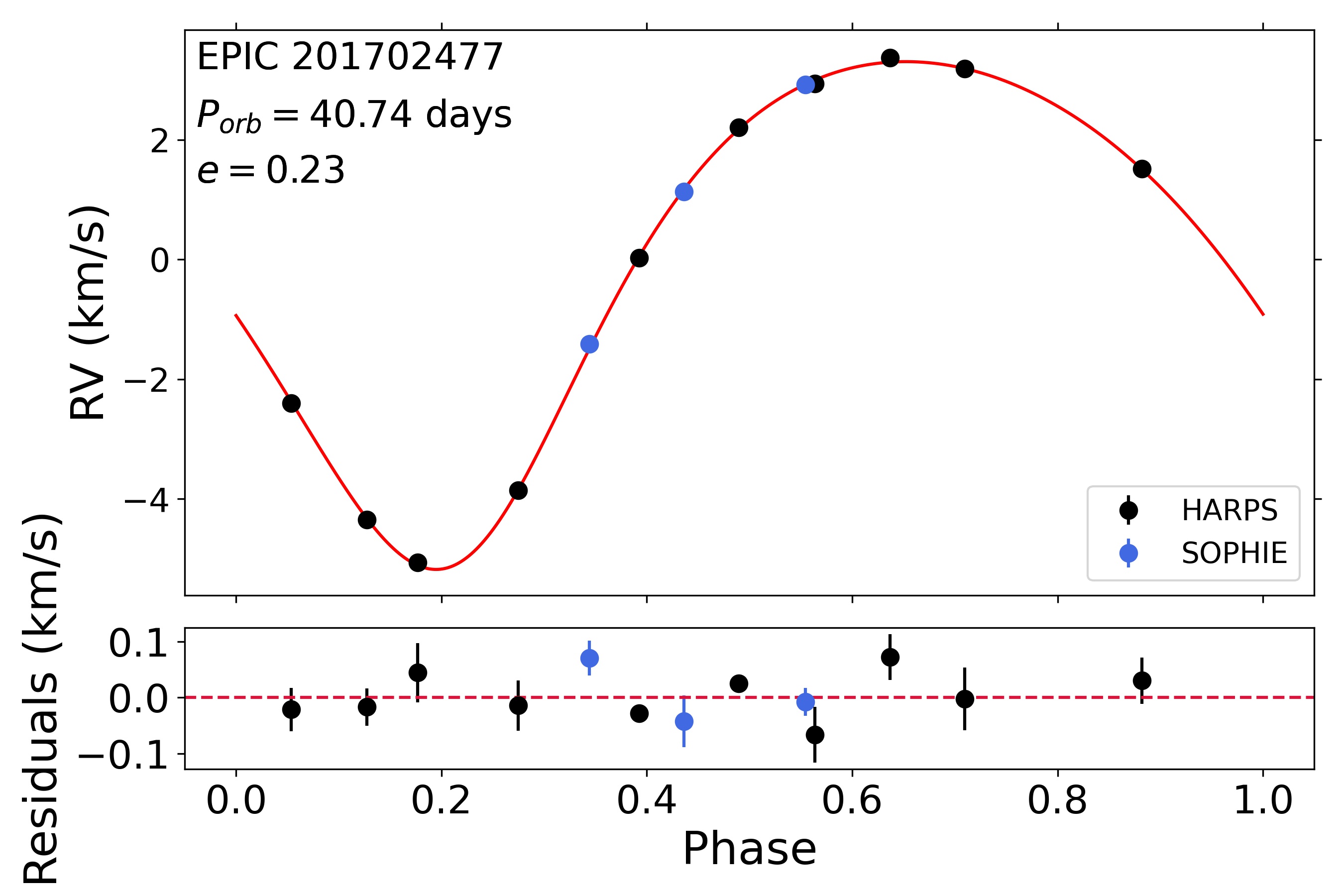}
    \includegraphics[width=0.33\textwidth, trim={0.0cm 0.0cm 0.0cm 0.0cm}]{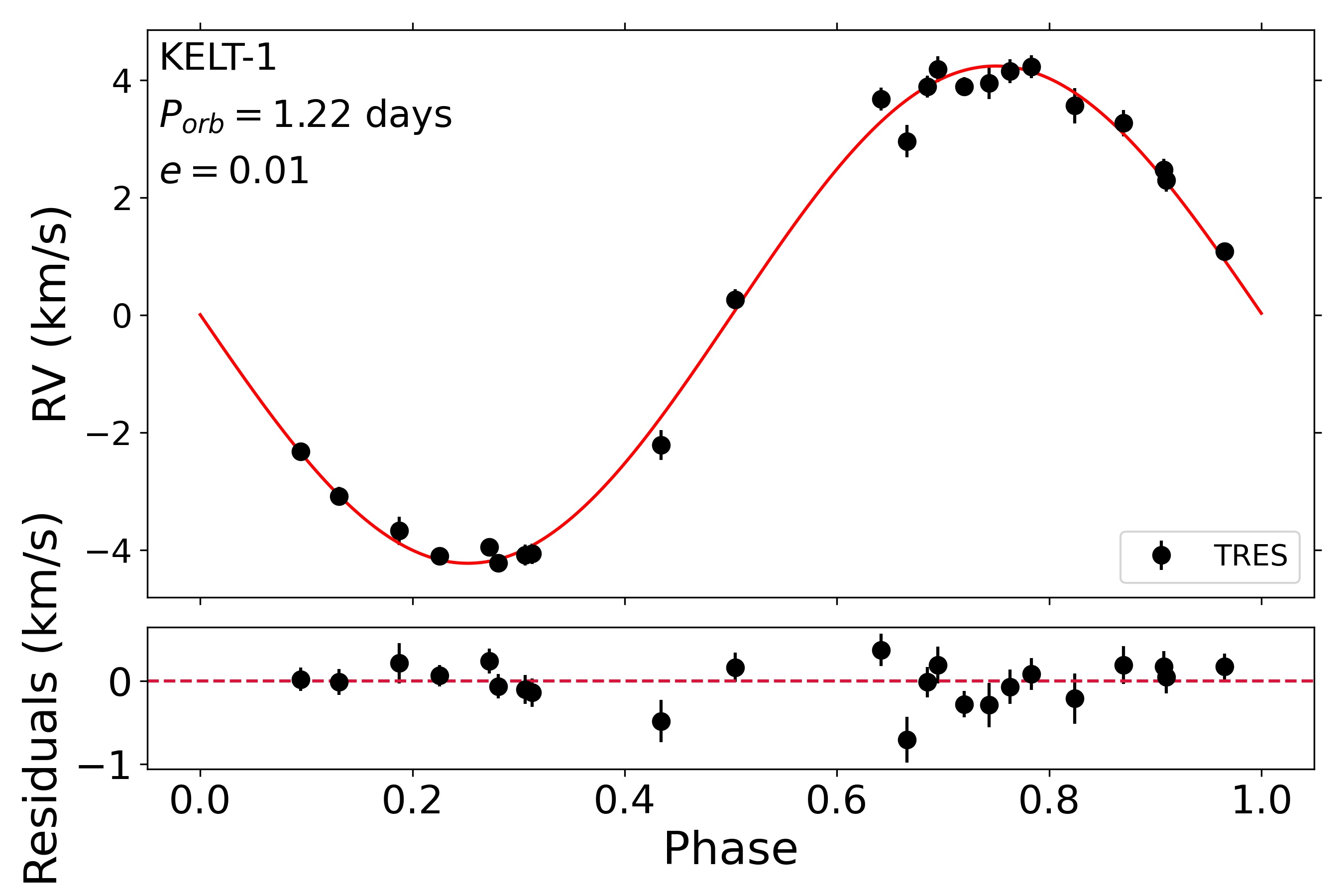}
    \includegraphics[width=0.33\textwidth, trim={0.0cm 0.0cm 0.0cm 0.0cm}]{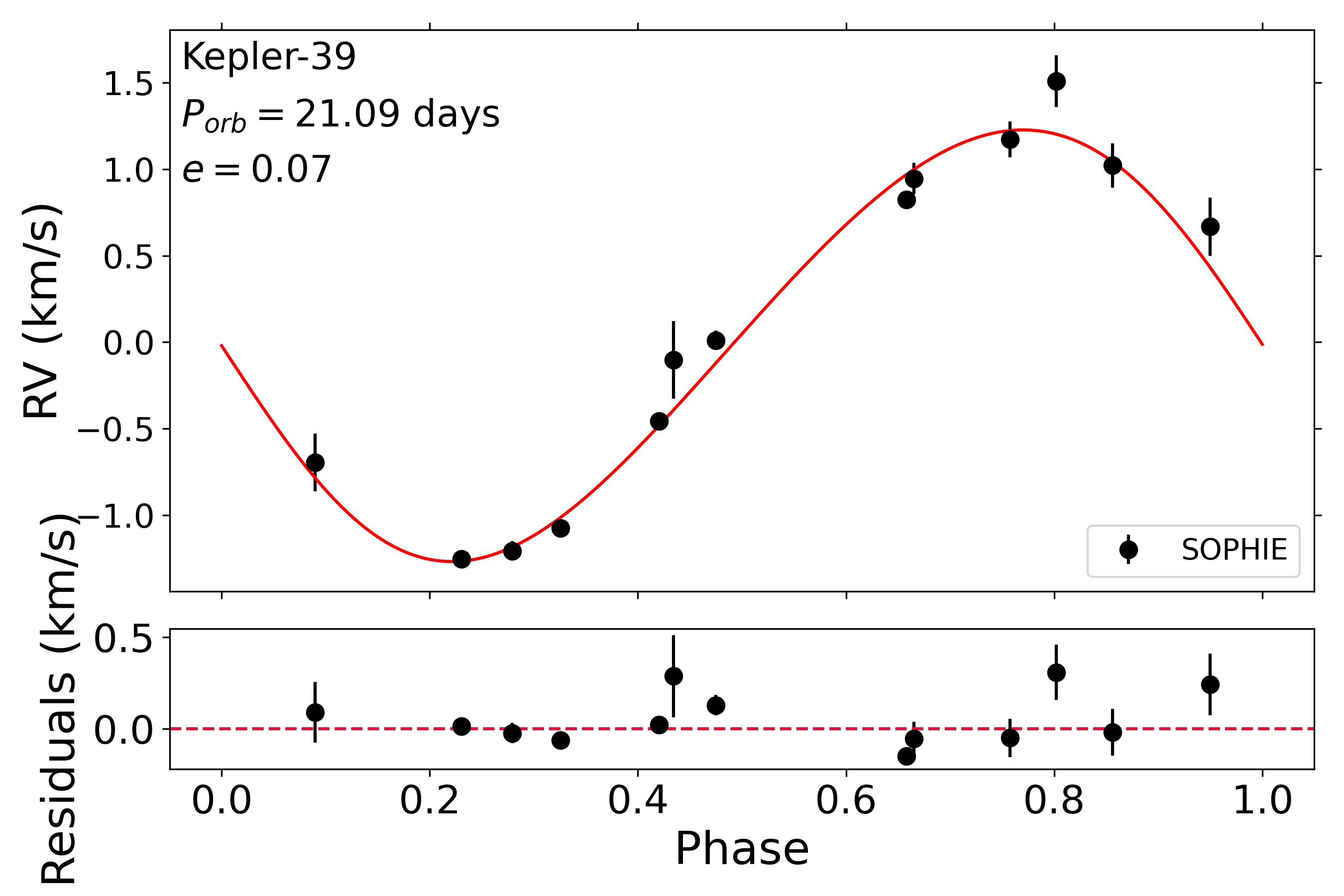}
    \includegraphics[width=0.33\textwidth, trim={0.0cm 0.0cm 0.0cm 0.0cm}]{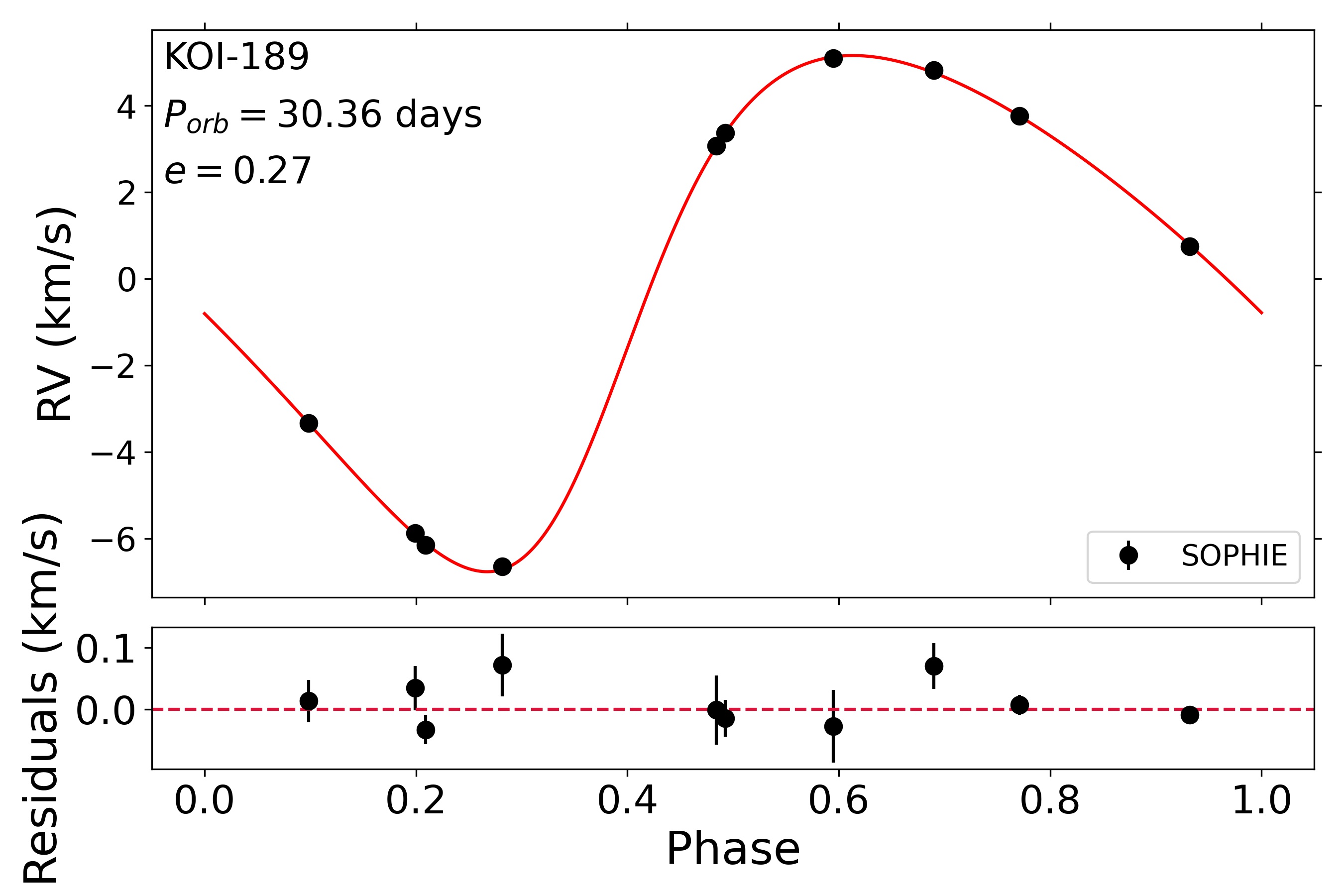}
    \includegraphics[width=0.33\textwidth, trim={0.0cm 0.0cm 0.0cm 0.0cm}]{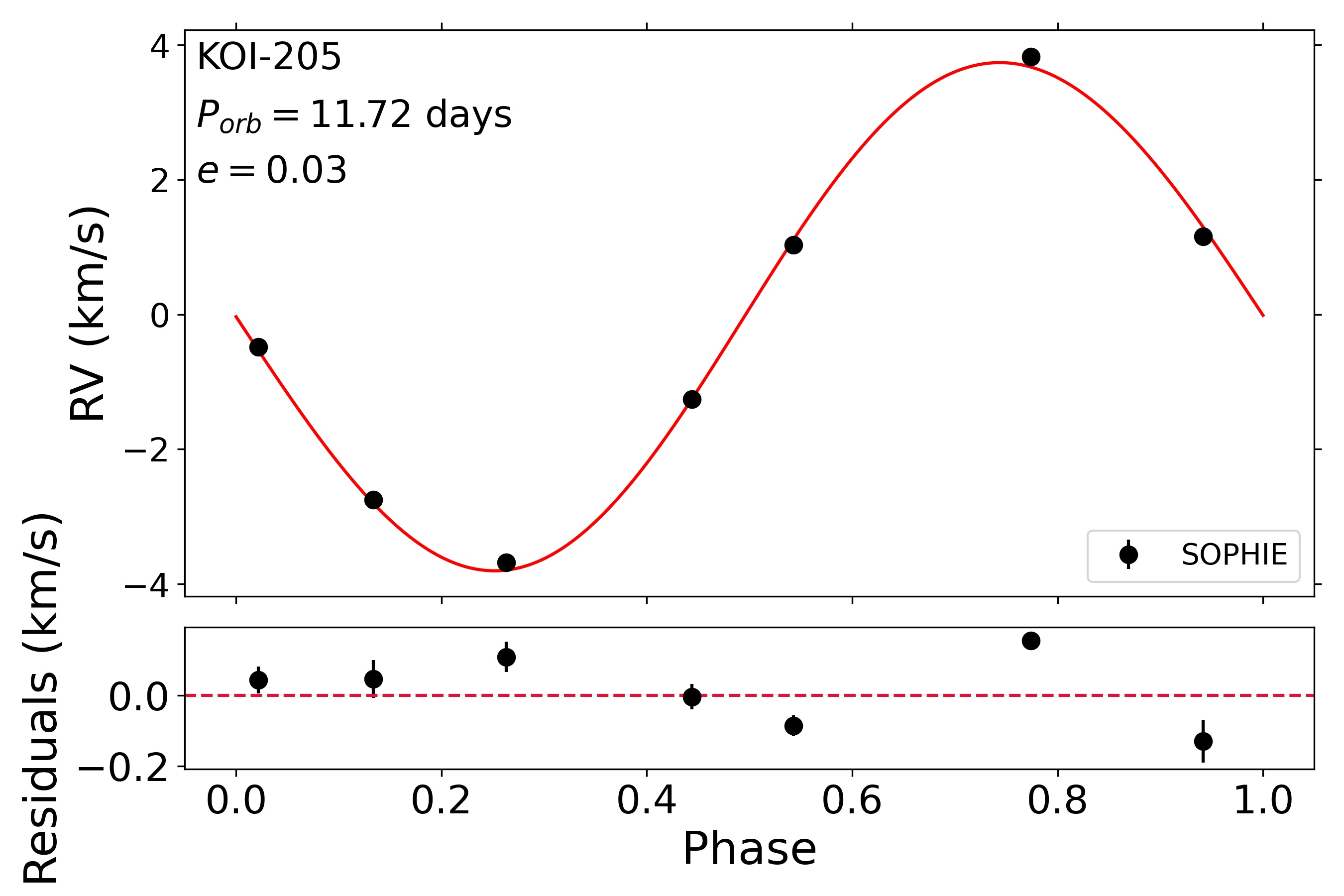}
    \includegraphics[width=0.33\textwidth, trim={0.0cm 0.0cm 0.0cm 0.0cm}]{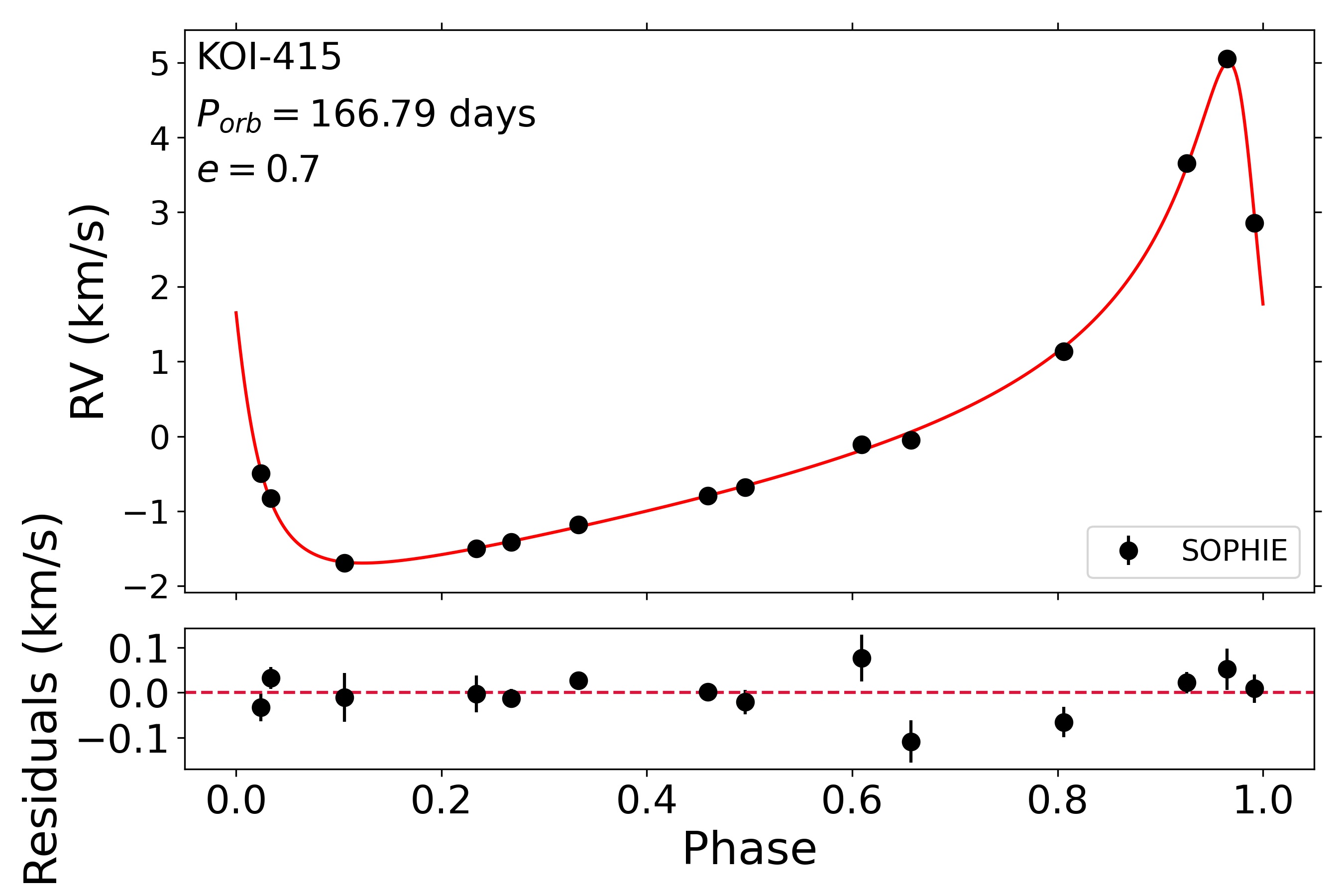}
    \includegraphics[width=0.33\textwidth, trim={0.0cm 0.0cm 0.0cm 0.0cm}]{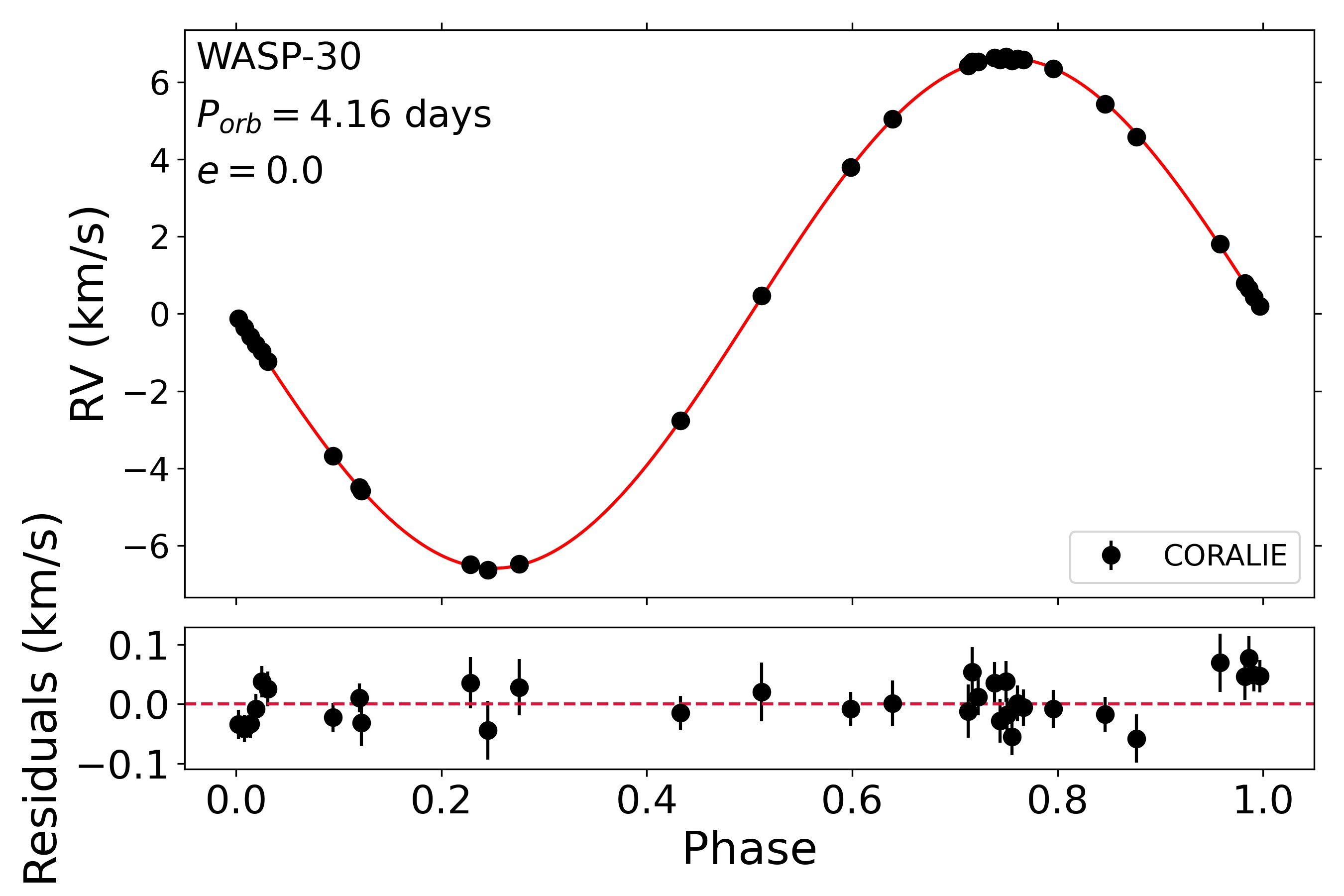}
    \includegraphics[width=0.33\textwidth, trim={0.0cm 0.0cm 0.0cm 0.0cm}]{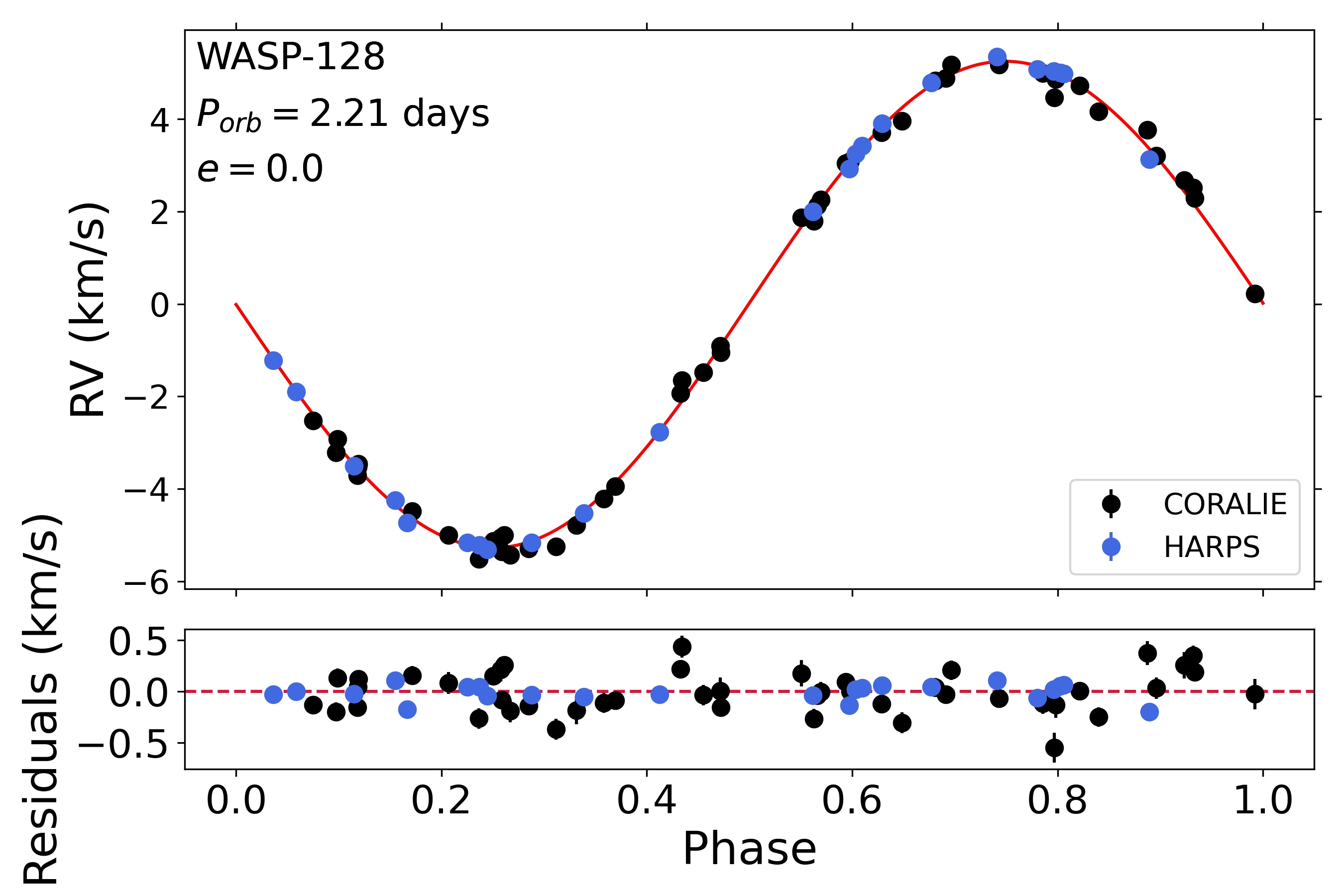}
    \caption{Relative radial velocity measurements for the systems analysed in this work. The red curve indicates the best-fit orbital solution. These data are sourced from their respective discovery works with the exception of KOI-189, which has had the BJD values altered due to a typographical error in the original work. Table \ref{tab:koi189_rvs} gives the corrected BJD values.}
    \label{fig:rvs}
\end{figure*}

Full tables describing priors and results are given in the Appendix section. Figure \ref{fig:seds} shows the SED models, Figure \ref{fig:transits} shows the transit models, and Figure \ref{fig:rvs} shows the RV orbital solutions.

\subsection{Treating the AD 3116 system}
Unlike every other system in this study, the AD 3116 system contains an M-dwarf primary star that is situated within a star cluster, Praesepe. This means that I may use prior information on the age of the system, $\tau = 617 \pm 17$ Myr from \cite{gossage2018}, and that I must invoke M-dwarf mass--magnitude relations from \cite{mann2019}. I use Equation 4 from \cite{mann2019} to set a mass prior based on the absolute $M_{K_S}$ magnitude, which I convert to from the apparent magnitude shown in Table \ref{tab:magnitudes_1}. This stellar mass prior is set in addition to the general priors mentioned in Section \ref{sec:exofast}.

\begin{table}
\centering
 \caption[]{Comparison of transiting brown dwarf radius values ($\rm R_b$) of this work to pre-Gaia DR2 work (except for WASP-128b, which was published previously using Gaia DR2 data). The $\Delta\sigma_{\rj}$ column quantifies the difference between the previous radius measurements and the measurements using data from Gaia DR3 in this work. Values that are most consistent with each other have $\Delta\sigma_{\rj} < 1.0\sigma$. A visual representation of these results are given in the Appendix.} \label{tab:comparison}
 \begin{tabular}{lcccc}
 \hline
 Object name & Previous work ($\rj$) & This work ($\rj$) & $\Delta\sigma_{\rj}$\\
  \hline
 AD 3116b & $1.02\pm 0.28 $        & $0.95\pm 0.07 $  & $0.98\sigma$ \\[1pt]
 CoRoT-3b & $1.01\pm 0.07 $        & $1.08\pm 0.05 $  & $1.44\sigma$\\[1pt] % 54
 CoRoT-15b & $1.12\pm 0.30 $       & $0.94\pm 0.12 $ &  $1.83\sigma$ \\[1pt]
 EPIC 201702477b & $0.76\pm 0.07 $ & $0.83\pm 0.04 $ & $1.84\sigma$ \\[1pt]
 KELT-1b & $1.11\pm 0.03 $         & $1.13\pm 0.03 $   & $0.66\sigma$ \\[1pt] %\ 57
 Kepler-39b & $1.22\pm 0.12$       & $1.07\pm 0.03$ & $5.67\sigma$\\[1pt] % 56
 KOI-189b & $1.00\pm 0.02 $        & $0.99\pm 0.02 $  & $0.49\sigma$ \\[1pt] % 54, 55
 KOI-205b & $0.81\pm 0.02 $        & $0.87\pm 0.02 $  & $2.61\sigma$ \\[1pt] % 54, 55
 KOI-415b & $0.79\pm 0.12 $        & $0.86\pm 0.03 $  & $2.80\sigma$ \\[1pt] % 54, 55
 WASP-30b & $0.95\pm 0.03 $        & $0.96\pm 0.03 $  & $0.43\sigma$ \\[1pt] % 69
 WASP-128b & $0.94\pm 0.02 $       & $0.96\pm 0.02 $ & $1.21\sigma$ \\[1pt] % 64
 \hline
 \end{tabular}
\end{table}

\begin{figure*}
    \centering
    \includegraphics[width=0.33\textwidth, trim={0.0cm 0.0cm 0.0cm 0.0cm}]{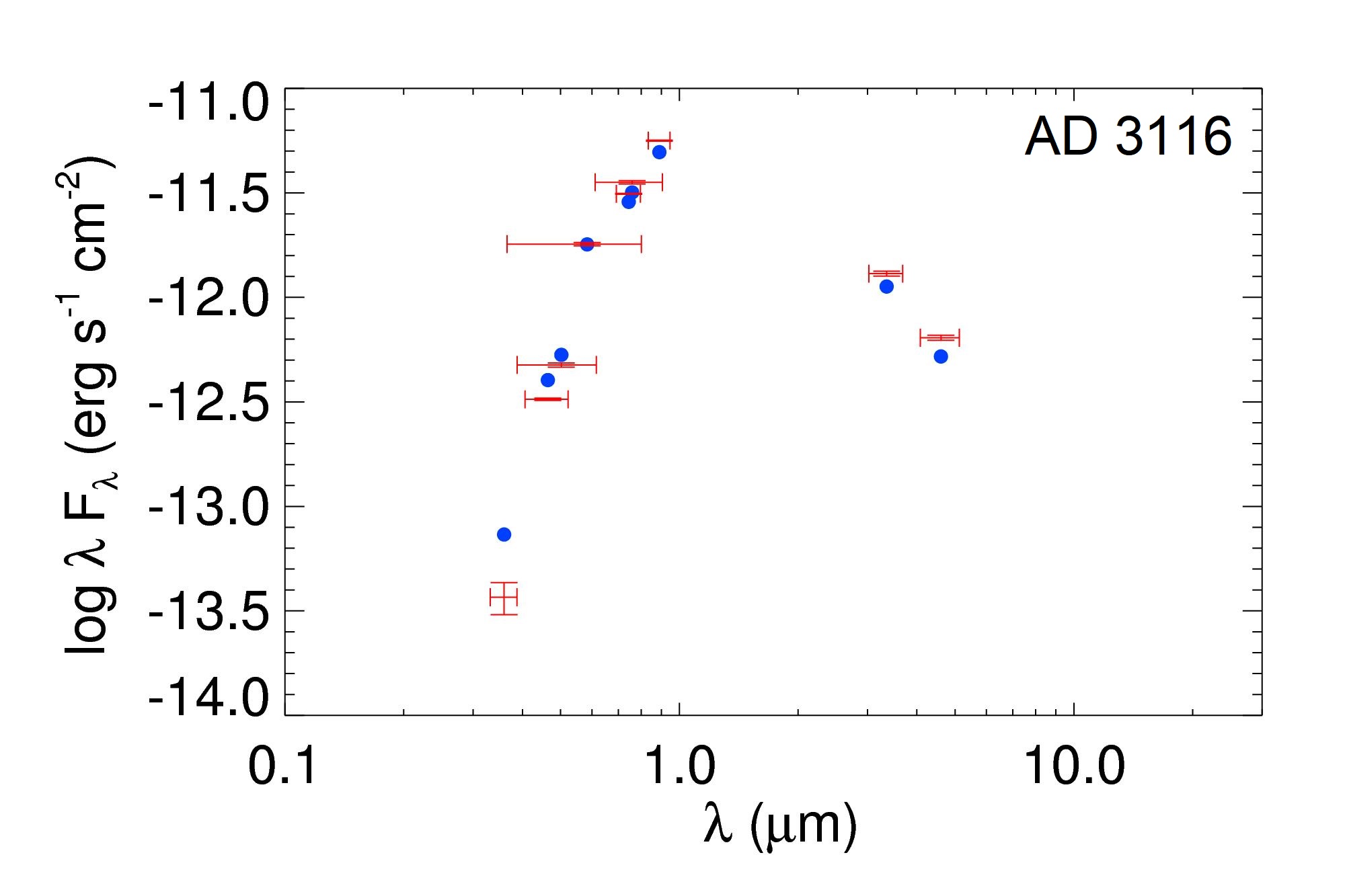}
    \includegraphics[width=0.33\textwidth, trim={0.0cm 0.0cm 0.0cm 0.0cm}]{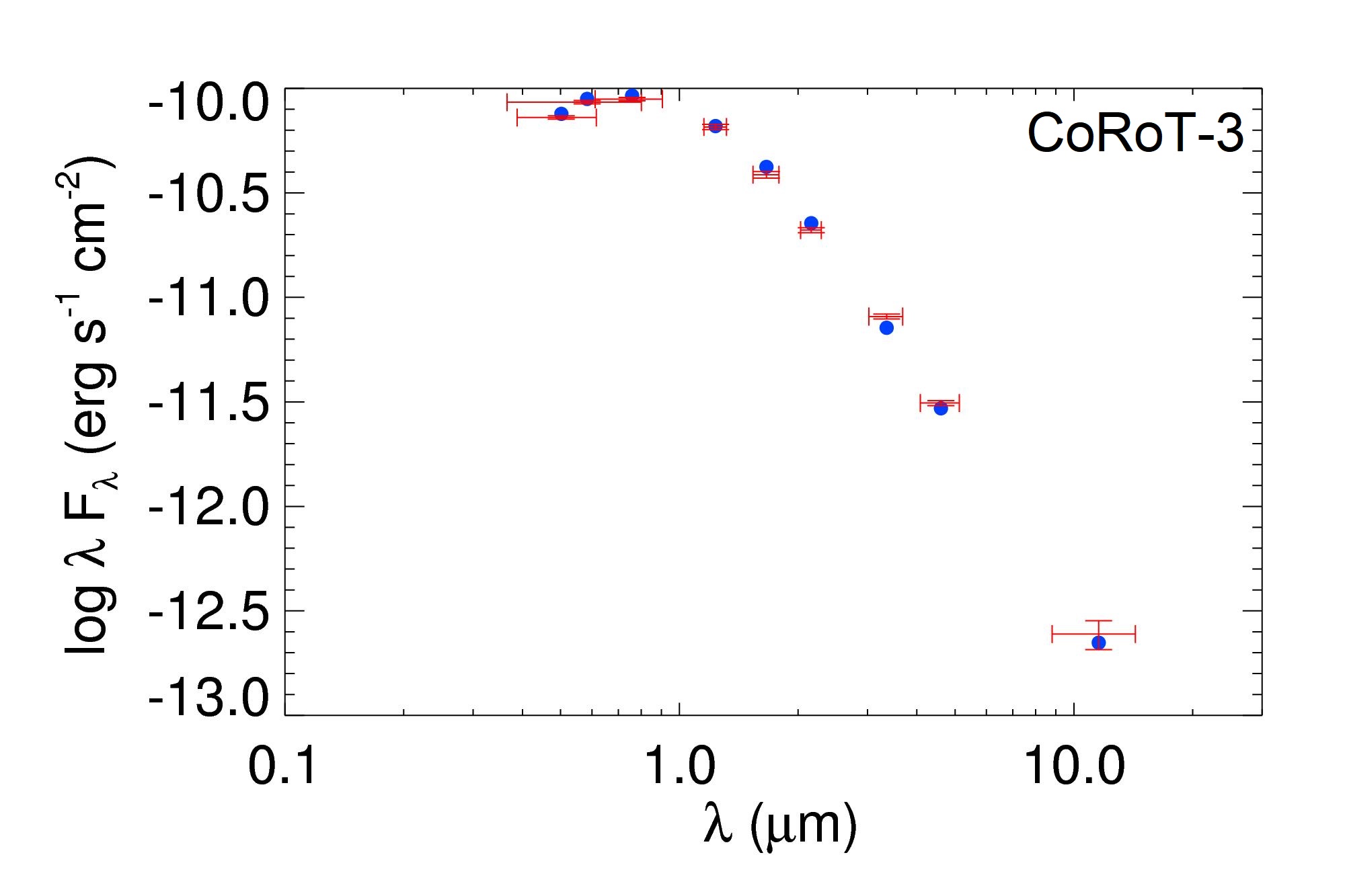}
    \includegraphics[width=0.33\textwidth, trim={0.0cm 0.0cm 0.0cm 0.0cm}]{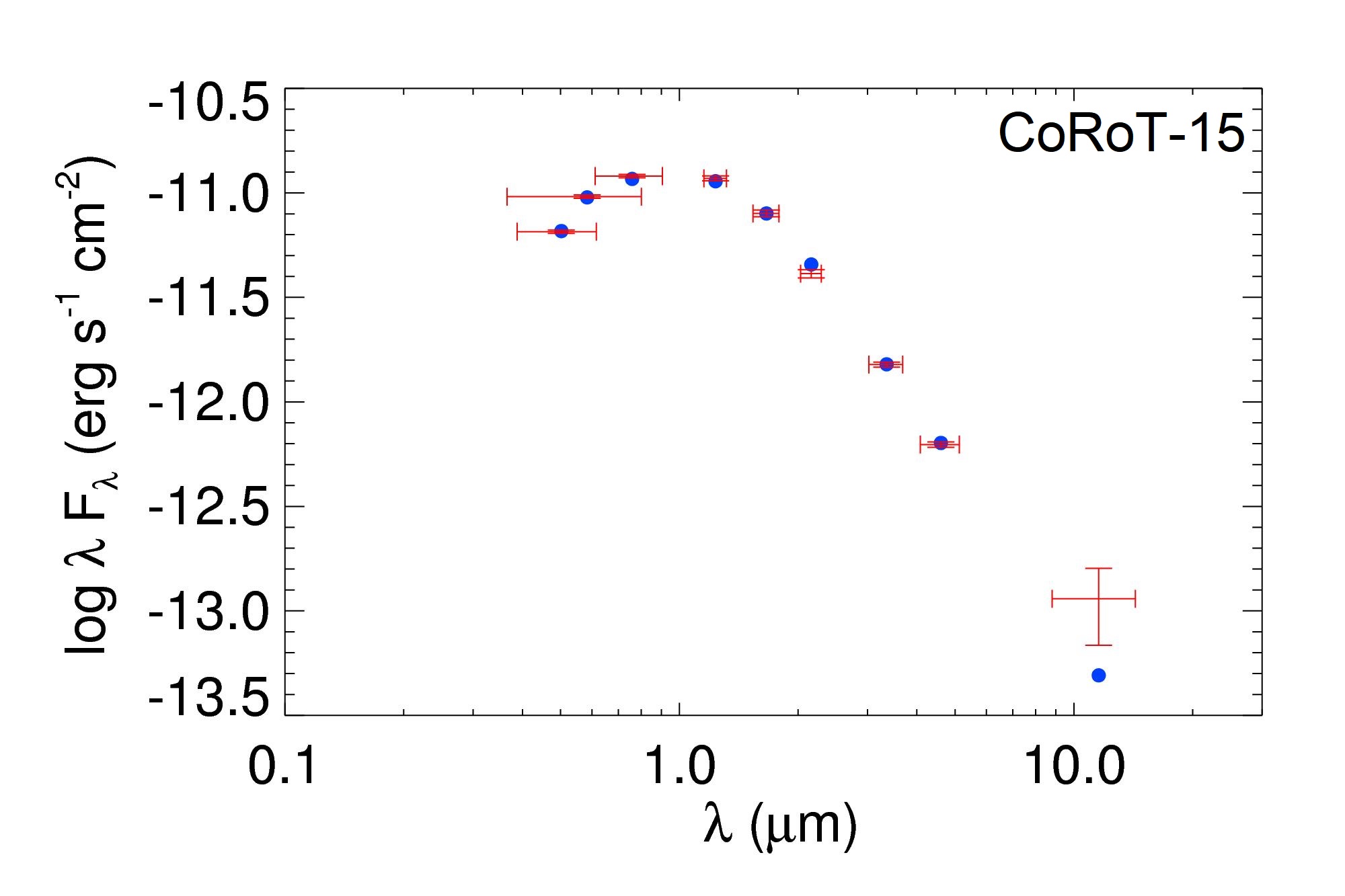}
    \includegraphics[width=0.33\textwidth, trim={0.0cm 0.0cm 0.0cm 0.0cm}]{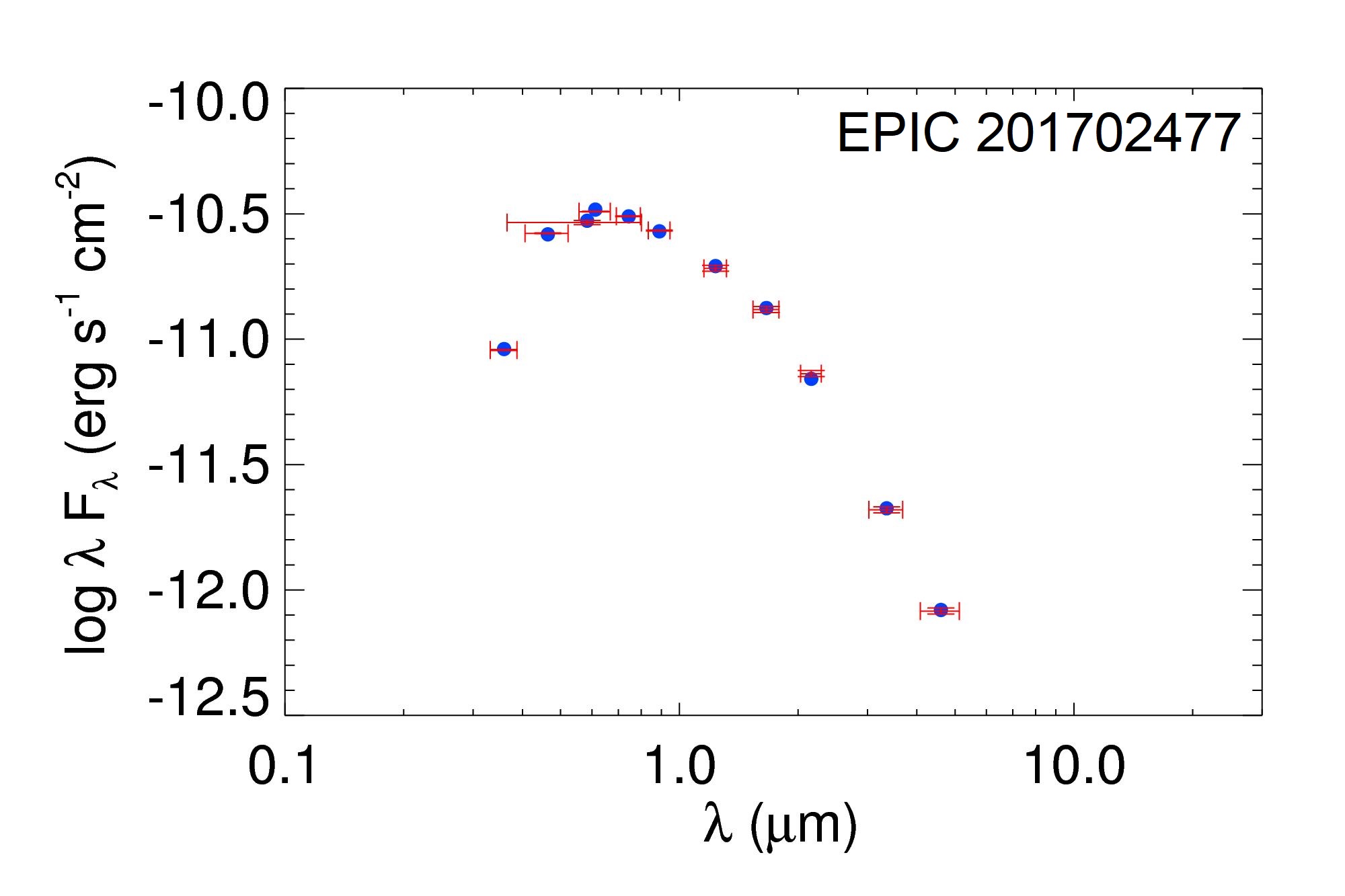}
    \includegraphics[width=0.33\textwidth, trim={0.0cm 0.0cm 0.0cm 0.0cm}]{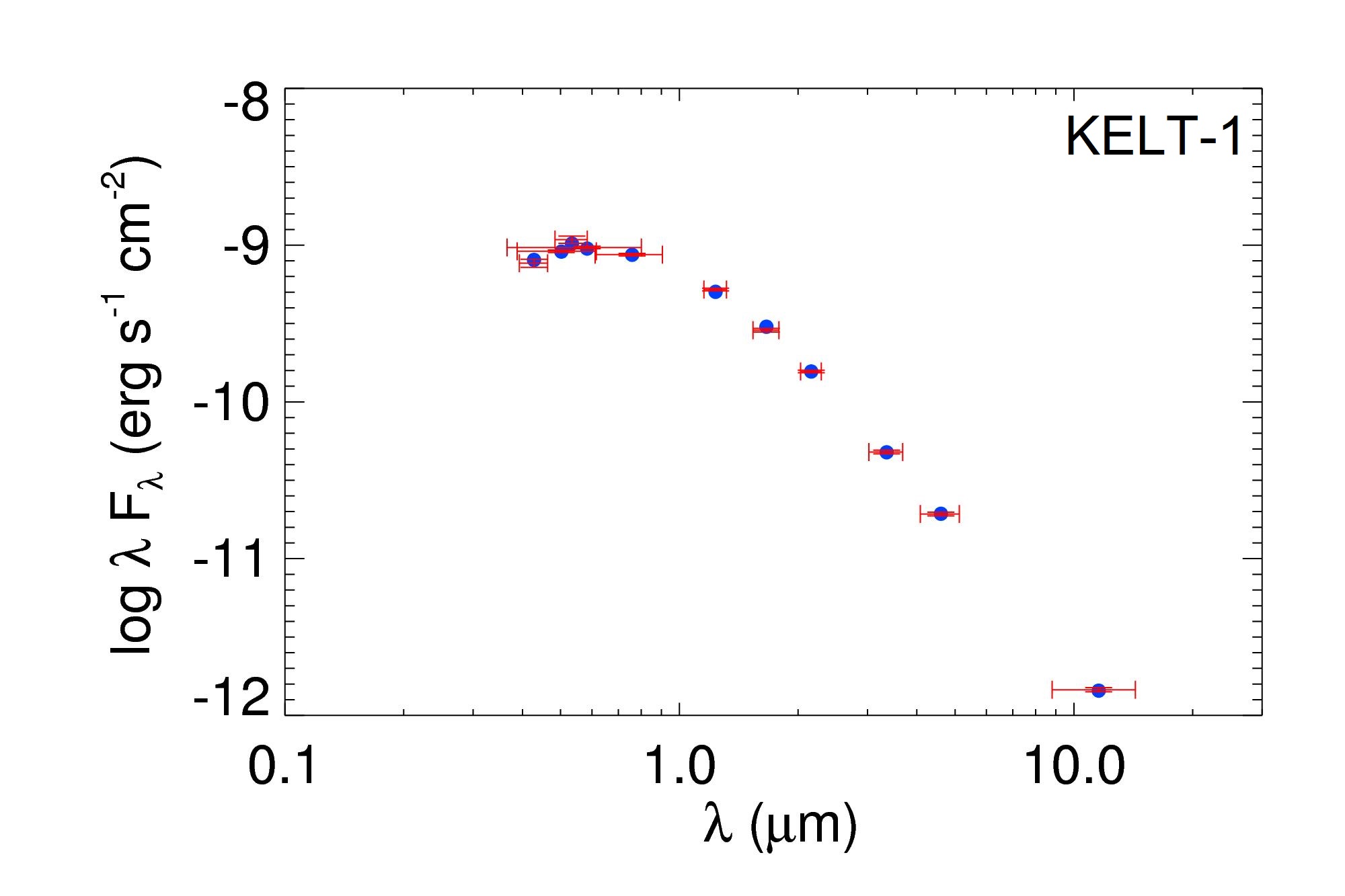}
    \includegraphics[width=0.33\textwidth, trim={0.0cm 0.0cm 0.0cm 0.0cm}]{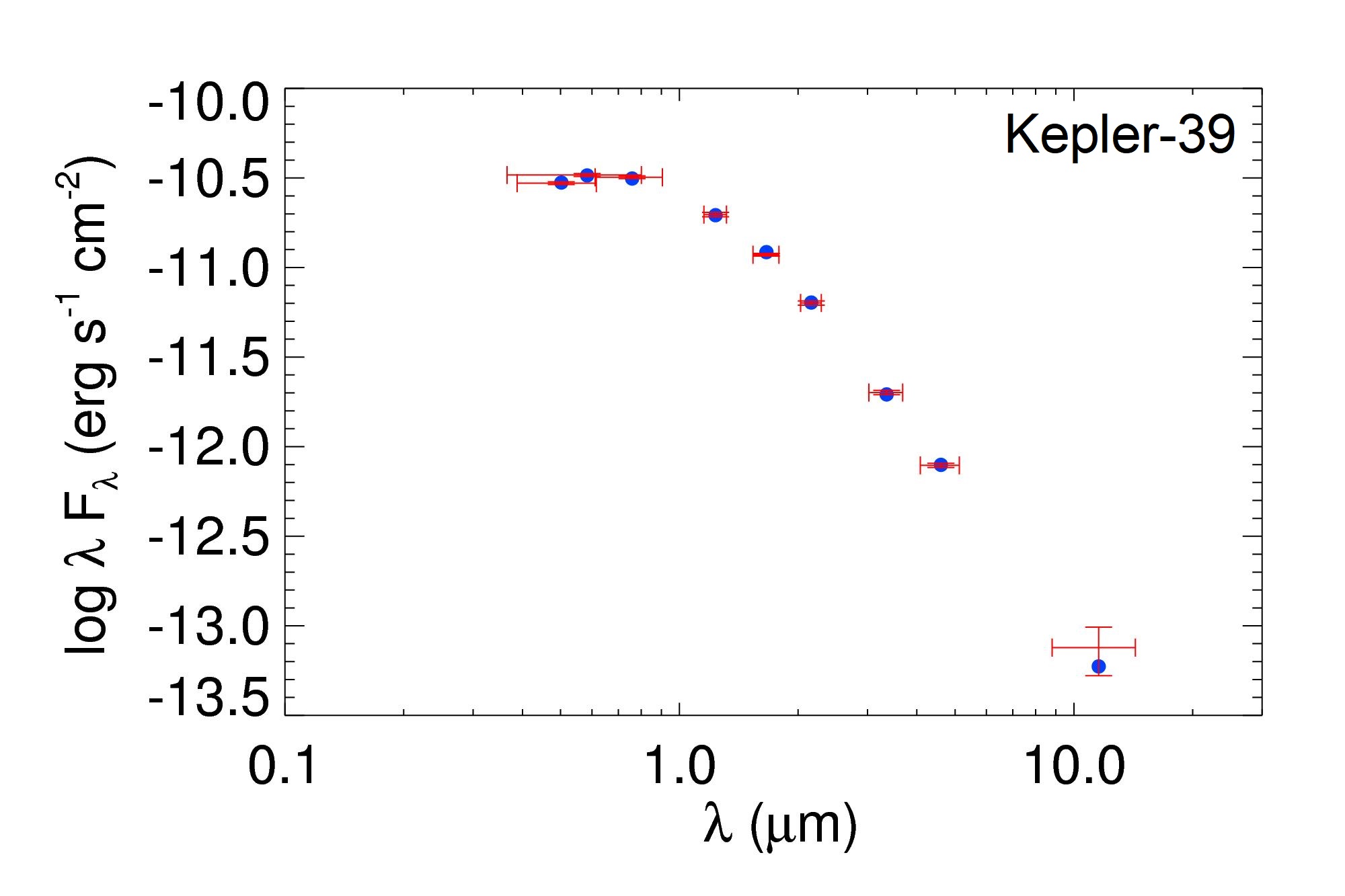}
    \includegraphics[width=0.33\textwidth, trim={0.0cm 0.0cm 0.0cm 0.0cm}]{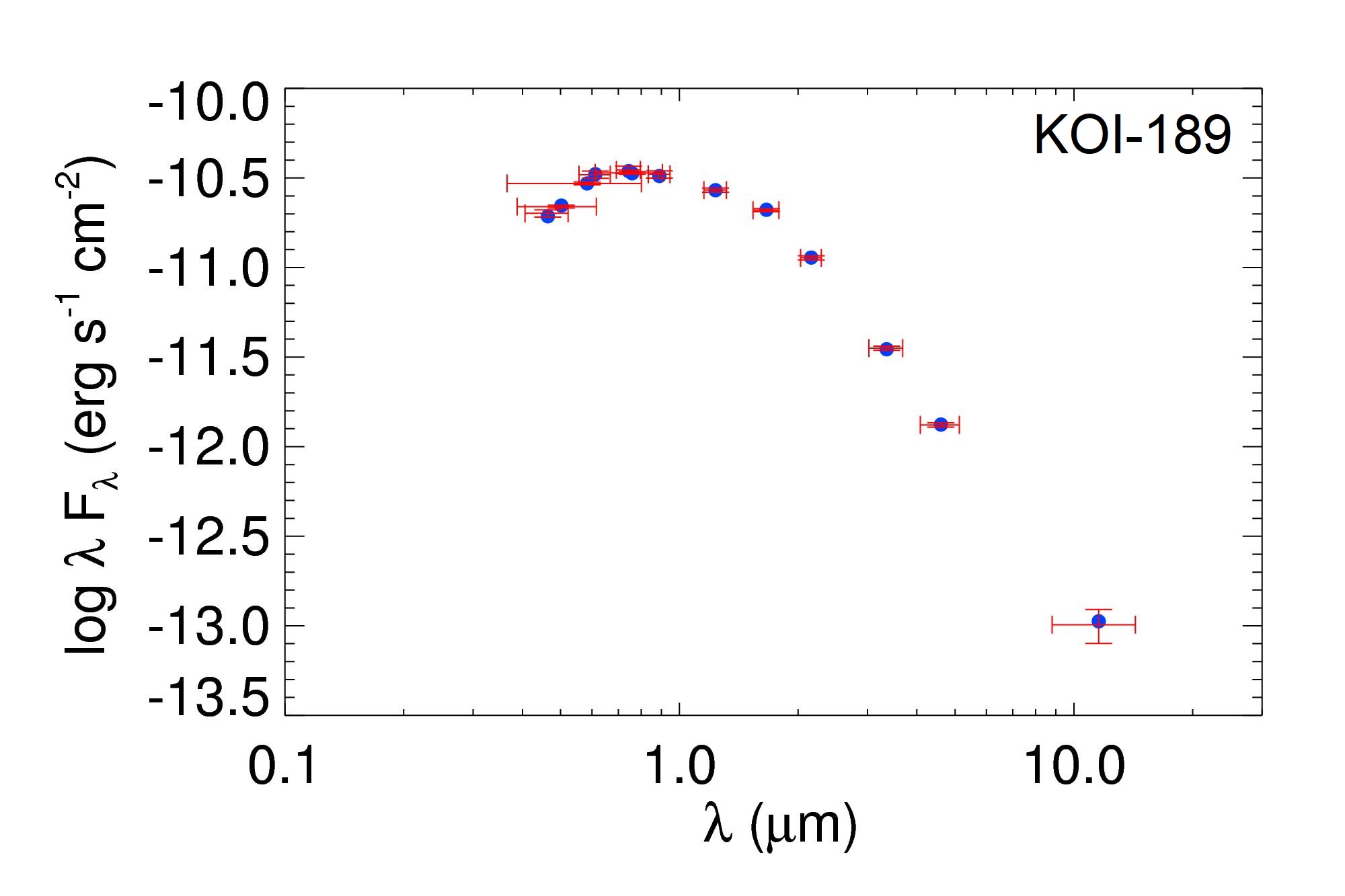}
    \includegraphics[width=0.33\textwidth, trim={0.0cm 0.0cm 0.0cm 0.0cm}]{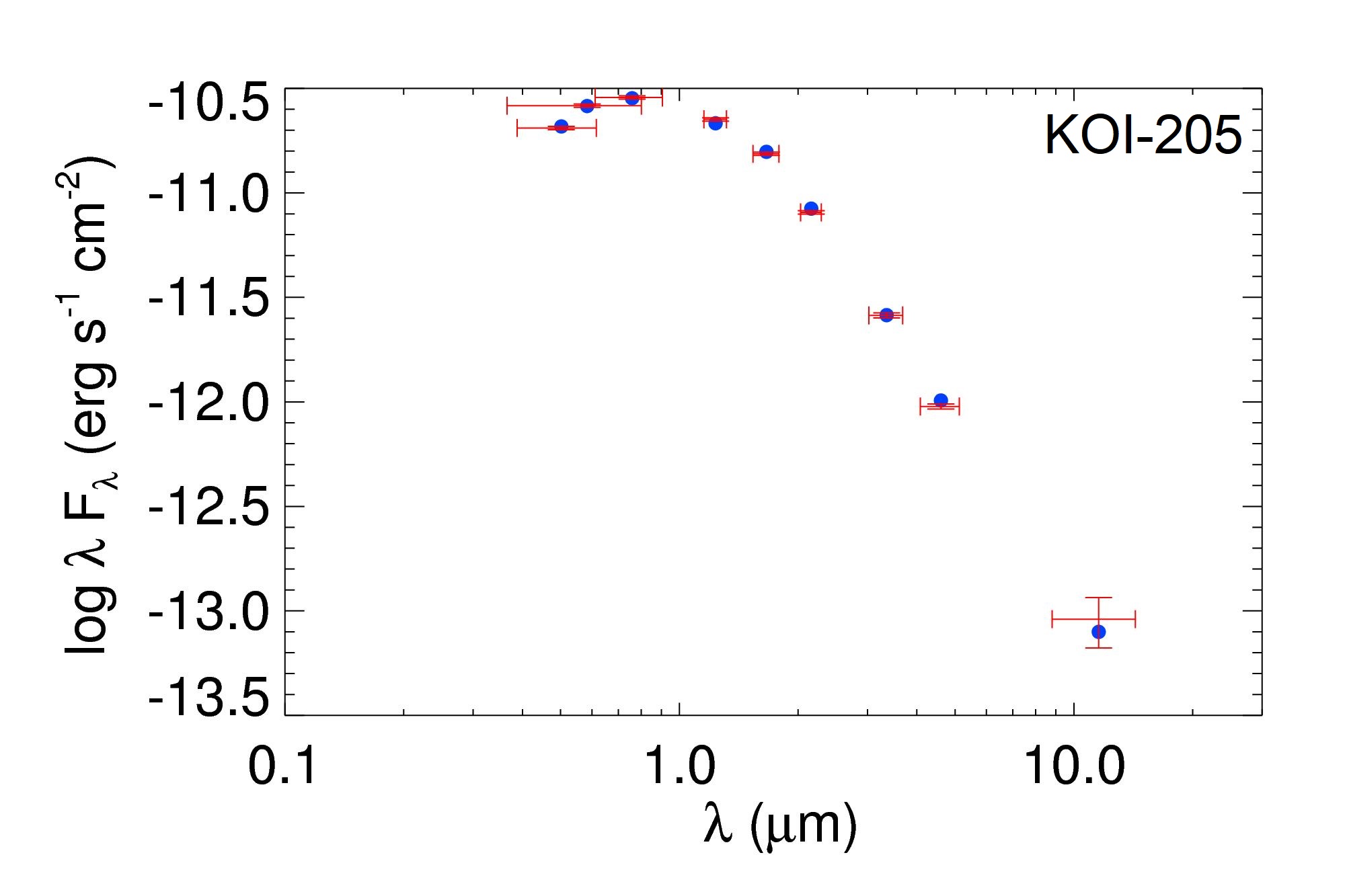}
    \includegraphics[width=0.33\textwidth, trim={0.0cm 0.0cm 0.0cm 0.0cm}]{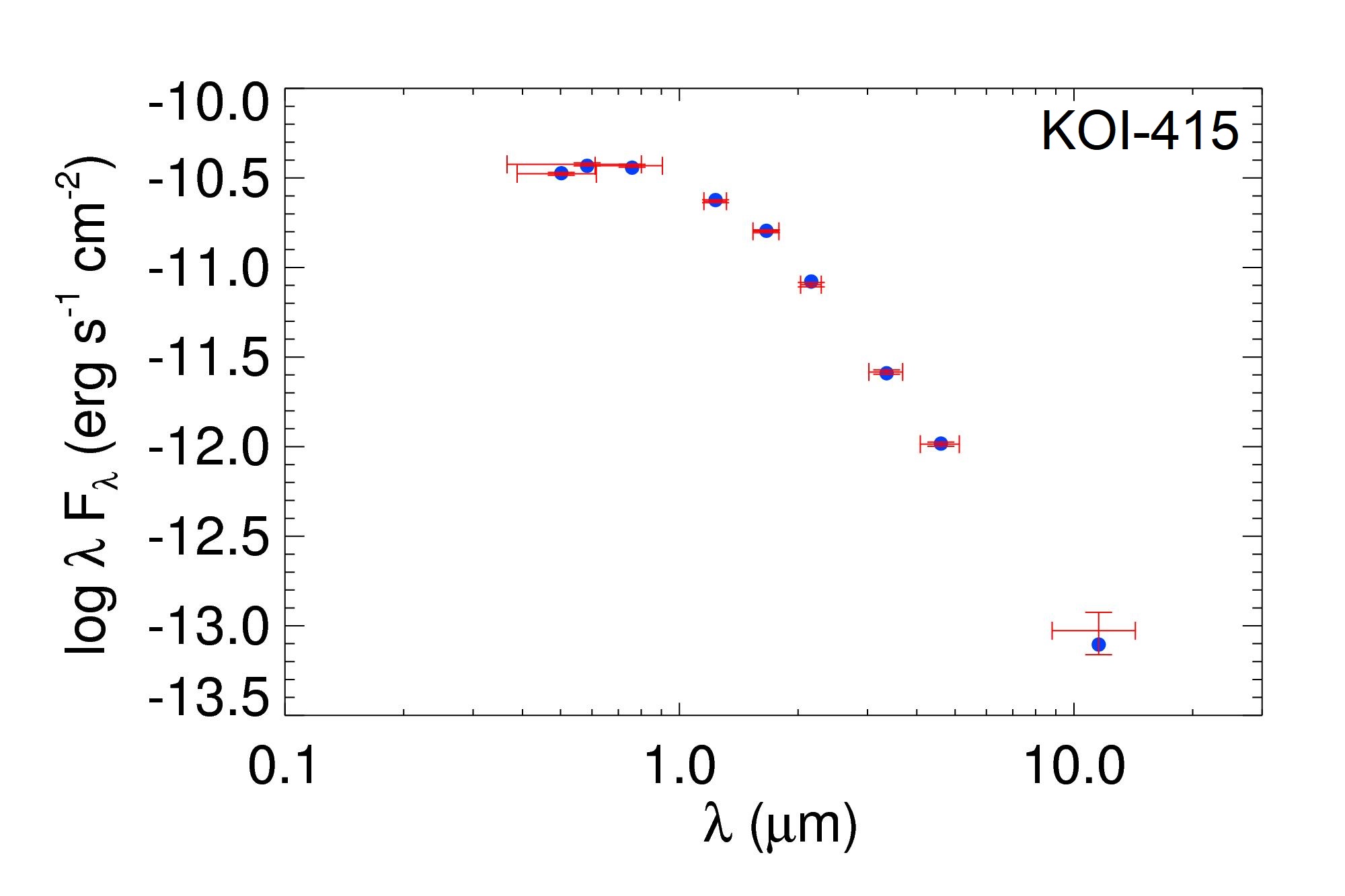}
    \includegraphics[width=0.33\textwidth, trim={0.0cm 0.0cm 0.0cm 0.0cm}]{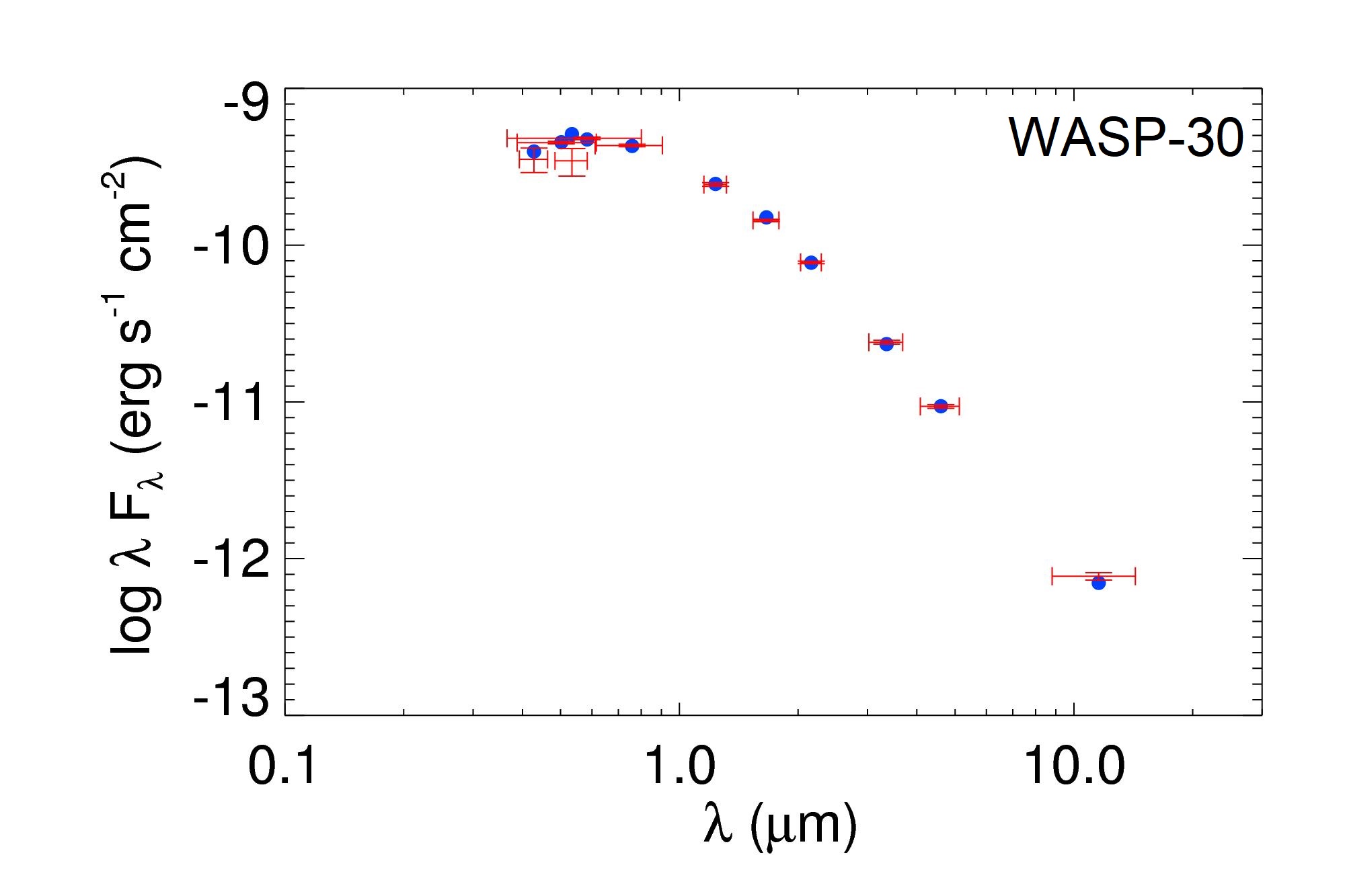}
    \includegraphics[width=0.33\textwidth, trim={0.0cm 0.0cm 0.0cm 0.0cm}]{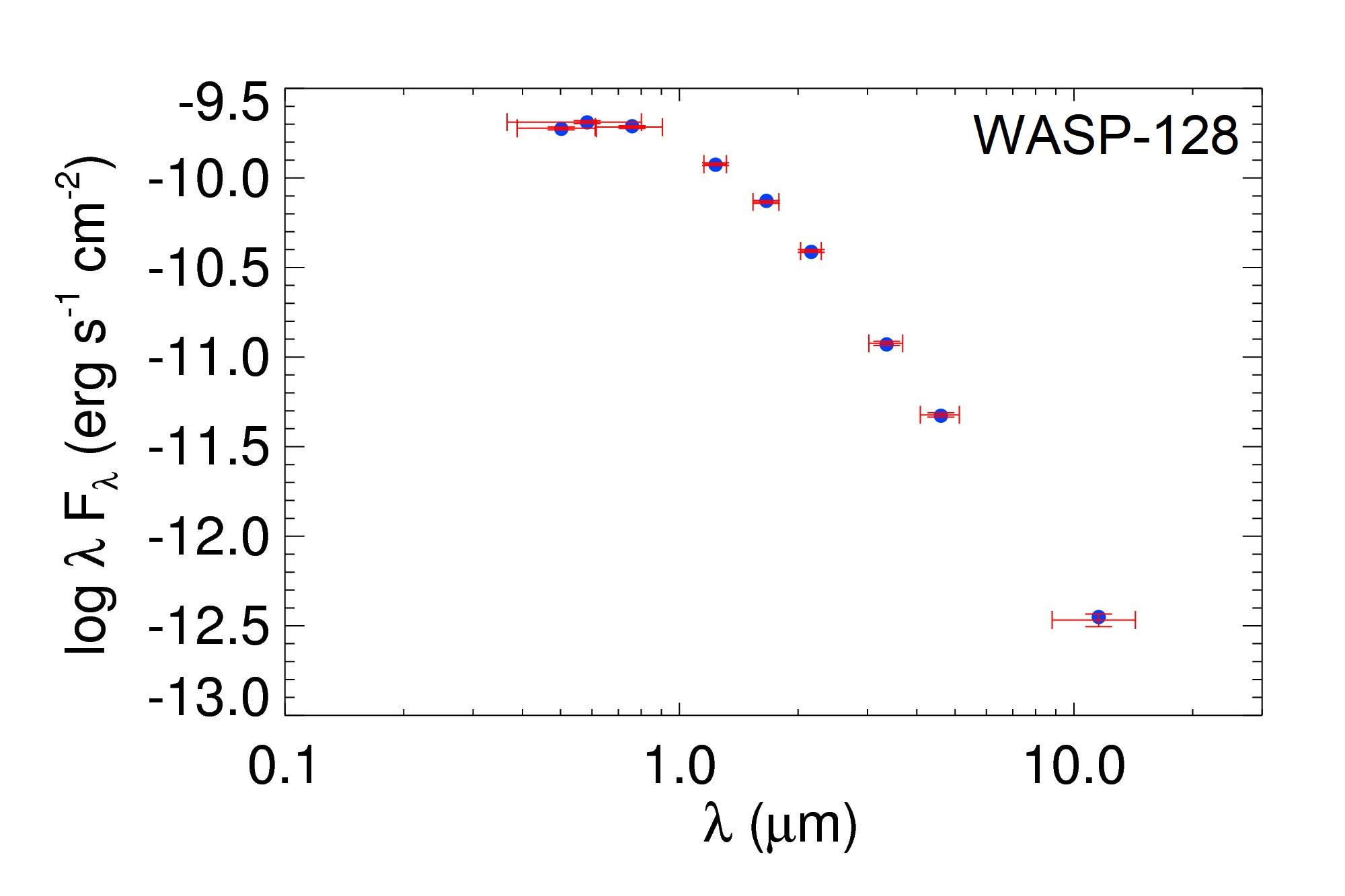}
    \caption{SEDs computed using the MIST models built into {\tt EXOFASTv2}. The blue dots represent the model values and the red crosses are the data from Table \ref{tab:magnitudes_1} and Table \ref{tab:magnitudes_2}. The most challenging SED to model is that of AD 3116, given that the host star's mass places it near the lower boundaries of what the MIST framework was designed for.}
    \label{fig:seds}
\end{figure*}

%%%%%%%%%%%%%%%%%%%%%%%%%%%%%%%%%%%%%%%%%%%%%%%%%%%%%%%%%%%%%%
%%%%%%%%%%%%%%%%%%%%%%%%%% DISCUSSION %%%%%%%%%%%%%%%%%%%%%%%%
%%%%%%%%%%%%%%%%%%%%%%%%%%%%%%%%%%%%%%%%%%%%%%%%%%%%%%%%%%%%%%
\section{Discussion}\label{sec:conclusion}
This work has seen an improvement in the mass--radius determinations for 7 transiting BDs: AD 3116b, CoRoT-3b, CoRoT-15b, EPIC 201702477b, Kepler-39b, KOI-205b, and KOI-415b. These systems show a reduction in their BD radius uncertainty, a $1\sigma$ or more change in their BD radius, or both of these, which changes the overall composition of the substellar mass--radius diagram (Figure \ref{fig:mass_radius_dr3}). For the systems whose mass--radius measurements I find to be consistent with previous works (KELT-1, KOI-189, WASP-30, WASP-128), a notable improvement to those are the updated transit timings with \tess data. These updated ephemerides will reduce the mid-transit time uncertainty for future works.

I focus on quantifying these improvements in terms of the BD radius as this is a characteristic by which BDs change significantly over their lifetimes and is also measurable via transit photometry. Now that astronomers are able to measure the radii of BDs to a precision of 3-5\%, it is important that substellar evolutionary models have a similar or better precision for the radii they predict. If these models can accurately and precisely predict the radius of a BD in scenarios where the age is known, then other radius-altering factors like metallicity and inflation via stellar irradiation can be more easily explored. This is especially relevant in cases where low-mass transiting BDs may experience inflation caused by irradiation \citep[e.g.][]{kelt1b, benni2020}. Table \ref{tab:comparison} provides a breakdown of the changes in BD radius values from previous works to this work. This focus on radius is not meant to imply that the mass of these BDs is unimportant, but in general, the mass of the BD is fairly static over the lifetime of the object and age--mass relations show little change compared to age--radius relations for these objects.

I also highlight the specific systems that show significant changes in agreement with age--radius predictions from substellar evolutionary models from \cite{ATMO2020}, referred to as the ATMO 2020 models, and the \cite{baraffe03} models, referred to as the COND03 models. These systems are Kepler-39, KOI-205, KOI-415, and EPIC 201702477. The changes in agreement are sometimes in favor of the models and sometimes not, depending on the system. Though other frameworks for substellar evolutionary models exist and take a different approach to BD radius evolution \citep[e.g.][]{sonora21}, I limit my comparisons to the ATMO 2020 and COND03 models for the sake of simplicity and to focus broadly on how these types of models match up against the transiting BD population, especially the oldest and smallest known transiting BDs.

\subsection{Specific changes to each brown dwarf system}
Here I will provide details for each system and, where appropriate, highlight the sources of significant changes in the radius or radius precision of these transiting BD systems. In several cases, my results are consistent with previous works, so those original works should be used when referencing the stellar and BD parameters for those systems. 

\subsubsection{AD 3116}
The original discovery paper for AD 3116 by \cite{ad3116} uses the parallax measurement to the star cluster, Praesepe, from Gaia DR1, which translates to a distance of $d = 182.8 \pm 1.7 \pm 14$ pc, where the second uncertainty measurement represents the observed radial spread of high-probability members of the cluster on the sky. In \cite{ad3116}, they choose $d = 182.8 \pm 14$ pc for this system and so are limited to a worse precision on the radius of the host star and radius of the BD. Using the relations from \cite{mann2019} along with the Gaia DR3 parallax measurement ($\varpi = 5.39\pm 0.14$ mas, $d=185.4\pm 4.8$ pc), I find an uncertainty on the stellar radius that is a factor of roughly 3 times smaller than that in \cite{ad3116}, which translates to a 3-4 times reduction in the BD radius uncertainty. I find a more precise radius measurement for AD 3116b to be $R_b=0.95\pm 0.07\rj$, which also improves its consistency with age--radius relations for BDs between 500-600 Myr from ATMO 2020 and COND03. The mass of the BD that I find, $M_b = 54.6 \pm 6.8\mj$, is slightly less precise, but still consistent with the BD mass found in \cite{ad3116} ($M_b = 54.2 \pm 4.3\mj$).

\subsubsection{CoRoT-3}
CoRoT-3 is the first known transiting BD and in \cite{corot3b}, they report a stellar distance and radius of $680 \pm 160$ pc and $R_\star = 1.56 \pm 0.09\rst$, respectively. The distance derived from Gaia DR3 in this work is $764 \pm 10$ pc, which effectively translates to a larger host star ($R_\star=1.66\pm 0.08\rst$) at a further distance compared to the original findings. The larger measurement for the host yields a larger radius measurement for the BD of $R_b = 1.08 \pm 0.05\rj$. I find that the mass $M_b=22.3\pm 1.0\mj$ of CoRoT-3b is consistent with the value in the original work. Given its mass--radius position in Figure \ref{fig:mass_radius_dr3}, CoRoT-3b seems to be among the group of inflated transiting BDs, like GPX-1b \citep{benni2020} and KELT-1b \citep{kelt1b}, where those systems have strong evidence for radius inflation due to their proximity to their host star.

\subsubsection{CoRoT-15}
The second ever known transiting BD, CoRoT-15b initially held one of the largest radius uncertainty estimates for transiting BDs at 27\% ($R_b = 1.12 \pm 0.30\rj$). With such a large radius uncertainty, this BD was not a particularly useful test of substellar isochrones in the mass--radius diagram. \cite{corot15b} estimate the distance to CoRoT-15 to be $d = 1270 \pm 300$ pc, which is slightly closer than the distance derived from Gaia DR3 of $d = 1367 \pm 130$ pc. However, the primary source of the large radius uncertainty on the BD in \cite{corot15b} is the stellar radius uncertainty, which is reduced by a factor of 2 in this work. I attribute this to improved stellar evolutionary models from MIST since the original publication of CoRoT-15 in 2011. I find the revised radius for CoRoT-15b is $R_b=0.94\pm 0.12\rj$. Though improved, the radius uncertainty is still somewhat large, which makes the use of this transiting BD as a test for substellar mass--radius models challenging at best.

\subsubsection{EPIC 201702477}
The discovery work for this system by \cite{bayliss16} does not explicitly state a parallax or distance to the host star, so the provenance in the differences between my stellar radius measurement ($R_\star = 1.01\rst$) and  the measurement from \cite{bayliss16} ($R_\star = 0.90\rst$) is not directly traceable to an inaccuracy in one of those parameters. Following this change in stellar radius measurement, I find that the transiting BD in this system increases from $R_b=0.76\rj$ to $R_b=0.83\rj$. This change in radius ultimately does not affect the interpretation offered by \cite{bayliss16} that this system is old at $8.8\pm 4.1$ Gyr, but it does clarify that it is more likely suited as a test of 4-5 Gyr substellar models rather than models examining BDs as old as 8 Gyr. My own age estimate for EPIC 201702477 at $6.4\pm 4.6$ Gyr does not have any better precision than that of \cite{bayliss16}.

\subsubsection{KELT-1}
The mass--radius determination made by \cite{kelt1b} stands up to a reanalysis of the KELT-1 system using Gaia DR3. That is to say, I find the original mass--radius value of KELT-1b to be consistent with the values I present here. The original estimate of the distance to KELT-1 nearly exactly matches that from Gaia DR3, so the stellar luminosity and radius measurements were accurate in the original work. As \cite{kelt1b} conclude, KELT-1b is a transiting BD whose radius has been inflated as a result of the proximity to its host star. I also note that the KELT-1 system has been extensively studied \citep[e.g.][]{beatty2017, beatty2019, parviainen2022} in large part due to the accessibility of the host star (it is bright at $V=10.6$) and the strongly detected secondary eclipses of the BD. In the most recent study of KELT-1, \cite{parviainen2022} find that the secondary eclipse depth varies greatly between the \textit{TESS} and CHEOPS \citep{cheops} mission data, despite these data being taken in largely-overlapping bandpasses. A discussion and analysis of the secondary eclipse of KELT-1b would be redundant here and is beyond the scope of this work. Nonetheless, this aspect of the KELT-1 system is worthy of acknowledgement.

This work, \cite{kelt1b}, and \cite{parviainen2022} find a radius for KELT-1b to be $R_b=1.116\pm 0.038\rj$, $R_b=1.129\pm 0.031\rj$, and $R_b=1.138\pm 0.010\rj$, respectively.

\subsubsection{Kepler-39}
The largest discrepancy in BD radius between this work and previous works is seen the Kepler-39 system. The BD in the Kepler-39 system sees a $10\%$ decrease in radius from $R_b = 1.22\pm 0.12\rj$ to $R_b = 1.07\pm 0.03\rj$. This is a result of the stellar radius decreasing by a similar amount as a result of the application of Gaia DR3 parallax measurements to the system and not any particular improvement to the analysis of the transit events of the BD. The discovery paper \citep{bouchy11} and the follow up study \citep{kepler39} both discuss an over-inflated radius of the transiting BD when compared to the COND03 models by \cite{baraffe03}, but this no longer seems to be a point of concern with the updated radius that I find. This smaller radius is also consistent with the a non- or less-inflated BD at a relatively long 21-day orbital period receiving little stellar flux compared to its shorter-period, equal-mass counterparts in the mass--radius diagram. For example, this is in contrast to KELT-1b, which orbits a similar star as Kepler-39b, but at a much closer 1.2-day period.

\subsubsection{KOI-189}
Aside from the publishing error in the RV data, \cite{diaz14} present a mass--radius determination of KOI-189b consistent with my own here. \cite{diaz14} emphasize that KOI-189b straddles the traditional line between BDs and low-mass stars and I support that conclusion here where I show KOI-189b to be at $M_b=80.4 \pm 2.5\mj$ and a radius $R_b = 0.99\pm 0.02\rj$.

\subsubsection{KOI-205}
\begin{figure}
    \centering
    \includegraphics[width=0.42\textwidth, trim={1.0cm 0.5cm 1.0cm 0.0cm}]{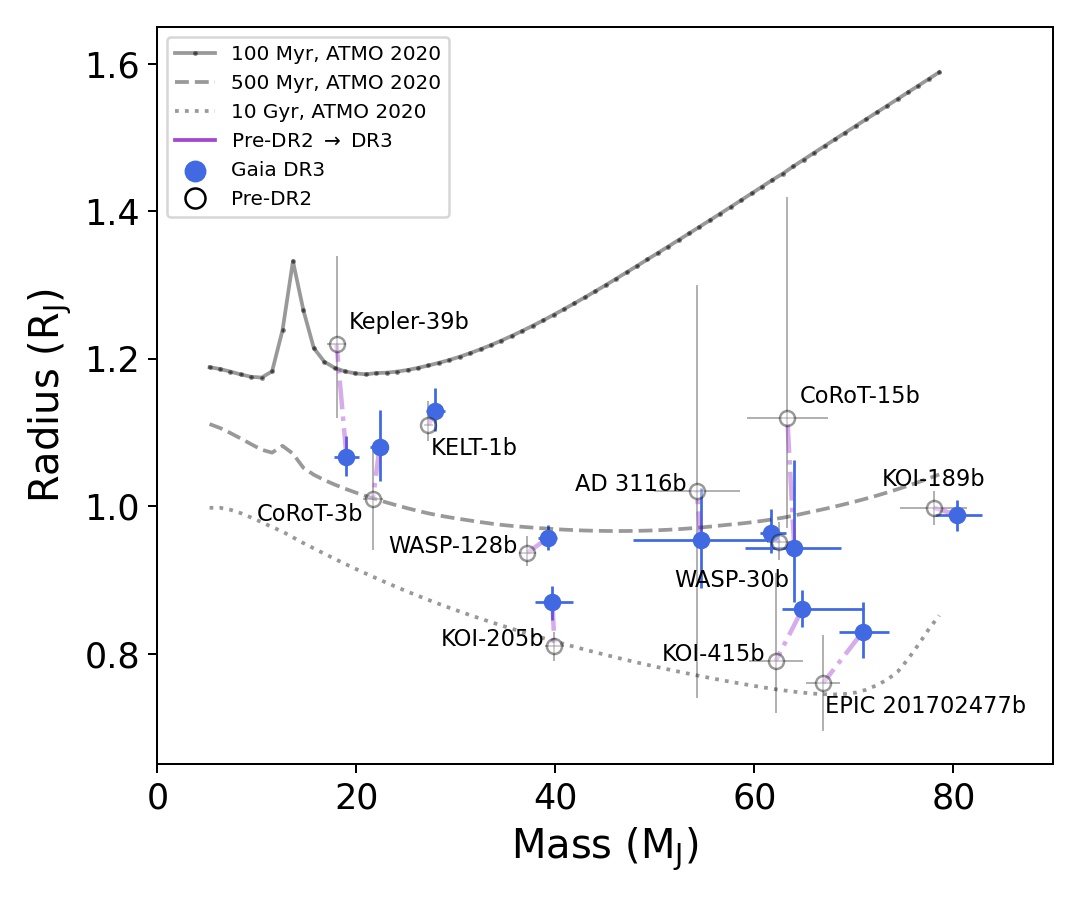}
    \includegraphics[width=0.42\textwidth, trim={1.0cm 0.5cm 1.5cm 0.0cm}]{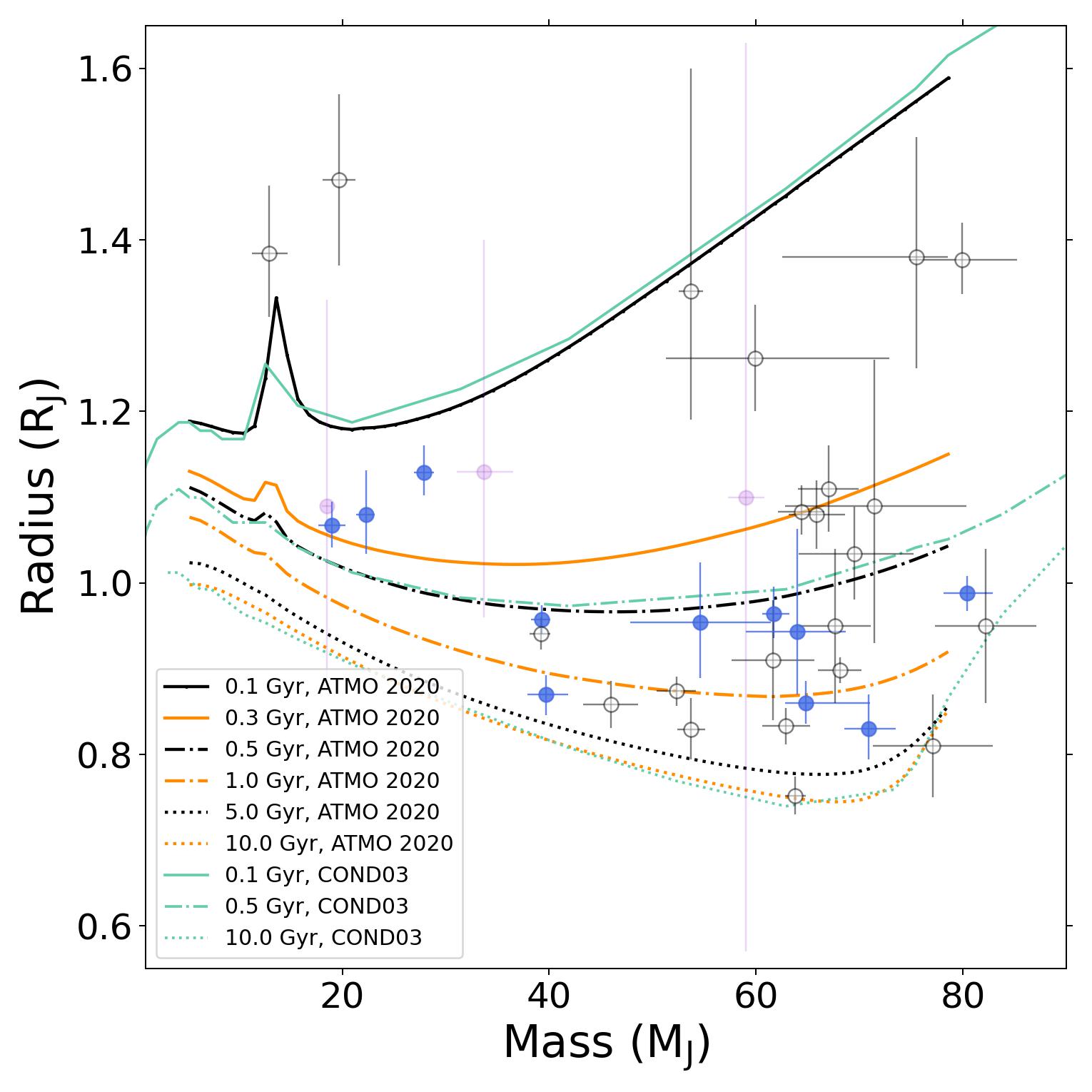}
    \caption{\textit{Top}: Mass--radius diagram showing comparison between previous works and this work. The purple line connects the measurements made by previous works (open circles) to this work (blue filled circles). Another visual breakdown of this comparison is shown in Figures \ref{fig:pdfs_radius} and \ref{fig:pdfs_mass}. \textit{Bottom}: The complete known transiting BD population as of August 2022. This excludes eclipsing binary BDs and BDs that transit white dwarf hosts. The blue points show the systems reanalysed with Gaia DR3 information in this work and the faded purple points indicate transiting BDs with radius uncertainties $>20\%$. The substellar evolutionary models are the ATMO 2020 models showing 0.1 Gyr (solid black), 0.3 Gyr (solid orange), 0.5 Gyr (dash-dotted black), 1.0 Gyr (dash-dotted orange), 5.0 Gyr (dotted black), and 10.0 Gyr (dotted orange). The COND03 models shown are 0.1 Gyr (solid light blue), 0.5 Gyr (dash-dotted light blue), and 10.0 Gyr (dotted light blue).}
    \label{fig:mass_radius_dr3}
\end{figure}

In \cite{diaz13}, they find a stellar radius of $R_\star = 0.84 \pm 0.02\rst$ which is 7\% smaller than the radius I find, $R_\star = 0.90 \pm 0.03\rst$. Given that the distance \cite{diaz13} use, $d = 585 \pm 16$ pc, is consistent with the distance derived from Gaia DR3, $d = 597 \pm 6$ pc, the difference in stellar radius arises not from discrepant parallax measurement, but likely from the different stellar evolutionary models used in \cite{diaz13} \citep[the BT-Settl models by][]{bt-settl}. Both the MIST and BT-Settl frameworks cover a stellar mass range that includes KOI-205 at $M_\star = 0.88\mst$, though the BT-Settl models reach masses well into the LTY brown dwarf regime while the MIST models are lower bound at $M_\star=0.1{\rm M_\odot}$. Using the MIST models yields a larger estimate for the stellar radius, which increases the radius estimate for the BD from $R_b=0.81\pm 0.02\rj$ to $R_b=0.87\pm 0.02\rj$. Though a difference of $0.06\rj$ may seem relatively insignificant, such a change for a BD this small can noticeably impact the interpretation of the best-fitting substellar age--radius models from ATMO 2020 or COND03. At a given mass, 5-7\% changes in the radius at ages older than 1 Gyr can lead to different interpretations of the age by 2-5 Gyr (see the ATMO 2020 models in Figure \ref{fig:mass_radius_dr3}). 

\subsubsection{KOI-415}
This transiting BD has the longest orbital period known to date at 166 days. The original BD radius value of $R_b=0.79 \pm 0.12\rj$ made by \cite{moutou13} is inconsistent with my radius determination of $R_b=0.86\pm 0.03\rj$. My findings show a larger host star ($R_\star=1.36\pm 0.04\rst$ compared to $R_\star=1.25\pm 0.15\rst$) at a greater distance $d = 950 \pm 12$ pc than what \cite{moutou13} find, resulting in a larger transiting BD by several percent. \cite{moutou13} use the Yonsei-Yale stellar isochrones from 2004 \citep{old_yy} in their work, which may give rise to the radius discrepancies when compared to the MIST isochrones I make use of here.

\subsubsection{WASP-30}
The follow-up work by \cite{wasp30b} to the original discovery work by \cite{wasp30b_old} report a BD radius of $R_b=0.95 \pm 0.03\rj$, which is consistent with my own findings of a radius $R_b=0.96\pm 0.03\rj$. Both previous studies cite evidence of the WASP-30 system being as young as 500 Myr based on Li 6708\AA\, absorption seen the stellar spectrum. Though this age estimate may be supported by the age--radius relationships from the substellar ATMO 2020 models, stellar ages up to 2 Gyr cannot be ruled out in these previous studies or this one here.

\subsubsection{WASP-128}
The BD radius determination from \cite{wasp128b} is consistent with my own at $R_b = 0.96\pm 0.02\rj$, although I find the BD mass to be slightly higher at $M_b = 39.3\pm 1.0\mj$ compared to $M_b = 37.5 \pm 0.8\mj$ from the original study. The original work used parallax measurements from Gaia DR2, so it was expected that the final stellar and BD parameters between that work and this one would be fairly consistent with each other.

\subsection{Changes to the substellar mass--radius diagram}
A summary of the new BD parameters in this work is shown in Table \ref{tab:bdlist}. One notable change to the radius distribution of transiting BDs is the increase in radii for several of the smallest known transiting BDs. These are the BDs that once occupied radii between $0.75\rj < R_b < 0.85\rj$, which is a range of radii that is predicted to comprise the oldest transiting BDs; those that are older than 5 Gyr \citep{baraffe03, ATMO2020, sonora21}. Understanding these oldest and smallest BDs is as important as understanding young BDs (less than 500 Myr old) that rapidly contract because these old objects represent an extreme limit of BD evolution. Old BDs enable tests of theories for the smallest possible BDs and at which ages these smallest BD radii are expected to be reached. With this work, the subset of smallest transiting BDs that trace out the substellar age--radius models from 8-10 Gyr decreases from 5 objects to 2. Prior to this work, KOI-205b, KOI-415b, EPIC 201702477b, TOI-148b \citep{grieves2021}, and TOI-569b \citep{carmichael2020} were the transiting BDs that could comfortably be characterised as ``very old", but now only TOI-148b and TOI-569b remain in this region of age--radius space.

In each of the original studies for KOI-205b, KOI-415b, and EPIC 201702477b, and from my own stellar isochrone analysis in this work, the ages of the host stars in these systems are uncertain enough that ages as young as 5 Gyr are not ruled out at the $1\sigma$ level. This means that the stellar age (which is assumed to match the BD age) broadly makes the age--radius relationships between 5 Gyr and 10 Gyr consistent with this sample of 5 small transiting BDs, even when using Gaia DR3. However, the uncertainties in the ages for these systems can exceed 50\%, making them somewhat clumsy as tools to assess the accuracy of substellar age--radius models beyond 5 Gyr. This starkly contrasts the single-digit percent precision of the radius measurements for these objects.

These wide age uncertainties mean that, despite very precise radius measurements, this group of BDs older than 5 Gyr cannot be used to effectively test substellar age--radius models from 5-10 Gyr. However, establishing precise and accurate BD radii is a key step in testing these models and until the ages of these host stars can be more precisely determined, I can only speculatively discuss the implications of this reduction in the number of objects in the small radius region ($R_b<0.85\rj$) of the substellar mass--radius diagram.

These implications are: 1) that very old transiting BDs do not always reach the theoretical minimum radius predicted by these models or 2) that the minimum BD radius is not as small as previously thought and is a more complicated function of BD mass and composition. A caveat to this is that the ATMO 2020 models are not designed with the irradiated environments of transiting BDs in mind, however, similar implications still hold true for the COND03 \citep{baraffe03} and \cite{baraffe2015} models, which do consider stellar irradiation. Additionally, this discussion omits considerations of the transiting BD metallicity, which certainly affects the radius at a given age, but to a diminished degree for BDs older than 5 Gyr \citep{saumon08, sonora21}.

Regarding the AD 3116, CoRoT-3, CoRoT-15, and Kepler-39 systems, this work sees a marked improvement in the radius precision on the BDs in these systems. AD 3116b sees a factor of 4 improvement to its radius determination, making it much more consistent with substellar age--radius predictions for its age to be 500-600 Myr, which matches the age of the star cluster, Praesepe, that the system resides in. Kepler-39b moves from being a curiously radius-inflated transiting BD to a smaller object that is more inline with substellar model predictions. However, like with CoRoT-3b and CoRoT-15b, the host star to Kepler-39b does not have a particularly well-defined age like AD 3116 does, so I am limited in the depth of conclusions that may be drawn for these 3 systems.

\subsection{Summary}
This work is part of a larger effort to take the transiting brown dwarf community from a place of exciting individual boutique studies to a place where we more routinely perform population statistics of various characteristics of transiting brown dwarfs. To that end, I find that the following transiting BD systems have improved accuracy and precision on their BD radius as a result of the application of data from Gaia DR3: AD 3116, CoRoT-3, CoRoT-15, EPIC 201702477, Kepler-39, KOI-205, and KOI-415. These 7 systems represent a substantial fraction (20\%) of the known transiting BD population and their new radius measurements (all mass measurements were consistent between past works and this one) change the interpretation of the substellar mass--radius diagram as a whole. Fewer transiting BDs than previously thought occupy the radius space smaller than $R_b=0.85\rj$, which is a region of the mass--radius diagram where substellar evolutionary models predict the oldest BDs to be. The remaining 4 out of 11 transiting BDs, KELT-1b, KOI-189b, WASP-30b, and WASP-128b, show consistent mass--radius determinations with their previously published works.

%%%%%%%%%%%%%%%%%%%% ACKNOWLEDGEMENTS %%%%%%%%%%%%%%%%%%%%%
\section*{Acknowledgements}
I acknowledge the significant and thorough efforts of the members of the TESS Followup Program and the Science Processing Operations Center in making the TESS data readily accessible for the analysis in this work.

This work has made use of data from the European Space Agency (ESA) mission {\it Gaia} (\url{https://www.cosmos.esa.int/gaia}), processed by the {\it Gaia} Data Processing and Analysis Consortium (DPAC,
\url{https://www.cosmos.esa.int/web/gaia/dpac/consortium}). Funding for the DPAC has been provided by national institutions, in particular the institutions participating in the {\it Gaia} Multilateral Agreement.

I would also like to thank the reviewer at MNRAS for their diligent feedback on this work.

\section*{Data Availability}
The \tess light curve data underlying this article were accessed from the Mikulski Archive for Space Telescopes (MAST) at \url{https://mast.stsci.edu/portal/Mashup/Clients/Mast/Portal.html}. These data are easiest accessed via a light curve management software such at {\tt lightkurve} \citep{lightkurve}. The \tess data for AD 3116, CoRoT-3, KELT-1, KOI-189, WASP-30, and WASP-128 used for the analysis in this work will be made available in an online database accompanying this article. All other light curve and RV data are taken from their respective previous works with the exception of the RV data for the KOI-189 system, which will be made available with the new \tess data used. All SED, extinction, and magnitude data used in the analysis of the stellar and brown dwarf parameters are queried and properly formatted by {\tt EXOFASTv2} from their respective databases (see Section \ref{sec:exofast}). The installation and use instructions for {\tt EXOFASTv2} can be found at \url{https://github.com/jdeast/EXOFASTv2}.

\bibliographystyle{mnras}
\bibliography{citations}

\begin{thebibliography}{}
\makeatletter
\relax
\def\mn@urlcharsother{\let\do\@makeother \do\$\do\&\do\#\do\^\do\_\do\%\do\~}
\def\mn@doi{\begingroup\mn@urlcharsother \@ifnextchar [ {\mn@doi@}
  {\mn@doi@[]}}
\def\mn@doi@[#1]#2{\def\@tempa{#1}\ifx\@tempa\@empty \href
  {http://dx.doi.org/#2} {doi:#2}\else \href {http://dx.doi.org/#2} {#1}\fi
  \endgroup}
\def\mn@eprint#1#2{\mn@eprint@#1:#2::\@nil}
\def\mn@eprint@arXiv#1{\href {http://arxiv.org/abs/#1} {{\tt arXiv:#1}}}
\def\mn@eprint@dblp#1{\href {http://dblp.uni-trier.de/rec/bibtex/#1.xml}
  {dblp:#1}}
\def\mn@eprint@#1:#2:#3:#4\@nil{\def\@tempa {#1}\def\@tempb {#2}\def\@tempc
  {#3}\ifx \@tempc \@empty \let \@tempc \@tempb \let \@tempb \@tempa \fi \ifx
  \@tempb \@empty \def\@tempb {arXiv}\fi \@ifundefined
  {mn@eprint@\@tempb}{\@tempb:\@tempc}{\expandafter \expandafter \csname
  mn@eprint@\@tempb\endcsname \expandafter{\@tempc}}}

\bibitem[\protect\citeauthoryear{{Acton} et~al.,}{{Acton}
  et~al.}{2021}]{acton2021}
{Acton} J.~S.,  et~al., 2021, \mn@doi [\mnras] {10.1093/mnras/stab1459}, \href
  {https://ui.adsabs.harvard.edu/abs/2021MNRAS.505.2741A} {505, 2741}

\bibitem[\protect\citeauthoryear{{Allard}, {Homeier}  \& {Freytag}}{{Allard}
  et~al.}{2012}]{bt-settl}
{Allard} F.,  {Homeier} D.,   {Freytag} B.,  2012, \mn@doi [Philosophical
  Transactions of the Royal Society of London Series A]
  {10.1098/rsta.2011.0269}, \href
  {https://ui.adsabs.harvard.edu/abs/2012RSPTA.370.2765A} {370, 2765}

\bibitem[\protect\citeauthoryear{{Anderson} et~al.,}{{Anderson}
  et~al.}{2011}]{wasp30b_old}
{Anderson} D.~R.,  et~al., 2011, \mn@doi [\apjl] {10.1088/2041-8205/726/2/L19},
  \href {http://adsabs.harvard.edu/abs/2011ApJ...726L..19A} {726, L19}

\bibitem[\protect\citeauthoryear{{Artigau} et~al.,}{{Artigau}
  et~al.}{2021}]{artigau2021}
{Artigau} {\'E}.,  et~al., 2021, \mn@doi [\aj] {10.3847/1538-3881/ac096d},
  \href {https://ui.adsabs.harvard.edu/abs/2021AJ....162..144A} {162, 144}

\bibitem[\protect\citeauthoryear{{Baglin} et~al.,}{{Baglin}
  et~al.}{2006}]{corot}
{Baglin} A.,  et~al., 2006, in 36th COSPAR Scientific Assembly. p.~3749

\bibitem[\protect\citeauthoryear{{Baraffe}, {Chabrier}, {Barman}, {Allard}  \&
  {Hauschildt}}{{Baraffe} et~al.}{2003}]{baraffe03}
{Baraffe} I.,  {Chabrier} G.,  {Barman} T.~S.,  {Allard} F.,   {Hauschildt}
  P.~H.,  2003, \mn@doi [\aap] {10.1051/0004-6361:20030252}, \href
  {http://adsabs.harvard.edu/abs/2003A%26A...402..701B} {402, 701}

\bibitem[\protect\citeauthoryear{{Baraffe}, {Homeier}, {Allard}  \&
  {Chabrier}}{{Baraffe} et~al.}{2015}]{baraffe2015}
{Baraffe} I.,  {Homeier} D.,  {Allard} F.,   {Chabrier} G.,  2015, \mn@doi
  [\aap] {10.1051/0004-6361/201425481}, \href
  {https://ui.adsabs.harvard.edu/abs/2015A&A...577A..42B} {577, A42}

\bibitem[\protect\citeauthoryear{{Bayliss} et~al.,}{{Bayliss}
  et~al.}{2017}]{bayliss16}
{Bayliss} D.,  et~al., 2017, \mn@doi [\aj] {10.3847/1538-3881/153/1/15}, \href
  {http://adsabs.harvard.edu/abs/2017AJ....153...15B} {153, 15}

\bibitem[\protect\citeauthoryear{{Beatty} et~al.,}{{Beatty}
  et~al.}{2017}]{beatty2017}
{Beatty} T.~G.,  et~al., 2017, \mn@doi [\aj] {10.3847/1538-3881/aa7511}, \href
  {https://ui.adsabs.harvard.edu/abs/2017AJ....154...25B} {154, 25}

\bibitem[\protect\citeauthoryear{{Beatty}, {Marley}, {Gaudi}, {Col{\'o}n},
  {Fortney}  \& {Showman}}{{Beatty} et~al.}{2019}]{beatty2019}
{Beatty} T.~G.,  {Marley} M.~S.,  {Gaudi} B.~S.,  {Col{\'o}n} K.~D.,  {Fortney}
  J.~J.,   {Showman} A.~P.,  2019, \mn@doi [\aj] {10.3847/1538-3881/ab33fc},
  \href {https://ui.adsabs.harvard.edu/abs/2019AJ....158..166B} {158, 166}

\bibitem[\protect\citeauthoryear{{Benni}, {Burdanov}, {Sokov}, {Barkaoui},
  {Team}, {Team}  \& {Krushinsky}}{{Benni} et~al.}{2020}]{benni2020}
{Benni} P.,  {Burdanov} A.,  {Sokov} E.,  {Barkaoui} K.,  {Team} G.~F.,  {Team}
  S.,   {Krushinsky} V.,  2020, Journal of the American Association of Variable
  Star Observers (JAAVSO), \href
  {https://ui.adsabs.harvard.edu/abs/2020JAVSO..48..104B} {48, 104}

\bibitem[\protect\citeauthoryear{{Benz} et~al.,}{{Benz} et~al.}{2021}]{cheops}
{Benz} W.,  et~al., 2021, \mn@doi [Experimental Astronomy]
  {10.1007/s10686-020-09679-4}, \href
  {https://ui.adsabs.harvard.edu/abs/2021ExA....51..109B} {51, 109}

\bibitem[\protect\citeauthoryear{{Bonomo} et~al.,}{{Bonomo}
  et~al.}{2015}]{kepler39}
{Bonomo} A.~S.,  et~al., 2015, \mn@doi [\aap] {10.1051/0004-6361/201323042},
  \href {https://ui.adsabs.harvard.edu/\#abs/2015A&A...575A..85B} {575, A85}

\bibitem[\protect\citeauthoryear{{Borsato} et~al.,}{{Borsato}
  et~al.}{2021}]{borsato2021}
{Borsato} L.,  et~al., 2021, \mn@doi [\mnras] {10.1093/mnras/stab1782}, \href
  {https://ui.adsabs.harvard.edu/abs/2021MNRAS.506.3810B} {506, 3810}

\bibitem[\protect\citeauthoryear{{Borucki} et~al.,}{{Borucki}
  et~al.}{2010}]{kepler}
{Borucki} W.~J.,  et~al., 2010, \mn@doi [Science] {10.1126/science.1185402},
  \href {https://ui.adsabs.harvard.edu/abs/2010Sci...327..977B} {327, 977}

\bibitem[\protect\citeauthoryear{{Bouchy} et~al.,}{{Bouchy}
  et~al.}{2011a}]{corot15b}
{Bouchy} F.,  et~al., 2011a, \mn@doi [\aap] {10.1051/0004-6361/201015276},
  \href {http://adsabs.harvard.edu/abs/2011A%26A...525A..68B} {525, A68}

\bibitem[\protect\citeauthoryear{{Bouchy} et~al.,}{{Bouchy}
  et~al.}{2011b}]{bouchy11}
{Bouchy} F.,  et~al., 2011b, \mn@doi [\aap] {10.1051/0004-6361/201117095},
  \href {http://adsabs.harvard.edu/abs/2011A%26A...533A..83B} {533, A83}

\bibitem[\protect\citeauthoryear{{Burrows}, {Hubbard}, {Lunine}  \&
  {Liebert}}{{Burrows} et~al.}{2001}]{burrows01}
{Burrows} A.,  {Hubbard} W.~B.,  {Lunine} J.~I.,   {Liebert} J.,  2001, \mn@doi
  [Reviews of Modern Physics] {10.1103/RevModPhys.73.719}, \href
  {http://adsabs.harvard.edu/abs/2001RvMP...73..719B} {73, 719}

\bibitem[\protect\citeauthoryear{{Ca{\~n}as} et~al.,}{{Ca{\~n}as}
  et~al.}{2022}]{canas2022}
{Ca{\~n}as} C.~I.,  et~al., 2022, \mn@doi [\aj] {10.3847/1538-3881/ac415f},
  \href {https://ui.adsabs.harvard.edu/abs/2022AJ....163...89C} {163, 89}

\bibitem[\protect\citeauthoryear{{Carmichael}, {Latham}  \&
  {Vanderburg}}{{Carmichael} et~al.}{2019}]{carmichael19}
{Carmichael} T.~W.,  {Latham} D.~W.,   {Vanderburg} A.~M.,  2019, \mn@doi [\aj]
  {10.3847/1538-3881/ab245e}, \href
  {https://ui.adsabs.harvard.edu/abs/2019AJ....158...38C} {158, 38}

\bibitem[\protect\citeauthoryear{{Carmichael} et~al.,}{{Carmichael}
  et~al.}{2020}]{carmichael2020}
{Carmichael} T.~W.,  et~al., 2020, \mn@doi [\aj] {10.3847/1538-3881/ab9b84},
  \href {https://ui.adsabs.harvard.edu/abs/2020AJ....160...53C} {160, 53}

\bibitem[\protect\citeauthoryear{{Carmichael} et~al.,}{{Carmichael}
  et~al.}{2021}]{carmichael2021}
{Carmichael} T.~W.,  et~al., 2021, \mn@doi [\aj] {10.3847/1538-3881/abd4e1},
  \href {https://ui.adsabs.harvard.edu/abs/2021AJ....161...97C} {161, 97}

\bibitem[\protect\citeauthoryear{{Carmichael} et~al.,}{{Carmichael}
  et~al.}{2022}]{carmichael2022}
{Carmichael} T.~W.,  et~al., 2022, \mn@doi [\mnras] {10.1093/mnras/stac1666},
  \href {https://ui.adsabs.harvard.edu/abs/2022MNRAS.514.4944C} {514, 4944}

\bibitem[\protect\citeauthoryear{{Choi}, {Dotter}, {Conroy}, {Cantiello},
  {Paxton}  \& {Johnson}}{{Choi} et~al.}{2016}]{mist2}
{Choi} J.,  {Dotter} A.,  {Conroy} C.,  {Cantiello} M.,  {Paxton} B.,
  {Johnson} B.~D.,  2016, \mn@doi [\apj] {10.3847/0004-637X/823/2/102}, \href
  {http://adsabs.harvard.edu/abs/2016ApJ...823..102C} {823, 102}

\bibitem[\protect\citeauthoryear{{Csizmadia} et~al.,}{{Csizmadia}
  et~al.}{2015}]{corot33b}
{Csizmadia} S.,  et~al., 2015, \mn@doi [\aap] {10.1051/0004-6361/201526763},
  \href {http://adsabs.harvard.edu/abs/2015A%26A...584A..13C} {584, A13}

\bibitem[\protect\citeauthoryear{{David}, {Hillenbrand}, {Gillen}, {Cody},
  {Howell}, {Isaacson}  \& {Livingston}}{{David} et~al.}{2019}]{david19_bd}
{David} T.~J.,  {Hillenbrand} L.~A.,  {Gillen} E.,  {Cody} A.~M.,  {Howell}
  S.~B.,  {Isaacson} H.~T.,   {Livingston} J.~H.,  2019, \mn@doi [\apj]
  {10.3847/1538-4357/aafe09}, \href
  {https://ui.adsabs.harvard.edu/\#abs/2019ApJ...872..161D} {872, 161}

\bibitem[\protect\citeauthoryear{{Deleuil} et~al.,}{{Deleuil}
  et~al.}{2008}]{corot3b}
{Deleuil} M.,  et~al., 2008, \mn@doi [\aap] {10.1051/0004-6361:200810625},
  \href {http://adsabs.harvard.edu/abs/2008A%26A...491..889D} {491, 889}

\bibitem[\protect\citeauthoryear{{Demarque}, {Woo}, {Kim}  \& {Yi}}{{Demarque}
  et~al.}{2004}]{old_yy}
{Demarque} P.,  {Woo} J.-H.,  {Kim} Y.-C.,   {Yi} S.~K.,  2004, \mn@doi [\apjs]
  {10.1086/424966}, \href
  {https://ui.adsabs.harvard.edu/abs/2004ApJS..155..667D} {155, 667}

\bibitem[\protect\citeauthoryear{{D{\'{\i}}az} et~al.,}{{D{\'{\i}}az}
  et~al.}{2013}]{diaz13}
{D{\'{\i}}az} R.~F.,  et~al., 2013, \mn@doi [\aap]
  {10.1051/0004-6361/201321124}, \href
  {http://adsabs.harvard.edu/abs/2013A%26A...551L...9D} {551, L9}

\bibitem[\protect\citeauthoryear{{D{\'{\i}}az} et~al.,}{{D{\'{\i}}az}
  et~al.}{2014}]{diaz14}
{D{\'{\i}}az} R.~F.,  et~al., 2014, \mn@doi [\aap]
  {10.1051/0004-6361/201424406}, \href
  {http://adsabs.harvard.edu/abs/2014A%26A...572A.109D} {572, A109}

\bibitem[\protect\citeauthoryear{{Dotter}}{{Dotter}}{2016}]{mist1}
{Dotter} A.,  2016, \mn@doi [\apjs] {10.3847/0067-0049/222/1/8}, \href
  {http://adsabs.harvard.edu/abs/2016ApJS..222....8D} {222, 8}

\bibitem[\protect\citeauthoryear{{Eastman} et~al.,}{{Eastman}
  et~al.}{2019}]{eastman2019}
{Eastman} J.~D.,  et~al., 2019, arXiv e-prints, \href
  {https://ui.adsabs.harvard.edu/abs/2019arXiv190709480E} {p. arXiv:1907.09480}

\bibitem[\protect\citeauthoryear{{Gaia Collaboration} et~al.,}{{Gaia
  Collaboration} et~al.}{2022}]{dr3}
{Gaia Collaboration} et~al., 2022, arXiv e-prints, \href
  {https://ui.adsabs.harvard.edu/abs/2022arXiv220800211G} {p. arXiv:2208.00211}

\bibitem[\protect\citeauthoryear{{Gillen}, {Hillenbrand}, {David}, {Aigrain},
  {Rebull}, {Stauffer}, {Cody}  \& {Queloz}}{{Gillen} et~al.}{2017}]{ad3116}
{Gillen} E.,  {Hillenbrand} L.~A.,  {David} T.~J.,  {Aigrain} S.,  {Rebull} L.,
   {Stauffer} J.,  {Cody} A.~M.,   {Queloz} D.,  2017, \mn@doi [\apj]
  {10.3847/1538-4357/aa84b3}, \href
  {https://ui.adsabs.harvard.edu/\#abs/2017ApJ...849...11G} {849, 11}

\bibitem[\protect\citeauthoryear{{Gossage}, {Conroy}, {Dotter}, {Choi},
  {Rosenfield}, {Cargile}  \& {Dolphin}}{{Gossage} et~al.}{2018}]{gossage2018}
{Gossage} S.,  {Conroy} C.,  {Dotter} A.,  {Choi} J.,  {Rosenfield} P.,
  {Cargile} P.,   {Dolphin} A.,  2018, \mn@doi [\apj]
  {10.3847/1538-4357/aad0a0}, \href
  {https://ui.adsabs.harvard.edu/abs/2018ApJ...863...67G} {863, 67}

\bibitem[\protect\citeauthoryear{{Grieves} et~al.,}{{Grieves}
  et~al.}{2021}]{grieves2021}
{Grieves} N.,  et~al., 2021, \mn@doi [\aap] {10.1051/0004-6361/202141145},
  \href {https://ui.adsabs.harvard.edu/abs/2021A&A...652A.127G} {652, A127}

\bibitem[\protect\citeauthoryear{{Hatzes} \& {Rauer}}{{Hatzes} \&
  {Rauer}}{2015}]{hatzes15}
{Hatzes} A.~P.,  {Rauer} H.,  2015, \mn@doi [\apjl]
  {10.1088/2041-8205/810/2/L25}, \href
  {http://adsabs.harvard.edu/abs/2015ApJ...810L..25H} {810, L25}

\bibitem[\protect\citeauthoryear{{Hod{\v z}i{\'c}} et~al.,}{{Hod{\v z}i{\'c}}
  et~al.}{2018}]{wasp128b}
{Hod{\v z}i{\'c}} V.,  et~al., 2018, \mn@doi [\mnras] {10.1093/mnras/sty2512},
  \href {http://adsabs.harvard.edu/abs/2018MNRAS.481.5091H} {481, 5091}

\bibitem[\protect\citeauthoryear{{Howell} et~al.,}{{Howell} et~al.}{2014}]{k2}
{Howell} S.~B.,  et~al., 2014, \mn@doi [\pasp] {10.1086/676406}, \href
  {https://ui.adsabs.harvard.edu/abs/2014PASP..126..398H} {126, 398}

\bibitem[\protect\citeauthoryear{{Irwin} et~al.,}{{Irwin}
  et~al.}{2010}]{irwin10}
{Irwin} J.,  et~al., 2010, \mn@doi [\apj] {10.1088/0004-637X/718/2/1353}, \href
  {http://adsabs.harvard.edu/abs/2010ApJ...718.1353I} {718, 1353}

\bibitem[\protect\citeauthoryear{{Irwin} et~al.,}{{Irwin}
  et~al.}{2018}]{irwin18}
{Irwin} J.~M.,  et~al., 2018, \mn@doi [\aj] {10.3847/1538-3881/aad9a3}, \href
  {http://adsabs.harvard.edu/abs/2018AJ....156..140I} {156, 140}

\bibitem[\protect\citeauthoryear{{Jackman} et~al.,}{{Jackman}
  et~al.}{2019}]{jackman2019}
{Jackman} J. A.~G.,  et~al., 2019, \mn@doi [\mnras] {10.1093/mnras/stz2496},
  \href {https://ui.adsabs.harvard.edu/abs/2019MNRAS.489.5146J} {489, 5146}

\bibitem[\protect\citeauthoryear{{Jenkins} et~al.,}{{Jenkins}
  et~al.}{2016}]{jenkins_2016}
{Jenkins} J.~M.,  et~al., 2016, in Software and Cyberinfrastructure for
  Astronomy IV. p. 99133E, \mn@doi{10.1117/12.2233418}

\bibitem[\protect\citeauthoryear{{Johnson} et~al.,}{{Johnson}
  et~al.}{2011}]{johnson11_bd}
{Johnson} J.~A.,  et~al., 2011, \mn@doi [\apj] {10.1088/0004-637X/730/2/79},
  \href {http://adsabs.harvard.edu/abs/2011ApJ...730...79J} {730, 79}

\bibitem[\protect\citeauthoryear{{Lightkurve Collaboration}
  et~al.,}{{Lightkurve Collaboration} et~al.}{2018}]{lightkurve}
{Lightkurve Collaboration} et~al., 2018, {Lightkurve: Kepler and TESS time
  series analysis in Python} (\mn@eprint {ascl} {1812.013})

\bibitem[\protect\citeauthoryear{{Lindegren} et~al.,}{{Lindegren}
  et~al.}{2021}]{dr3_correction}
{Lindegren} L.,  et~al., 2021, \mn@doi [\aap] {10.1051/0004-6361/202039653},
  \href {https://ui.adsabs.harvard.edu/abs/2021A&A...649A...4L} {649, A4}

\bibitem[\protect\citeauthoryear{{Mann} et~al.,}{{Mann}
  et~al.}{2019}]{mann2019}
{Mann} A.~W.,  et~al., 2019, \mn@doi [\apj] {10.3847/1538-4357/aaf3bc}, \href
  {https://ui.adsabs.harvard.edu/abs/2019ApJ...871...63M} {871, 63}

\bibitem[\protect\citeauthoryear{{Marley} et~al.,}{{Marley}
  et~al.}{2021}]{sonora21}
{Marley} M.~S.,  et~al., 2021, \mn@doi [\apj] {10.3847/1538-4357/ac141d}, \href
  {https://ui.adsabs.harvard.edu/abs/2021ApJ...920...85M} {920, 85}

\bibitem[\protect\citeauthoryear{{Moutou} et~al.,}{{Moutou}
  et~al.}{2013}]{moutou13}
{Moutou} C.,  et~al., 2013, \mn@doi [\aap] {10.1051/0004-6361/201322201}, \href
  {http://adsabs.harvard.edu/abs/2013A%26A...558L...6M} {558, L6}

\bibitem[\protect\citeauthoryear{{Nowak} et~al.,}{{Nowak}
  et~al.}{2017}]{nowak17}
{Nowak} G.,  et~al., 2017, \mn@doi [\aj] {10.3847/1538-3881/aa5cb6}, \href
  {http://adsabs.harvard.edu/abs/2017AJ....153..131N} {153, 131}

\bibitem[\protect\citeauthoryear{{Palle} et~al.,}{{Palle}
  et~al.}{2021}]{palle2021}
{Palle} E.,  et~al., 2021, \mn@doi [\aap] {10.1051/0004-6361/202039937}, \href
  {https://ui.adsabs.harvard.edu/abs/2021A&A...650A..55P} {650, A55}

\bibitem[\protect\citeauthoryear{{Parviainen} et~al.,}{{Parviainen}
  et~al.}{2022}]{parviainen2022}
{Parviainen} H.,  et~al., 2022, \mn@doi [\aap] {10.1051/0004-6361/202244117},
  \href {https://ui.adsabs.harvard.edu/abs/2022A&A...668A..93P} {668, A93}

\bibitem[\protect\citeauthoryear{{Paxton} et~al.,}{{Paxton}
  et~al.}{2015}]{mist3}
{Paxton} B.,  et~al., 2015, \mn@doi [\apjs] {10.1088/0067-0049/220/1/15}, \href
  {http://adsabs.harvard.edu/abs/2015ApJS..220...15P} {220, 15}

\bibitem[\protect\citeauthoryear{{Pepper} et~al.,}{{Pepper}
  et~al.}{2007}]{kelt}
{Pepper} J.,  et~al., 2007, \mn@doi [\pasp] {10.1086/521836}, \href
  {https://ui.adsabs.harvard.edu/abs/2007PASP..119..923P} {119, 923}

\bibitem[\protect\citeauthoryear{{Persson} et~al.,}{{Persson}
  et~al.}{2019}]{persson2019}
{Persson} C.~M.,  et~al., 2019, \mn@doi [\aap] {10.1051/0004-6361/201935505},
  \href {https://ui.adsabs.harvard.edu/abs/2019A&A...628A..64P} {628, A64}

\bibitem[\protect\citeauthoryear{{Phillips} et~al.,}{{Phillips}
  et~al.}{2020}]{ATMO2020}
{Phillips} M.~W.,  et~al., 2020, \mn@doi [\aap] {10.1051/0004-6361/201937381},
  \href {https://ui.adsabs.harvard.edu/abs/2020A&A...637A..38P} {637, A38}

\bibitem[\protect\citeauthoryear{{Pollacco} et~al.,}{{Pollacco}
  et~al.}{2006}]{wasp}
{Pollacco} D.~L.,  et~al., 2006, \mn@doi [\pasp] {10.1086/508556}, \href
  {https://ui.adsabs.harvard.edu/abs/2006PASP..118.1407P} {118, 1407}

\bibitem[\protect\citeauthoryear{{Psaridi} et~al.,}{{Psaridi}
  et~al.}{2022}]{psaridi2022}
{Psaridi} A.,  et~al., 2022, \mn@doi [\aap] {10.1051/0004-6361/202243454},
  \href {https://ui.adsabs.harvard.edu/abs/2022A&A...664A..94P} {664, A94}

\bibitem[\protect\citeauthoryear{{Ricker} et~al.,}{{Ricker}
  et~al.}{2015}]{tess}
{Ricker} G.~R.,  et~al., 2015, \mn@doi [Journal of Astronomical Telescopes,
  Instruments, and Systems] {10.1117/1.JATIS.1.1.014003}, \href
  {http://adsabs.harvard.edu/abs/2015JATIS...1a4003R} {1, 014003}

\bibitem[\protect\citeauthoryear{{Saumon} \& {Marley}}{{Saumon} \&
  {Marley}}{2008}]{saumon08}
{Saumon} D.,  {Marley} M.~S.,  2008, \mn@doi [\apj] {10.1086/592734}, \href
  {https://ui.adsabs.harvard.edu/abs/2008ApJ...689.1327S} {689, 1327}

\bibitem[\protect\citeauthoryear{{Schanche} et~al.,}{{Schanche}
  et~al.}{2020}]{schanche2020}
{Schanche} N.,  et~al., 2020, \mn@doi [\mnras] {10.1093/mnras/staa2848}, \href
  {https://ui.adsabs.harvard.edu/abs/2020MNRAS.499..428S} {499, 428}

\bibitem[\protect\citeauthoryear{{Schlafly} \& {Finkbeiner}}{{Schlafly} \&
  {Finkbeiner}}{2011}]{av_priors}
{Schlafly} E.~F.,  {Finkbeiner} D.~P.,  2011, \mn@doi [\apj]
  {10.1088/0004-637X/737/2/103}, \href
  {https://ui.adsabs.harvard.edu/abs/2011ApJ...737..103S} {737, 103}

\bibitem[\protect\citeauthoryear{{Sebastian} et~al.,}{{Sebastian}
  et~al.}{2022}]{corot34}
{Sebastian} D.,  et~al., 2022, \mn@doi [\mnras] {10.1093/mnras/stac2131}, \href
  {https://ui.adsabs.harvard.edu/abs/2022MNRAS.516..636S} {516, 636}

\bibitem[\protect\citeauthoryear{{Siverd} et~al.,}{{Siverd}
  et~al.}{2012}]{kelt1b}
{Siverd} R.~J.,  et~al., 2012, \mn@doi [\apj] {10.1088/0004-637X/761/2/123},
  \href {http://adsabs.harvard.edu/abs/2012ApJ...761..123S} {761, 123}

\bibitem[\protect\citeauthoryear{{Skrutskie} et~al.,}{{Skrutskie}
  et~al.}{2006}]{2MASS}
{Skrutskie} M.~F.,  et~al., 2006, \mn@doi [\aj] {10.1086/498708}, \href
  {https://ui.adsabs.harvard.edu/abs/2006AJ....131.1163S} {131, 1163}

\bibitem[\protect\citeauthoryear{{Smith} et~al.,}{{Smith}
  et~al.}{2012}]{smith2012_pdc}
{Smith} J.~C.,  et~al., 2012, \mn@doi [\pasp] {10.1086/667697}, \href
  {https://ui.adsabs.harvard.edu/abs/2012PASP..124.1000S} {124, 1000}

\bibitem[\protect\citeauthoryear{{Spiegel}, {Burrows}  \& {Milsom}}{{Spiegel}
  et~al.}{2011}]{spiegel2011}
{Spiegel} D.~S.,  {Burrows} A.,   {Milsom} J.~A.,  2011, \mn@doi [\apj]
  {10.1088/0004-637X/727/1/57}, \href
  {https://ui.adsabs.harvard.edu/abs/2011ApJ...727...57S} {727, 57}

\bibitem[\protect\citeauthoryear{{Stassun} et~al.,}{{Stassun}
  et~al.}{2018}]{stassun18}
{Stassun} K.~G.,  et~al., 2018, \mn@doi [\aj] {10.3847/1538-3881/aad050}, \href
  {https://ui.adsabs.harvard.edu/abs/2018AJ....156..102S} {156, 102}

\bibitem[\protect\citeauthoryear{{Stumpe}, {Smith}, {Catanzarite}, {Van Cleve},
  {Jenkins}, {Twicken}  \& {Girouard}}{{Stumpe} et~al.}{2014}]{stumpe2014_pdc}
{Stumpe} M.~C.,  {Smith} J.~C.,  {Catanzarite} J.~H.,  {Van Cleve} J.~E.,
  {Jenkins} J.~M.,  {Twicken} J.~D.,   {Girouard} F.~R.,  2014, \mn@doi [\pasp]
  {10.1086/674989}, \href
  {https://ui.adsabs.harvard.edu/abs/2014PASP..126..100S} {126, 100}

\bibitem[\protect\citeauthoryear{{Subjak} et~al.,}{{Subjak}
  et~al.}{2020}]{subjak2019}
{Subjak} J.,  et~al., 2020, \mn@doi [\aj] {10.3847/1538-3881/ab7245}, \href
  {https://ui.adsabs.harvard.edu/abs/2020AJ....159..151S} {159, 151}

\bibitem[\protect\citeauthoryear{{Triaud} et~al.,}{{Triaud}
  et~al.}{2013}]{wasp30b}
{Triaud} A.~H.~M.~J.,  et~al., 2013, \mn@doi [\aap]
  {10.1051/0004-6361/201219643}, \href
  {https://ui.adsabs.harvard.edu/abs/2013A&A...549A..18T} {549, A18}

\bibitem[\protect\citeauthoryear{{Wright} et~al.,}{{Wright}
  et~al.}{2010}]{WISE}
{Wright} E.~L.,  et~al., 2010, \mn@doi [\aj] {10.1088/0004-6256/140/6/1868},
  \href {https://ui.adsabs.harvard.edu/abs/2010AJ....140.1868W} {140, 1868}

\bibitem[\protect\citeauthoryear{Zhou et~al.,}{Zhou et~al.}{2019}]{zhou19}
Zhou G.,  et~al., 2019, \mn@doi [The Astronomical Journal]
  {10.3847/1538-3881/aaf1bb}, 157, 31

\makeatother
\end{thebibliography}
%%%%%%%%%%%%%%%%%%%%%%%%%%%%%%%%%%%%%%%%%%%%%%%%%%

\clearpage

\onecolumn

\appendix

\section{Summary of transiting brown dwarf parameters}\label{app:params}
%%% Table of known transiting/eclipsing brown dwarfs
\begin{table*}
\setlength{\tabcolsep}{4pt}
\centering
\caption[]{List of published transiting brown dwarfs and very low-mass stars as of August 2022. The original discovery paper or the most recent paper is shown in the right-most column. The systems analysed in this work are in boldface in this column.} \label{tab:bdlist}
\begin{tabular}{cccccccccc}
\hline
~~~Name & $P$ (days) & $\rm M_{BD}/M_J$ & $\rm R_{BD}/R_J$& e & $\rm M_\star/\mst$ & $\rm R_\star/\rst$ & $\rm T_{eff} (K)$ & [Fe/H] & Ref. \\
\hline
 HATS-70b & 1.888 & $12.9\pm 1.8$ & $1.38\pm 0.08$ & $<0.18$ & $1.78 \pm 0.12$ & $1.88\pm 0.07$ & $7930\pm 820$ & $+0.04 \pm 0.11$ & 1\\
 TOI-1278b & 14.476 & $18.5 \pm 0.5$ & $1.09 \pm 0.24$ & $0.013 \pm 0.004$ & $0.54 \pm 0.02$ & $0.57 \pm 0.01$ & $3799 \pm 42$ & $+0.00 \pm 0.09$ & 2 \\ 
 \textbf{Kepler-39b} & 21.087 & $19.0 \pm 1.3$ & $1.07 \pm 0.03$ & $0.066 \pm 0.036$ & $1.20 \pm 0.09$ & $1.26 \pm 0.03$ & $6290 \pm 120$ & $+0.07 \pm 0.14$ & 3, \textbf{34}\\
 GPX-1b & 1.745 & $19.7 \pm 1.6$ & $1.47 \pm 0.10$ & 0 (adopted) & $1.68 \pm 0.10$ & $1.56 \pm 0.10$ & $7000 \pm 200$ & $+0.35 \pm 0.10$ & 4\\
 \textbf{CoRoT-3b} & 4.256 & $22.3 \pm 1.0$ & $1.08 \pm 0.05$ & $ 0.012\pm 0.01$ & $1.44 \pm 0.09$ & $1.66 \pm 0.08$ & $6710 \pm 140$ & $-0.04 \pm 0.08$ & 5, \textbf{34}\\
 \textbf{KELT-1b} & 1.217 & $ 27.9 \pm 1.0$ & $1.13 \pm 0.03$ & $<0.0015$ & $1.39 \pm 0.07$ & $1.49 \pm 0.04$ & $6520 \pm 94$ & $+0.07 \pm 0.08$ & 6, \textbf{34}\\
 NLTT 41135b & 2.889 & $33.7 \pm 2.8$ &  $1.13 \pm 0.27$ & $<0.02$ & $0.19 \pm 0.03$ & $0.21 \pm 0.02$ & $3230 \pm 130$ & $-0.25 \pm 0.25$ & 7\\
 \textbf{WASP-128b} & 2.209 & $39.3 \pm 1.0$ & $0.96 \pm 0.02$ & $<0.007$ & $1.21 \pm 0.05$ & $1.18 \pm 0.02$ & $6081 \pm 78$ & $+0.12 \pm 0.09$ & 8, \textbf{34}\\
 CWW 89Ab & 5.293 & $39.2 \pm 1.1$ & $0.94 \pm 0.02$ & $0.189 \pm 0.002$ & $1.10 \pm 0.05$ & $1.03 \pm 0.02$ & $5755 \pm 49$ & $+0.20 \pm 0.09$ & 9, 10 \\
 \textbf{KOI-205b} & 11.720 & $39.7 \pm 2.1$ & $0.87 \pm 0.02$ & $0.026\pm 0.018$ & $0.88 \pm 0.05$ & $0.90 \pm 0.03$ & $5238 \pm 87$ & $+0.15 \pm 0.14$ & 11, \textbf{34}\\
 TOI-1406b & 10.574 & $46.0\pm 2.7$ & $0.86\pm 0.03$ & 0 (adopted) & $1.18 \pm 0.09$ & $1.35\pm 0.03$ & $6290\pm 100$ & $-0.08 \pm 0.09$ & 12\\
 EPIC 212036875b & 5.170 & $52.3 \pm 1.9$ & $0.87 \pm 0.02$ & $0.132 \pm 0.004$ & $1.29 \pm 0.07$ & $1.50 \pm 0.03$ & $6238 \pm 60$ & $+0.01 \pm 0.10$ &  10, 13 \\
 TOI-503b & 3.677 & $53.7 \pm 1.2$ & $1.34\pm 0.26$ & 0 (adopted) & $1.80 \pm 0.06$ & $1.70 \pm 0.05$ & $7650 \pm 160$ & $+0.61 \pm 0.07$ & 14 \\
 TOI-852b & 4.946 & $53.7 \pm 1.4$ & $0.83 \pm 0.04$ & 0 (adopted) & $1.29 \pm 0.03$ & $1.72 \pm 0.04$ & $5750 \pm 76$ & $+0.34 \pm 0.08$ & 15\\
 \textbf{AD 3116b} &  1.983 & $54.6 \pm 6.8$ & $0.95 \pm 0.07$ & $0.144 \pm 0.045$ & $0.28 \pm 0.03$ & $0.29 \pm 0.02$ & $3165 \pm 106$ & $+0.17 \pm 0.09$ & 16, \textbf{34}\\
 CoRoT-33b & 5.819 & $59.0 \pm 1.8$ & $1.10 \pm 0.53$ & $0.070 \pm 0.002$ & $0.86 \pm 0.04$ & $0.94 \pm 0.14$ & $5225 \pm 80$ & $+0.44 \pm 0.10$ & 17\\
 RIK 72b & 97.760 & $59.2 \pm 6.8$ & $3.10 \pm 0.31$ & $0.146 \pm 0.012$ & $0.44 \pm 0.04$ & $0.96 \pm 0.10$ & $3349 \pm 142$ & - & 18\\
 TOI-811b & 25.166 & $59.9 \pm 13.0$ & $1.26 \pm 0.06$ & $0.407\pm 0.046$ & $1.32 \pm 0.05$ & $1.27 \pm 0.06$ & $6107 \pm 76$ & $+0.40 \pm 0.09$ & 15\\ 
 \textbf{WASP-30b} & 4.157 & $61.7 \pm 1.5$ & $0.96 \pm 0.03$ & $<0.004$ & $1.23 \pm 0.05$ & $ 1.42 \pm 0.04$ & $6208 \pm 85$ & $+0.03 \pm 0.09$ & 19, \textbf{34}\\
 TOI-263b & 0.557 & $61.6 \pm 4.0$ & $0.91 \pm 0.07$ & $0.017\pm 0.003$ & $0.44 \pm 0.03$ & $0.44 \pm 0.03$ & $3471 \pm 100$ & $+0.00 \pm 0.09$ & 20\\ 
 LHS 6343c & 12.713 & $62.9 \pm 2.3$ & $0.83 \pm 0.02$ & $0.056 \pm 0.032$ & $0.37\pm 0.01$ & $0.38\pm 0.01$ & $3750 \pm 125$ & $+0.02 \pm 0.19$ & 21\\
 TOI-569b & 6.556 & $63.8 \pm 1.0$ & $0.75 \pm 0.02$ & 0 (adopted) & $1.21 \pm 0.03$ & $1.48 \pm 0.03$ & $5705 \pm 76$ & $+0.40 \pm 0.08$ & 12\\ 
 TOI-2119b & 7.201 & $64.4 \pm 2.3$ & $1.083 \pm 0.03$ & $0.339\pm 0.015$ & $0.53 \pm 0.02$ & $ 0.50 \pm 0.01$ & $3647 \pm 96$ & $+0.08 \pm 0.08$ & 22, 33\\ 
 \textbf{CoRoT-15b} & 3.060 & $64.0 \pm 5.0$ & $0.94  \pm 0.12$ & $0.042\pm 0.039$ & $1.29 \pm 0.12$ & $1.40 \pm 0.14$ & $6340 \pm 190$ & $+0.09 \pm 0.19$ & 23, \textbf{34}\\
 \textbf{KOI-415b} & 166.788 & $64.8 \pm 6.8$ & $0.86 \pm 0.03$ & $0.700 \pm 0.003$ & $0.93 \pm 0.14$ & $1.36 \pm 0.04$ & $5783 \pm 84$ & $-0.24 \pm 0.10$ & 24, \textbf{34}\\
 TOI-1982b & 17.172 & $65.9 \pm 2.8$ & $1.08 \pm 0.04$ & $0.272\pm 0.014$ & $1.41 \pm 0.08$ & $1.51 \pm 0.05$ & $6325 \pm 110$ & $-0.10 \pm 0.09$ & 25\\ 
 TOI-629b & 8.718 & $67.0 \pm 3.0$ & $1.11 \pm 0.05$ & $0.298\pm 0.008$ & $2.16 \pm 0.13$ & $2.37 \pm 0.11$ & $9100 \pm 200$ & $+0.10 \pm 0.15$ & 25\\ 
 TOI-2543b & 7.543 & $67.6 \pm 3.5$ & $0.95 \pm 0.09$ & $0.009\pm 0.003$ & $1.29 \pm 0.08$ & $1.86 \pm 0.15$ & $6060 \pm 82$ & $-0.28 \pm 0.10$ & 25\\
 LP 261-75b & 1.882 & $68.1 \pm 2.1$ & $0.90 \pm 0.02$ & $<0.007$ & $0.30 \pm 0.02$ & $0.31 \pm 0.01$ & $3100 \pm 50$ & - & 26\\
 NGTS-19b & 17.840 & $69.5\pm 5.7$ & $1.03 \pm 0.06$ & $0.377 \pm 0.006$ & $0.81\pm 0.04$ & $0.90\pm 0.04$ & $4716\pm 39$ & - & 29\\
 \textbf{EPIC 201702477b} & 40.737 & $70.9 \pm 2.6$ & $0.83 \pm 0.04$ & $0.228 \pm 0.003$ & $0.95 \pm 0.05$ & $1.01 \pm 0.03$ & $5542 \pm 62$ & $+0.01 \pm 0.03$ & 27, \textbf{34}\\
 CoRoT-34b & 2.119 & $71.4 \pm 8.9$ & $1.09 \pm 0.17$ & 0 (adopted) & $1.66 \pm 0.15$ & $1.85 \pm 0.29$ & $7820 \pm 160$ & $-0.02 \pm 0.20$ & 28\\
 NGTS-7Ab & 0.676 & $75.5\pm 13.7$ & $1.38\pm 0.14$ & 0 (adopted) & $0.48\pm 0.13$ & $0.61\pm 0.06$ & $3359\pm 106$ & - & 30\\
 TOI-148b & 4.870 & $77.1 \pm 5.8$ & $0.81 \pm 0.06$ & $<0.01$ & $0.97 \pm 0.12$ & $1.20 \pm 0.07$ & $5900 \pm 140$ & $-0.24 \pm 0.25$ & 31\\ 
 TOI-587b & 8.042 & $79.9 \pm 5.3$ & $1.38 \pm 0.04$ & 0 (adopted) & $2.32 \pm 0.11$ & $2.10 \pm 0.06$ & $9780 \pm 200$ & $+0.00 \pm 0.10$ & 31\\
 \textbf{KOI-189b} & 30.360 & $80.4 \pm 2.3$ & $0.99 \pm 0.02$ & $0.274 \pm 0.004$ & $0.80 \pm 0.04$ & $0.77 \pm 0.01$ & $4973 \pm 84$ & $+0.04 \pm 0.14$ & 32, \textbf{34}\\
 TOI-746b & 10.980 & $82.2 \pm 4.9$ & $0.95 \pm 0.09$ & $0.199\pm 0.003$ & $0.94 \pm 0.13$ & $0.97 \pm 0.07$ & $5690 \pm 140$ & $-0.20 \pm 0.23$ & 31\\ 
\hline
\end{tabular}
 \begin{list}{}{}
 \item[References:]  1 - \cite{zhou19}, 2 - \cite{artigau2021},  3 - \cite{kepler39}, 4 - \cite{benni2020}, 5 - \cite{corot3b}, 6 - \cite{kelt1b},  7 - \cite{irwin10}, 8 - \cite{wasp128b}, 9 - \cite{nowak17}, 10 - \cite{carmichael19}, 11 - \cite{diaz13}, 12 - \cite{carmichael2020}, 13 - \cite{persson2019}, 14 - \cite{subjak2019}, 15 - \cite{carmichael2021}, 16 - \cite{ad3116}, 17 - \cite{corot33b}, 18 - \cite{david19_bd}, 19 - \cite{wasp30b}, 20 - \cite{palle2021}, 21 - \cite{johnson11_bd}, 22 - \cite{carmichael2022}, 23 - \cite{corot15b}, 24 - \cite{moutou13}, 25 - \cite{psaridi2022}, 26 - \cite{irwin18}, 27 - \cite{bayliss16}, 28 - \cite{corot34}, 29 - \cite{acton2021}, 30 - \cite{jackman2019}, 31 - \cite{grieves2021}, 32 - \cite{diaz14}, 33 - \cite{canas2022}, {\bf 34 - this work}

 \end{list}
\end{table*}

\clearpage

\begin{figure*}
    \centering
    \includegraphics[width=0.95\textwidth, trim={1.0cm 0.0cm 0.0cm 0.0cm}]{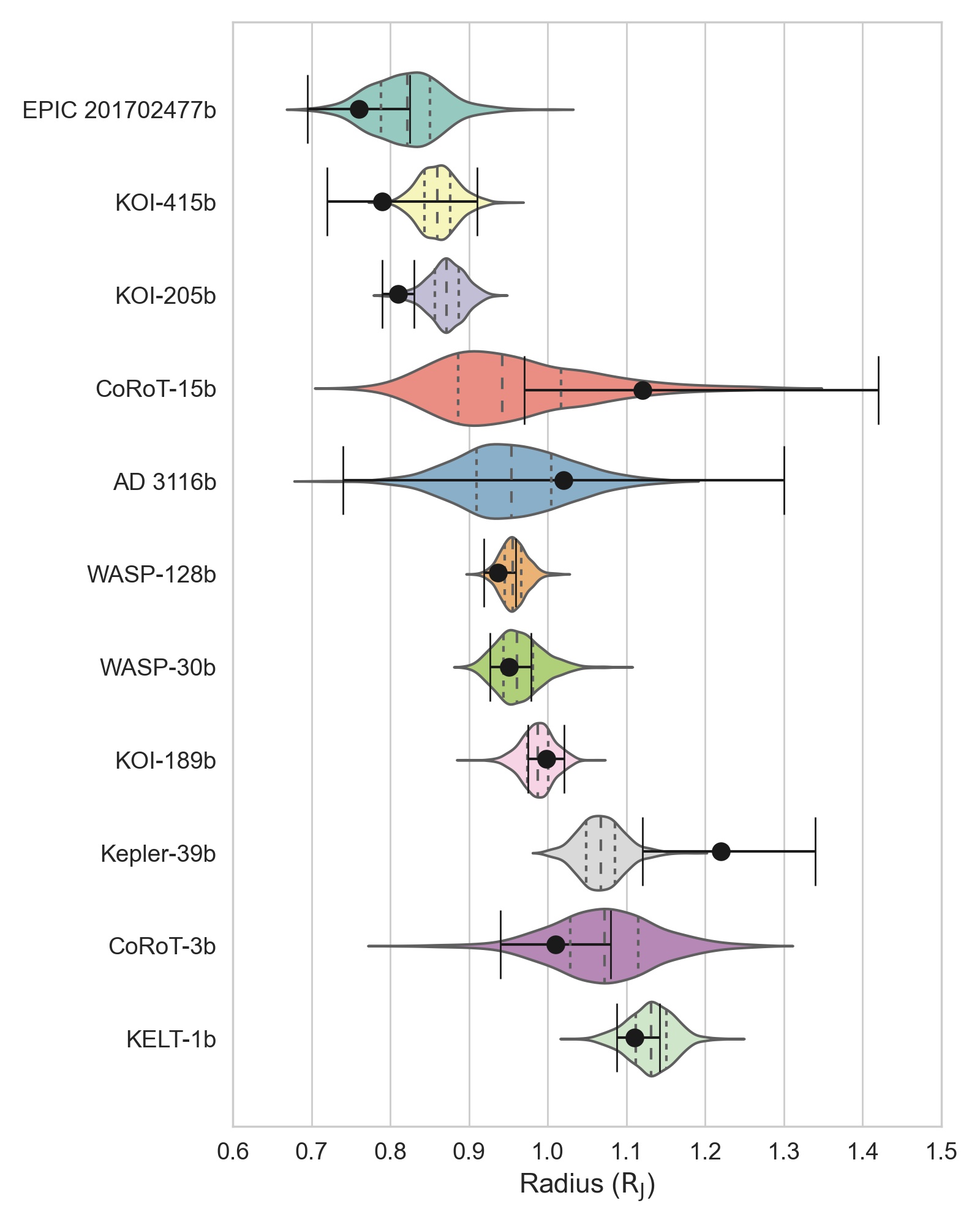}
    \caption{Brown dwarf radius posterior distribution functions from {\tt EXOFASTv2}. Dashed lines show the median values from this work and dotted lines show the $1\sigma$ range from this work. The black points show these same metrics but for previous works.}
    \label{fig:pdfs_radius}
\end{figure*}

\clearpage

\begin{figure*}
    \centering
    \includegraphics[width=0.95\textwidth, trim={1.0cm 0.0cm 0.0cm 0.0cm}]{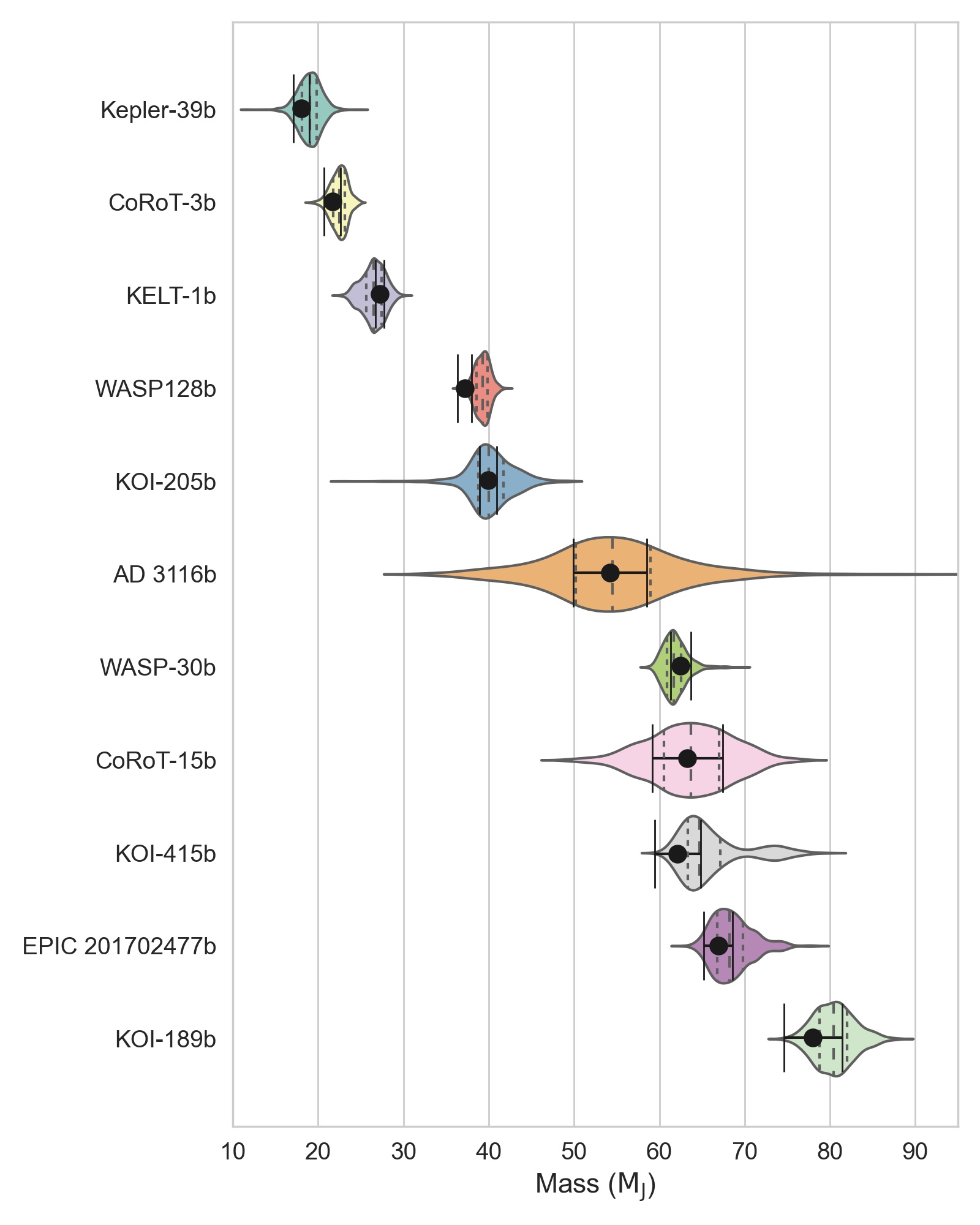}
    \caption{Brown dwarf mass posterior distribution functions from {\tt EXOFASTv2}. Dashed lines show the median values from this work and dotted lines show the $1\sigma$ range from this work. The black points show these same metrics but for previous works.}
    \label{fig:pdfs_mass}
\end{figure*}

\clearpage

\section{Global fits with {\tt EXOFASTv2}}

\begin{table*}
\centering
\caption[]{MIST median values and 68\% confidence interval for AD 3116, created using {\tt EXOFASTv2} commit number f8f3437. Here, $\mathcal{U}$[a,b] is the uniform prior bounded between $a$ and $b$, and $\mathcal{G}[a,b]$ is a Gaussian prior of mean $a$ and width $b$. I adopt the age of Praesepe from \cite{gossage2018} and use a mass prior from \cite{mann2019}.} \label{tab:exofast_ad3116}

\begin{tabular}{lcccc}
~~~Parameter & Units & Priors & Values\\
\hline
\multicolumn{2}{l}{\bf Stellar Parameters:}&\smallskip\\
~~~~$M_*$\dotfill &Mass (\mst)\dotfill & $\mathcal{G}[0.265,0.05]$ &$0.280^{+0.030}_{-0.032}$\\
~~~~$R_*$\dotfill &Radius (\rst)\dotfill & - &$0.288^{+0.021}_{-0.020}$\\
~~~~$L_*$\dotfill &Luminosity (\lst)\dotfill & - &$0.00751^{+0.00071}_{-0.00073}$\\
~~~~$\rho_*$\dotfill &Density (cgs)\dotfill & - &$16.6^{+2.5}_{-2.3}$\\
~~~~$\log{g}$\dotfill &Surface gravity (cgs)\dotfill & -  &$4.967^{+0.039}_{-0.042}$\\
~~~~$T_{\rm eff}$\dotfill &Effective Temperature (K)\dotfill & $\mathcal{G}[3191,100]$ &$3165^{+96}_{-106}$\\
~~~~$[{\rm Fe/H}]$\dotfill &Metallicity (dex)\dotfill & $\mathcal{G}[0.14,0.09]$ &$0.165^{+0.092}_{-0.094}$\\
~~~~$Age_{\rm Praesepe}$\dotfill &Age (Gyr)\dotfill & \cite{gossage2018} &$0.617\pm 0.017$\\
~~~~$A_V$\dotfill &V-band extinction (mag)\dotfill & $\mathcal{U}[0,0.079]$ &$0.041^{+0.026}_{-0.028}$ \\
~~~~$\sigma_{SED}$\dotfill &SED photometry error scaling \dotfill & - &$15.6^{+10.}_{-6.2}$\\
~~~~$\varpi$\dotfill &Parallax (mas)\dotfill & $\mathcal{G}[5.3142, 0.1397]$ &$5.39\pm0.14$\\
~~~~$d$\dotfill &Distance (pc)\dotfill & - & $185.4^{+4.8}_{-4.6}$\\
\hline
\multicolumn{1}{l}{\bf Brown Dwarf Parameters:}& \\
~~~~$M_b$\dotfill &Mass (\mj)\dotfill  & - &$54.6^{+6.8}_{-6.9}$\\
~~~~$R_b$\dotfill &Radius (\rj)\dotfill & - & $0.954^{+0.070}_{-0.065}$\\
~~~~$P_{\rm orb}$\dotfill & Period (days)\dotfill & - &$1.98279482\pm0.00000030$\\
~~~~$T_C$\dotfill &Time of conjunction (\bjdtdb)\dotfill & - &$2457815.294810\pm 0.000077$\\
~~~~$a$\dotfill &Semi-major axis (AU)\dotfill  & - &$0.02140^{+0.00071}_{-0.00079}$\\
~~~~$i$\dotfill &Orbital inclination (Degrees)\dotfill & - &$89.48^{+0.36}_{-0.47}$\\
~~~~$e$\dotfill &Eccentricity \dotfill  & - &$0.144^{+0.045}_{-0.025}$\\
~~~~$ecos{\omega_*}$\dotfill & \dotfill & - &$0.128^{+0.028}_{-0.025}$\\
~~~~$esin{\omega_*}$\dotfill & \dotfill & - &$0.059^{+0.055}_{-0.048}$\\
~~~~$\fave$\dotfill &Incident Flux (\fluxcgs)\dotfill & - &$0.0218^{+0.0015}_{-0.0014}$\\
~~~~$T_{eq}$\dotfill &Equilibrium temperature (K)\dotfill  & - &$559.9^{+10.}_{-10.0}$\\
~~~~$K$\dotfill &RV semi-amplitude ($\rm m\, s^{-1}$)\dotfill  & - &$18800^{+1400}_{-1800}$\\
~~~~$R_b/R_*$\dotfill &Radius of planet in stellar radii \dotfill  & - &$0.3406^{+0.0044}_{-0.0041}$\\
~~~~$a/R_*$\dotfill &Semi-major axis in stellar radii \dotfill  & - &$16.00^{+0.80}_{-0.83}$\\
~~~~$\delta$\dotfill &Transit depth (fraction)\dotfill  & - &$0.1160^{+0.0030}_{-0.0028}$\\
~~~~$T_{14}$\dotfill &Total transit duration (days)\dotfill & - & $0.04892^{+0.00064}_{-0.00056}$\\
~~~~$\tau$\dotfill &Ingress/egress transit duration (days)\dotfill  & - &$0.01258^{+0.00076}_{-0.00024}$\\
~~~~$b$\dotfill &Transit Impact parameter \dotfill  & - &$0.134^{+0.12}_{-0.093}$\\
~~~~$logg_b$\dotfill &Surface gravity \dotfill & -  &$5.179^{+0.059}_{-0.080}$\\
~~~~$M_b\sin i$\dotfill &Minimum mass (\mj)\dotfill & -  &$54.6^{+6.8}_{-6.9}$\\
~~~~$M_b/M_*$\dotfill &Mass ratio \dotfill & -  &$0.188^{+0.019}_{-0.023}$\\
\hline
\multicolumn{2}{l}{Wavelength Parameters:}&&Kepler\smallskip\\
~~~~$u_{1}$\dotfill &linear limb-darkening coeff \dotfill &-&$0.393^{+0.054}_{-0.055}$\\
~~~~$u_{2}$\dotfill &quadratic limb-darkening coeff \dotfill &-&$0.327\pm0.053$\\
\multicolumn{2}{l}{Transit Parameters:}&&Kepler UT 2015-06-05\smallskip\\
~~~~$\sigma^{2}$\dotfill &Added Variance \dotfill &-&$0.00003014^{+0.00000064}_{-0.00000062}$\\
~~~~$F_0$\dotfill &Baseline flux \dotfill &-&$1.000397\pm0.000082$\\
\multicolumn{2}{l}{Telescope Parameters:}&&HIRES\smallskip\\
~~~~$\gamma_{\rm rel}$\dotfill &Relative RV Offset ($\rm m\, s^{-1}$)\dotfill &-&$34710^{+990}_{-880}$\\
~~~~$\sigma_J$\dotfill &RV Jitter ($\rm m\, s^{-1}$)\dotfill &-&$1760^{+2100}_{-990}$\\

\hline
\end{tabular}
\end{table*}

\begin{table*}
\centering
\caption[]{MIST median values and 68\% confidence interval for CoRoT-3, created using {\tt EXOFASTv2} commit number f8f3437. Here, $\mathcal{U}$[a,b] is the uniform prior bounded between $a$ and $b$, and $\mathcal{G}[a,b]$ is a Gaussian prior of mean $a$ and width $b$.} \label{tab:exofast_corot3}

\begin{tabular}{lcccc}
~~~Parameter & Units & Priors & Values\\
\hline
\multicolumn{2}{l}{\bf Stellar Parameters:}&\smallskip\\
~~~~$M_*$\dotfill &Mass (\mst)\dotfill & - &$1.437^{+0.081}_{-0.086}$\\
~~~~$R_*$\dotfill &Radius (\rst)\dotfill & - &$1.661^{+0.076}_{-0.065}$\\
~~~~$L_*$\dotfill &Luminosity (\lst)\dotfill & - &$5.02^{+0.57}_{-0.53}$\\
~~~~$\rho_*$\dotfill &Density (cgs)\dotfill & - &$0.444^{+0.061}_{-0.057}$\\
~~~~$\log{g}$\dotfill &Surface gravity (cgs)\dotfill & -  &$4.156^{+0.039}_{-0.044}$\\
~~~~$T_{\rm eff}$\dotfill &Effective Temperature (K)\dotfill & $\mathcal{G}[6740, 150]$ &$6710\pm 140$\\
~~~~$[{\rm Fe/H}]$\dotfill &Metallicity (dex)\dotfill & $\mathcal{G}[-0.02,0.1]$ &$-0.037^{+0.100}_{-0.099}$\\
~~~~$Age$\dotfill &Age (Gyr)\dotfill & - &$1.58^{+0.80}_{-0.50}$\\
~~~~$A_V$\dotfill &V-band extinction (mag)\dotfill & $\mathcal{U}[0.0,0.9663]$ &$0.68^{+0.20}_{-0.34}$ \\
~~~~$\sigma_{SED}$\dotfill &SED photometry error scaling \dotfill & - &$3.05^{+1.9}_{-0.92}$\\
~~~~$\varpi$\dotfill &Parallax (mas)\dotfill & $\mathcal{G}[1.2716,0.0177]$ &$1.307^{+0.018}_{-0.017}$\\
~~~~$d$\dotfill &Distance (pc)\dotfill & - & $764\pm10$\\
\hline
\multicolumn{1}{l}{\bf Brown Dwarf Parameters:}& \\
~~~~$M_b$\dotfill &Mass (\mj)\dotfill  & - &$22.32^{+0.84}_{-1.00}$\\
~~~~$R_b$\dotfill &Radius (\rj)\dotfill & - & $1.080^{+0.051}_{-0.046}$\\
~~~~$P_{\rm orb}$\dotfill & Period (days)\dotfill & - &$4.25636^{+0.00028}_{-0.00023}$\\
~~~~$T_C$\dotfill &Time of conjunction (\bjdtdb)\dotfill & -  &$2459744.5649^{+0.0025}_{-0.0030}$\\
~~~~$a$\dotfill &Semi-major axis (AU)\dotfill  & - &$0.0583^{+0.0011}_{-0.0012}$\\
~~~~$i$\dotfill &Orbital inclination (Degrees)\dotfill & - &$85.52^{+0.57}_{-0.60}$\\
~~~~$e$\dotfill &Eccentricity \dotfill  & - &$0.0120^{+0.018}_{-0.0087}$\\
~~~~$ecos{\omega_*}$\dotfill & \dotfill & - &$-0.0022^{+0.0051}_{-0.0080}$\\
~~~~$esin{\omega_*}$\dotfill & \dotfill & - &$0.003^{+0.020}_{-0.010}$\\
~~~~$\fave$\dotfill &Incident Flux (\fluxcgs)\dotfill & - &$2.01^{+0.21}_{-0.18}$\\
~~~~$T_{eq}$\dotfill &Equilibrium temperature (K)\dotfill  & - &$1726^{+44}_{-41}$\\
~~~~$K$\dotfill &RV semi-amplitude ($\rm m\, s^{-1}$)\dotfill  & - &$2173^{+32}_{-45}$\\
~~~~$R_b/R_*$\dotfill &Radius of planet in stellar radii \dotfill  & - &$0.06677^{+0.00062}_{-0.00034}$\\
~~~~$a/R_*$\dotfill &Semi-major axis in stellar radii \dotfill  & - &$7.56^{+0.33}_{-0.34}$\\
~~~~$\delta$\dotfill &Transit depth (fraction)\dotfill  & - &$0.004458^{+0.000083}_{-0.000045}$\\
~~~~$T_{14}$\dotfill &Total transit duration (days)\dotfill & - &$0.1597\pm0.0026$\\
~~~~$\tau$\dotfill &Ingress/egress transit duration (days)\dotfill  & - &$0.0149^{+0.0016}_{-0.0014}$\\
~~~~$b$\dotfill &Transit Impact parameter \dotfill  & - &$0.588^{+0.049}_{-0.058}$\\
~~~~$logg_b$\dotfill &Surface gravity \dotfill & -  &$4.675^{+0.041}_{-0.042}$\\
~~~~$M_b\sin i$\dotfill &Minimum mass (\mj)\dotfill & -  &$22.25^{+0.84}_{-1.0}$\\
~~~~$M_b/M_*$\dotfill &Mass ratio \dotfill & -  &$0.01485^{+0.00040}_{-0.00047}$\\
\hline
\multicolumn{2}{l}{Wavelength Parameters:}& TESS&CoRoT\smallskip\\
~~~~$A_D$\dotfill &Dilution from neighboring stars \dotfill &$0.417^{+0.041}_{-0.026}$ &-\\
~~~~$u_{1}$\dotfill &linear limb-darkening coeff \dotfill &$0.190^{+0.053}_{-0.055}$&$0.260^{+0.052}_{-0.050}$\\
~~~~$u_{2}$\dotfill &quadratic limb-darkening coeff \dotfill &$0.302^{+0.040}_{-0.020}$&$0.300^{+0.046}_{-0.045}$\\
\multicolumn{2}{l}{Transit Parameters:}&TESS UT 2022-07-09&CoRoT UT 2007-06-06\smallskip\\
~~~~$\sigma^{2}$\dotfill &Added Variance \dotfill &$-0.00000332^{+0.00000053}_{-0.00000020}$&$-0.000202591^{+0.000000027}_{-0.000000025}$\\
~~~~$F_0$\dotfill &Baseline flux \dotfill &$1.000112^{+0.000038}_{-0.000041}$&$1.000138\pm0.000018$\\
\multicolumn{2}{l}{Telescope Parameters:}&&CORALIE\smallskip\\
~~~~$\gamma_{\rm rel}$\dotfill &Relative RV Offset ($\rm m\, s^{-1}$)\dotfill &-&$-56659^{+95}_{-100}$\\
~~~~$\sigma_J$\dotfill &RV Jitter ($\rm m\, s^{-1}$)\dotfill &-&$190^{+160}_{-140}$\\

\multicolumn{2}{l}{Telescope Parameters:}&&HARPS\smallskip\\
~~~~$\gamma_{\rm rel}$\dotfill &Relative RV Offset ($\rm m\, s^{-1}$)\dotfill &-&$-56156^{+52}_{-48}$\\
~~~~$\sigma_J$\dotfill &RV Jitter ($\rm m\, s^{-1}$)\dotfill &-&$73^{+230}_{-74}$\\

\multicolumn{2}{l}{Telescope Parameters:}&&SOHPIE\smallskip\\
~~~~$\gamma_{\rm rel}$\dotfill &Relative RV Offset ($\rm m\, s^{-1}$)\dotfill &-&$-56205^{+39}_{-43}$\\
~~~~$\sigma_J$\dotfill &RV Jitter ($\rm m\, s^{-1}$)\dotfill &-&$82^{+60}_{-45}$\\

\multicolumn{2}{l}{Telescope Parameters:}&&Sandiford\smallskip\\
~~~~$\gamma_{\rm rel}$\dotfill &Relative RV Offset ($\rm m\, s^{-1}$)\dotfill &-&$-55500^{+1300}_{-1400}$\\
~~~~$\sigma_J$\dotfill &RV Jitter ($\rm m\, s^{-1}$)\dotfill &-&$1700^{+9400}_{-1300}$\\

\multicolumn{2}{l}{Telescope Parameters:}&&TLS\smallskip\\
~~~~$\gamma_{\rm rel}$\dotfill &Relative RV Offset ($\rm m\, s^{-1}$)\dotfill &-&$-55300^{+3500}_{-2100}$\\
~~~~$\sigma_J$\dotfill &RV Jitter ($\rm m\, s^{-1}$)\dotfill &-&$2500^{+24000}_{-2300}$\\
\hline
\end{tabular}
\end{table*}

\begin{table*}
\centering
\caption[]{MIST median values and 68\% confidence interval for CoRoT-15, created using {\tt EXOFASTv2} commit number f8f3437. Here, $\mathcal{U}$[a,b] is the uniform prior bounded between $a$ and $b$, and $\mathcal{G}[a,b]$ is a Gaussian prior of mean $a$ and width $b$.} \label{tab:exofast_corot15}

\begin{tabular}{lcccc}
~~~Parameter & Units & Priors & Values\\
\hline
\multicolumn{2}{l}{\bf Stellar Parameters:}&\smallskip\\
~~~~$M_*$\dotfill &Mass (\mst)\dotfill & - &$1.29^{+0.11}_{-0.12}$\\
~~~~$R_*$\dotfill &Radius (\rst)\dotfill & - &$1.401^{+0.14}_{-0.094}$\\
~~~~$L_*$\dotfill &Luminosity (\lst)\dotfill & - &$2.90^{+0.70}_{-0.51}$\\
~~~~$\rho_*$\dotfill &Density (cgs)\dotfill & - &$0.66^{+0.13}_{-0.15}$\\
~~~~$\log{g}$\dotfill &Surface gravity (cgs)\dotfill & -  &$4.252^{+0.051}_{-0.076}$\\
~~~~$T_{\rm eff}$\dotfill &Effective Temperature (K)\dotfill & $\mathcal{G}[6350,200]$ &$6340^{+190}_{-180}$\\
~~~~$[{\rm Fe/H}]$\dotfill &Metallicity (dex)\dotfill & $\mathcal{G}[0.10,0.2]$ &$0.09\pm0.19$\\
~~~~$Age$\dotfill &Age (Gyr)\dotfill & - &$2.1^{+2.0}_{-1.3}$\\
~~~~$A_V$\dotfill &V-band extinction (mag)\dotfill & $\mathcal{U}[0,3.243]$ &$1.23^{+0.30}_{-0.31}$ \\
~~~~$\sigma_{SED}$\dotfill &SED photometry error scaling \dotfill & - &$1.60^{+1.1}_{-0.52}$\\
~~~~$\varpi$\dotfill &Parallax (mas)\dotfill & $\mathcal{G}[0.6750,0.0422]$ &$0.731^{+0.055}_{-0.064}$\\
~~~~$d$\dotfill &Distance (pc)\dotfill & - & $1367^{+130}_{-95}$\\
\hline
\multicolumn{1}{l}{\bf Brown Dwarf Parameters:}& \\
~~~~$M_b$\dotfill &Mass (\mj)\dotfill  & - &$64.0^{+4.7}_{-5.0}$\\
~~~~$R_b$\dotfill &Radius (\rj)\dotfill & - & $0.943^{+0.12}_{-0.073}$\\
~~~~$P_{\rm orb}$\dotfill & Period (days)\dotfill & - &$3.06012^{+0.00022}_{-0.00029}$\\
~~~~$T_C$\dotfill &Time of conjunction (\bjdtdb)\dotfill & - &$2454768.8571\pm0.0013$\\
~~~~$a$\dotfill &Semi-major axis (AU)\dotfill  & - &$0.0456^{+0.0012}_{-0.0015}$\\
~~~~$i$\dotfill &Orbital inclination (Degrees)\dotfill & - &$87.0^{+1.9}_{-1.8}$\\
~~~~$e$\dotfill &Eccentricity \dotfill  & - &$0.042^{+0.039}_{-0.029}$\\
~~~~$ecos{\omega_*}$\dotfill & \dotfill & - &$0.029^{+0.042}_{-0.030}$\\
~~~~$esin{\omega_*}$\dotfill & \dotfill & - &$-0.002^{+0.023}_{-0.027}$\\
~~~~$\fave$\dotfill &Incident Flux (\fluxcgs)\dotfill & - &$1.90^{+0.39}_{-0.28}$\\
~~~~$T_{eq}$\dotfill &Equilibrium temperature (K)\dotfill  & - &$1702^{+81}_{-67}$\\
~~~~$K$\dotfill &RV semi-amplitude ($\rm m\, s^{-1}$)\dotfill  & - &$7350\pm300$\\
~~~~$R_b/R_*$\dotfill &Radius of planet in stellar radii \dotfill  & - &$0.0694^{+0.0021}_{-0.0019}$\\
~~~~$a/R_*$\dotfill &Semi-major axis in stellar radii \dotfill  & - &$6.98^{+0.42}_{-0.58}$\\
~~~~$\delta$\dotfill &Transit depth (fraction)\dotfill  & - &$0.00481^{+0.00030}_{-0.00026}$\\
~~~~$T_{14}$\dotfill &Total transit duration (days)\dotfill & - &$0.1407^{+0.0031}_{-0.0030}$\\
~~~~$\tau$\dotfill &Ingress/egress transit duration (days)\dotfill  & - &$0.0105^{+0.0026}_{-0.0013}$\\
~~~~$b$\dotfill &Transit Impact parameter \dotfill  & - &$0.37^{+0.17}_{-0.22}$\\
~~~~$logg_b$\dotfill &Surface gravity \dotfill & -  &$5.249^{+0.066}_{-0.096}$\\
~~~~$M_b\sin i$\dotfill &Minimum mass (\mj)\dotfill & -  &$63.8^{+4.7}_{-5.0}$\\
~~~~$M_b/M_*$\dotfill &Mass ratio \dotfill & -  &$0.0476^{+0.0027}_{-0.0025}$\\
\hline
\multicolumn{2}{l}{Wavelength Parameters:}&&CoRoT\smallskip\\
~~~~$u_{1}$\dotfill &linear limb-darkening coeff \dotfill &-&$0.325^{+0.054}_{-0.053}$\\
~~~~$u_{2}$\dotfill &quadratic limb-darkening coeff \dotfill &-&$0.308^{+0.051}_{-0.050}$\\
\multicolumn{2}{l}{Transit Parameters:}&&CoRoT UT 2008-10-14 \smallskip\\
~~~~$\sigma^{2}$\dotfill &Added Variance \dotfill &-&$0.00000656^{+0.00000017}_{-0.00000016}$\\
~~~~$F_0$\dotfill &Baseline flux \dotfill &-&$0.999387\pm0.000044$\\
\multicolumn{2}{l}{Telescope Parameters:}&&HARPS\smallskip\\
~~~~$\gamma_{\rm rel}$\dotfill &Relative RV Offset ($\rm m\, s^{-1}$)\dotfill &-&$2270^{+180}_{-200}$\\
~~~~$\sigma_J$\dotfill &RV Jitter ($\rm m\, s^{-1}$)\dotfill &-&$400^{+340}_{-270}$\\
\multicolumn{2}{l}{Transit Parameters:}&&CoRoT\smallskip\\
\multicolumn{2}{l}{Telescope Parameters:}&&HIRES\smallskip\\
~~~~$\gamma_{\rm rel}$\dotfill &Relative RV Offset ($\rm m\, s^{-1}$)\dotfill &-&$1130^{+430}_{-360}$\\
~~~~$\sigma_J$\dotfill &RV Jitter ($\rm m\, s^{-1}$)\dotfill &-&$700^{+960}_{-400}$\\
\hline
\end{tabular}
\end{table*}

\begin{table*}
\centering
\caption[]{MIST median values and 68\% confidence interval for EPIC 201702477, created using {\tt EXOFASTv2} commit number f8f3437. Here, $\mathcal{U}$[a,b] is the uniform prior bounded between $a$ and $b$, and $\mathcal{G}[a,b]$ is a Gaussian prior of mean $a$ and width $b$.} \label{tab:exofast_ep201}

\begin{tabular}{lcccc}
~~~Parameter & Units & Priors & Values\\
\hline
\multicolumn{2}{l}{\bf Stellar Parameters:}&\smallskip\\
~~~~$M_*$\dotfill &Mass (\mst)\dotfill & - &$0.951^{+0.050}_{-0.051}$\\
~~~~$R_*$\dotfill &Radius (\rst)\dotfill & - &$1.013^{+0.032}_{-0.031}$\\
~~~~$L_*$\dotfill &Luminosity (\lst)\dotfill & - &$0.872^{+0.046}_{-0.043}$\\
~~~~$\rho_*$\dotfill &Density (cgs)\dotfill & - &$1.28^{+0.14}_{-0.12}$\\
~~~~$\log{g}$\dotfill &Surface gravity (cgs)\dotfill & -  &$4.404^{+0.033}_{-0.031}$\\
~~~~$T_{\rm eff}$\dotfill &Effective Temperature (K)\dotfill & $\mathcal{G}[5517,100]$ &$5542^{+62}_{-58}$\\
~~~~$[{\rm Fe/H}]$\dotfill &Metallicity (dex)\dotfill & $\mathcal{G}[-0.164,0.1]$ &$0.014^{+0.033}_{-0.023}$\\
~~~~$Age$\dotfill &Age (Gyr)\dotfill & - &$6.4^{+4.6}_{-6.3}$\\
~~~~$A_V$\dotfill &V-band extinction (mag)\dotfill & $\mathcal{U}[0,0.06076]$ &$0.035^{+0.018}_{-0.021}$\\
~~~~$\sigma_{SED}$\dotfill &SED photometry error scaling \dotfill & - &$3.27^{+1.2}_{-0.72}$\\
~~~~$\varpi$\dotfill &Parallax (mas)\dotfill & $\mathcal{G}[1.1944,0.0289]$ &$1.245^{+0.028}_{-0.029}$\\
~~~~$d$\dotfill &Distance (pc)\dotfill & - & $802^{+19}_{-18}$\\
\hline
\multicolumn{1}{l}{\bf Brown Dwarf Parameters:}& \\
~~~~$M_b$\dotfill &Mass (\mj)\dotfill  & - &$70.9^{+2.6}_{-2.4}$\\
~~~~$R_b$\dotfill &Radius (\rj)\dotfill & - & $0.830^{+0.040}_{-0.036}$\\
~~~~$P_{\rm orb}$\dotfill & Period (days)\dotfill & - &$40.73746^{+0.00045}_{-0.00047}$\\
~~~~$T_C$\dotfill &Time of conjunction (\bjdtdb)\dotfill & - &$2456893.0175\pm0.0011$\\
~~~~$a$\dotfill &Semi-major axis (AU)\dotfill  & - &$0.2332^{+0.0039}_{-0.0041}$\\
~~~~$i$\dotfill &Orbital inclination (Degrees)\dotfill & - &$89.007^{+0.046}_{-0.043}$\\
~~~~$e$\dotfill &Eccentricity \dotfill  & - &$0.2282^{+0.0031}_{-0.0033}$\\
~~~~$ecos{\omega_*}$\dotfill & \dotfill & - &$-0.2192^{+0.0024}_{-0.0025}$\\
~~~~$esin{\omega_*}$\dotfill & \dotfill & - &$-0.0627^{+0.0076}_{-0.0068}$\\
~~~~$\fave$\dotfill &Incident Flux (\fluxcgs)\dotfill & - &$0.0208\pm0.0011$\\
~~~~$T_{eq}$\dotfill &Equilibrium temperature (K)\dotfill  & - &$557.2^{+7.4}_{-7.7}$\\
~~~~$K$\dotfill &RV semi-amplitude ($\rm m\, s^{-1}$)\dotfill  & - &$4251^{+29}_{-33}$\\
~~~~$R_b/R_*$\dotfill &Radius of planet in stellar radii \dotfill  & - &$0.0843^{+0.0018}_{-0.0017}$\\
~~~~$a/R_*$\dotfill &Semi-major axis in stellar radii \dotfill  & - &$49.4^{+1.7}_{-1.5}$\\
~~~~$\delta$\dotfill &Transit depth (fraction)\dotfill  & - &$0.00711^{+0.00031}_{-0.00028}$\\
~~~~$T_{14}$\dotfill &Total transit duration (days)\dotfill & - &$0.1774^{+0.0038}_{-0.0037}$\\
~~~~$\tau$\dotfill &Ingress/egress transit duration (days)\dotfill  & - &$0.0485^{+0.0051}_{-0.0045}$\\
~~~~$b$\dotfill &Transit Impact parameter \dotfill  & - &$0.867^{+0.011}_{-0.012}$\\
~~~~$logg_b$\dotfill &Surface gravity \dotfill & -  &$5.405^{+0.042}_{-0.041}$\\
~~~~$M_b\sin i$\dotfill &Minimum mass (\mj)\dotfill & -  &$70.9^{+2.6}_{-2.4}$\\
~~~~$M_b/M_*$\dotfill &Mass ratio \dotfill & -  &$0.0712\pm0.0014$\\
\hline
\multicolumn{2}{l}{Wavelength Parameters:}&&Kepler\smallskip\\
~~~~$u_{1}$\dotfill &linear limb-darkening coeff \dotfill &-&$0.450^{+0.050}_{-0.046}$\\
~~~~$u_{2}$\dotfill &quadratic limb-darkening coeff \dotfill &-&$0.242\pm0.047$\\
\multicolumn{2}{l}{Wavelength Parameters:}&&r$^\prime$\smallskip\\
~~~~$u_{1}$\dotfill &linear limb-darkening coeff \dotfill &-&$0.429^{+0.035}_{-0.036}$\\
~~~~$u_{2}$\dotfill &quadratic limb-darkening coeff \dotfill &-&$0.250\pm0.036$\\
\multicolumn{2}{l}{Transit Parameters:}&&Kepler UT 2014-07-13 \smallskip\\
~~~~$\sigma^{2}$\dotfill &Added Variance \dotfill &-&$0.0000000857^{+0.0000000034}_{-0.0000000032}$\\
~~~~$F_0$\dotfill &Baseline flux \dotfill &-&$1.0000145^{+0.0000063}_{-0.0000062}$\\
\multicolumn{2}{l}{Transit Parameters:}&&CTIO UT 2015-03-15 (r$^\prime$)\smallskip\\
~~~~$\sigma^{2}$\dotfill &Added Variance \dotfill &-&$0.00000097^{+0.00000055}_{-0.00000045}$\\
~~~~$F_0$\dotfill &Baseline flux \dotfill &-&$1.00206\pm0.00022$\\
\multicolumn{2}{l}{Transit Parameters:}&& SAAO UT 2015-04-24 (r$^\prime$)\smallskip\\
~~~~$\sigma^{2}$\dotfill &Added Variance \dotfill &-&$-0.00000070^{+0.00000081}_{-0.00000066}$\\
~~~~$F_0$\dotfill &Baseline flux \dotfill &-&$1.00138^{+0.00025}_{-0.00024}$\\
\multicolumn{2}{l}{Telescope Parameters:}&&HARPS\smallskip\\
~~~~$\gamma_{\rm rel}$\dotfill &Relative RV Offset ($\rm m\, s^{-1}$)\dotfill &-&$34742^{+22}_{-24}$\\
~~~~$\sigma_J$\dotfill &RV Jitter ($\rm m\, s^{-1}$)\dotfill &-&$48^{+39}_{-23}$\\
\multicolumn{2}{l}{Telescope Parameters:}&&SOPHIE\smallskip\\
~~~~$\gamma_{\rm rel}$\dotfill &Relative RV Offset ($\rm m\, s^{-1}$)\dotfill &-&$34700^{+1100}_{-1600}$\\
~~~~$\sigma_J$\dotfill &RV Jitter ($\rm m\, s^{-1}$)\dotfill &-&$1400^{+11000}_{-1200}$\\
\hline
\end{tabular}
\end{table*}

\begin{table*}
\centering
\caption[]{MIST median values and 68\% confidence interval for KELT-1, created using {\tt EXOFASTv2} commit number f8f3437. Here, $\mathcal{U}$[a,b] is the uniform prior bounded between $a$ and $b$, and $\mathcal{G}[a,b]$ is a Gaussian prior of mean $a$ and width $b$.} \label{tab:exofast_kelt1}

\begin{tabular}{lcccc}
~~~Parameter & Units & Priors & Values\\
\hline
\multicolumn{2}{l}{\bf Stellar Parameters:}&\smallskip\\
~~~~$M_*$\dotfill &Mass (\mst)\dotfill & - &$1.390^{+0.065}_{-0.069}$\\
~~~~$R_*$\dotfill &Radius (\rst)\dotfill & - &$1.486^{+0.035}_{-0.032}$\\
~~~~$L_*$\dotfill &Luminosity (\lst)\dotfill & - &$3.60^{+0.19}_{-0.18}$\\
~~~~$\rho_*$\dotfill &Density (cgs)\dotfill & - &$0.600^{+0.031}_{-0.042}$\\
~~~~$\log{g}$\dotfill &Surface gravity (cgs)\dotfill & -  &$4.238^{+0.017}_{-0.024}$\\
~~~~$T_{\rm eff}$\dotfill &Effective Temperature (K)\dotfill & $\mathcal{G}[6512,100]$ &$6520\pm87$\\
~~~~$[{\rm Fe/H}]$\dotfill &Metallicity (dex)\dotfill & $\mathcal{G}[0.06,0.08]$ &$0.071^{+0.078}_{-0.077}$\\
~~~~$Age$\dotfill &Age (Gyr)\dotfill & - &$1.30^{+0.75}_{-0.58}$\\
~~~~$A_V$\dotfill &V-band extinction (mag)\dotfill & $\mathcal{U}[0,0.30287]$ &$0.139^{+0.093}_{-0.085}$\\
~~~~$\sigma_{SED}$\dotfill &SED photometry error scaling \dotfill & - &$3.17^{+1.3}_{-0.78}$\\
~~~~$\varpi$\dotfill &Parallax (mas)\dotfill & $\mathcal{G}[3.6589,0.0144]$ &$3.684\pm0.014$\\
~~~~$d$\dotfill &Distance (pc)\dotfill & - & $271.4^{+1.1}_{-1.0}$\\
\hline
\multicolumn{1}{l}{\bf Brown Dwarf Parameters:}& \\
~~~~$M_b$\dotfill &Mass (\mj)\dotfill  & - &$27.93^{+0.94}_{-0.99}$\\
~~~~$R_b$\dotfill &Radius (\rj)\dotfill & - & $1.129^{+0.031}_{-0.027}$\\
~~~~$P_{\rm orb}$\dotfill & Period (days)\dotfill & - &$1.217523^{+0.000013}_{-0.000014}$\\
~~~~$T_C$\dotfill &Time of conjunction (\bjdtdb)\dotfill & - &$2458326.21114\pm0.00019$\\
~~~~$a$\dotfill &Semi-major axis (AU)\dotfill  & - &$0.02506^{+0.00038}_{-0.00042}$\\
~~~~$i$\dotfill &Orbital inclination (Degrees)\dotfill & - &$87.0\pm1.9$\\
~~~~$e$\dotfill &Eccentricity \dotfill  & - &$0.0096^{+0.011}_{-0.0067}$\\
~~~~$ecos{\omega_*}$\dotfill & \dotfill & - &$0.0029^{+0.0092}_{-0.0053}$\\
~~~~$esin{\omega_*}$\dotfill & \dotfill & - &$-0.0001^{+0.0092}_{-0.0097}$\\
~~~~$\fave$\dotfill &Incident Flux (\fluxcgs)\dotfill & - &$7.80^{+0.39}_{-0.32}$\\
~~~~$T_{eq}$\dotfill &Equilibrium temperature (K)\dotfill  & - &$2421^{+30}_{-25}$\\
~~~~$K$\dotfill &RV semi-amplitude ($\rm m\, s^{-1}$)\dotfill  & - &$4210^{+58}_{-59}$\\
~~~~$R_b/R_*$\dotfill &Radius of planet in stellar radii \dotfill  & - &$0.07810\pm0.00070$\\
~~~~$a/R_*$\dotfill &Semi-major axis in stellar radii \dotfill  & - &$3.632^{+0.062}_{-0.087}$\\
~~~~$\delta$\dotfill &Transit depth (fraction)\dotfill  & - &$0.00610\pm0.00011$\\
~~~~$T_{14}$\dotfill &Total transit duration (days)\dotfill & - &$0.11494^{+0.00072}_{-0.00066}$\\
~~~~$\tau$\dotfill &Ingress/egress transit duration (days)\dotfill  & - &$0.00881^{+0.00056}_{-0.00029}$\\
~~~~$b$\dotfill &Transit Impact parameter \dotfill  & - &$0.19^{+0.11}_{-0.12}$\\
~~~~$logg_b$\dotfill &Surface gravity \dotfill & -  &$4.736^{+0.019}_{-0.025}$\\
~~~~$M_b\sin i$\dotfill &Minimum mass (\mj)\dotfill & -  &$27.88^{+0.94}_{-1.0}$\\
~~~~$M_b/M_*$\dotfill &Mass ratio \dotfill & -  &$0.01920^{+0.00044}_{-0.00041}$\\
\hline
\multicolumn{2}{l}{Wavelength Parameters:}&&i$^\prime$\smallskip\\
~~~~$u_{1}$\dotfill &linear limb-darkening coeff \dotfill &-&$0.189\pm0.042$\\
~~~~$u_{2}$\dotfill &quadratic limb-darkening coeff \dotfill &-&$0.295\pm0.047$\\
\multicolumn{2}{l}{Wavelength Parameters:}&&r$^\prime$\smallskip\\
~~~~$u_{1}$\dotfill &linear limb-darkening coeff \dotfill &-&$0.241\pm0.041$\\
~~~~$u_{2}$\dotfill &quadratic limb-darkening coeff \dotfill &-&$0.293^{+0.047}_{-0.046}$\\
\multicolumn{2}{l}{Wavelength Parameters:}&&z$^\prime$\smallskip\\
~~~~$u_{1}$\dotfill &linear limb-darkening coeff \dotfill &-&$0.186^{+0.043}_{-0.044}$\\
~~~~$u_{2}$\dotfill &quadratic limb-darkening coeff \dotfill &-&$0.328^{+0.048}_{-0.047}$\\
\multicolumn{2}{l}{Transit Parameters:}&&FLWO UT 2012-01-07 (i$^\prime$) \smallskip\\
~~~~$\sigma^{2}$\dotfill &Added Variance \dotfill &-&$-0.00000864^{+0.00000023}_{-0.00000021}$\\
~~~~$F_0$\dotfill &Baseline flux \dotfill &-&$1.00427\pm0.00011$\\
\multicolumn{2}{l}{Transit Parameters:}&&ULMO UT 2011-12-31 (r$^\prime$)\smallskip\\
~~~~$\sigma^{2}$\dotfill &Added Variance \dotfill &-&$0.000000312^{+0.00000011}_{-0.000000091}$\\
~~~~$F_0$\dotfill &Baseline flux \dotfill &-&$1.003066^{+0.000090}_{-0.000088}$\\
\multicolumn{2}{l}{Transit Parameters:}&& FLWO UT 2011-12-16 (z$^\prime$)\smallskip\\
~~~~$\sigma^{2}$\dotfill &Added Variance \dotfill &-&$-0.00001194^{+0.00000019}_{-0.00000017}$\\
~~~~$F_0$\dotfill &Baseline flux \dotfill &-&$1.00366\pm0.00010$\\
\multicolumn{2}{l}{Telescope Parameters:}&&TRES\smallskip\\
~~~~$\gamma_{\rm rel}$\dotfill &Relative RV Offset ($\rm m\, s^{-1}$)\dotfill &-&$4113^{+49}_{-52}$\\
~~~~$\sigma_J$\dotfill &RV Jitter ($\rm m\, s^{-1}$)\dotfill &-&$127^{+76}_{-96}$\\

\hline
\end{tabular}
\end{table*}

\begin{table*}
\centering
\caption[]{MIST median values and 68\% confidence interval for Kepler-39, created using {\tt EXOFASTv2} commit number f8f3437. Here, $\mathcal{U}$[a,b] is the uniform prior bounded between $a$ and $b$, and $\mathcal{G}[a,b]$ is a Gaussian prior of mean $a$ and width $b$.} \label{tab:exofast_kepler39}

\begin{tabular}{lcccc}
~~~Parameter & Units & Priors & Values\\
\hline
\multicolumn{2}{l}{\bf Stellar Parameters:}&\smallskip\\
~~~~$M_*$\dotfill &Mass (\mst)\dotfill & - &$1.201^{+0.071}_{-0.087}$\\
~~~~$R_*$\dotfill &Radius (\rst)\dotfill & - &$1.260^{+0.030}_{-0.029}$\\
~~~~$L_*$\dotfill &Luminosity (\lst)\dotfill & - &$2.24^{+0.17}_{-0.15}$\\
~~~~$\rho_*$\dotfill &Density (cgs)\dotfill & - &$0.844^{+0.080}_{-0.084}$\\
~~~~$\log{g}$\dotfill &Surface gravity (cgs)\dotfill & -  &$4.316^{+0.032}_{-0.039}$\\
~~~~$T_{\rm eff}$\dotfill &Effective Temperature (K)\dotfill & $\mathcal{G}[6260,140]$ &$6290\pm120$\\
~~~~$[{\rm Fe/H}]$\dotfill &Metallicity (dex)\dotfill & $\mathcal{G}[0.10,0.15]$ &$0.07\pm0.14$\\
~~~~$Age$\dotfill &Age (Gyr)\dotfill & - &$2.2^{+2.2}_{-1.5}$\\
~~~~$A_V$\dotfill &V-band extinction (mag)\dotfill & $\mathcal{U}[0,0.56017]$ &$0.14^{+0.17}_{-0.10}$\\
~~~~$\sigma_{SED}$\dotfill &SED photometry error scaling \dotfill & - &$1.22^{+0.74}_{-0.40}$\\
~~~~$\varpi$\dotfill &Parallax (mas)\dotfill & $\mathcal{G}[0.8910,0.0164]$ &$0.914\pm0.016$\\
~~~~$d$\dotfill &Distance (pc)\dotfill & - & $1093^{+20}_{-19}$\\
\hline
\multicolumn{1}{l}{\bf Brown Dwarf Parameters:}& \\
~~~~$M_b$\dotfill &Mass (\mj)\dotfill  & - &$19.0\pm1.3$\\
~~~~$R_b$\dotfill &Radius (\rj)\dotfill & - & $1.067^{+0.028}_{-0.026}$\\
~~~~$P_{\rm orb}$\dotfill & Period (days)\dotfill & - &$21.08701\pm0.00029$\\
~~~~$T_C$\dotfill &Time of conjunction (\bjdtdb)\dotfill & - &$2455141.29441\pm0.00036$\\
~~~~$a$\dotfill &Semi-major axis (AU)\dotfill  & - &$0.1596^{+0.0031}_{-0.0039}$\\
~~~~$i$\dotfill &Orbital inclination (Degrees)\dotfill & - &$89.59^{+0.28}_{-0.32}$\\
~~~~$e$\dotfill &Eccentricity \dotfill  & - &$0.066\pm0.036$\\
~~~~$ecos{\omega_*}$\dotfill & \dotfill & - &$-0.012^{+0.025}_{-0.027}$\\
~~~~$esin{\omega_*}$\dotfill & \dotfill & - &$0.057^{+0.038}_{-0.042}$\\
~~~~$\fave$\dotfill &Incident Flux (\fluxcgs)\dotfill & - &$0.1196^{+0.0087}_{-0.0079}$\\
~~~~$T_{eq}$\dotfill &Equilibrium temperature (K)\dotfill  & - &$853\pm15$\\
~~~~$K$\dotfill &RV semi-amplitude ($\rm m\, s^{-1}$)\dotfill  & - &$1231^{+66}_{-65}$\\
~~~~$R_b/R_*$\dotfill &Radius of planet in stellar radii \dotfill  & - &$0.08693^{+0.00075}_{-0.00059}$\\
~~~~$a/R_*$\dotfill &Semi-major axis in stellar radii \dotfill  & - &$27.21^{+0.83}_{-0.94}$\\
~~~~$\delta$\dotfill &Transit depth (fraction)\dotfill  & - &$0.00756^{+0.00013}_{-0.00010}$\\
~~~~$T_{14}$\dotfill &Total transit duration (days)\dotfill & - &$0.2478^{+0.0018}_{-0.0014}$\\
~~~~$\tau$\dotfill &Ingress/egress transit duration (days)\dotfill  & - &$0.02041^{+0.0020}_{-0.00065}$\\
~~~~$b$\dotfill &Transit Impact parameter \dotfill  & - &$0.18^{+0.15}_{-0.12}$\\
~~~~$logg_b$\dotfill &Surface gravity \dotfill & -  &$4.616^{+0.034}_{-0.037}$\\
~~~~$M_b\sin i$\dotfill &Minimum mass (\mj)\dotfill & -  &$19.0\pm1.3$\\
~~~~$M_b/M_*$\dotfill &Mass ratio \dotfill & -  &$0.01518^{+0.00094}_{-0.00090}$\\
\hline
\multicolumn{2}{l}{Wavelength Parameters:}&&Kepler\smallskip\\
~~~~$u_{1}$\dotfill &linear limb-darkening coeff \dotfill &-&$0.306\pm0.034$\\
~~~~$u_{2}$\dotfill &quadratic limb-darkening coeff \dotfill &-&$0.292\pm0.048$\\
\multicolumn{2}{l}{Transit Parameters:}&&Kepler UT 2012-01-12 \smallskip\\
~~~~$\sigma^{2}$\dotfill &Added Variance \dotfill &-&$0.0000000452\pm0.0000000018$\\
~~~~$F_0$\dotfill &Baseline flux \dotfill &-&$1.0000066^{+0.0000045}_{-0.0000044}$\\
\multicolumn{2}{l}{Telescope Parameters:}&&SOHPIE\smallskip\\
~~~~$\gamma_{\rm rel}$\dotfill &Relative RV Offset ($\rm m\, s^{-1}$)\dotfill &-&$-47^{+56}_{-47}$\\
~~~~$\sigma_J$\dotfill &RV Jitter ($\rm m\, s^{-1}$)\dotfill &-&$137^{+69}_{-47}$\\

\hline
\end{tabular}
\end{table*}

\begin{table*}
\centering
\caption[]{MIST median values and 68\% confidence interval for KOI-189, created using {\tt EXOFASTv2} commit number f8f3437. Here, $\mathcal{U}$[a,b] is the uniform prior bounded between $a$ and $b$, and $\mathcal{G}[a,b]$ is a Gaussian prior of mean $a$ and width $b$.} \label{tab:exofast_koi189}

\begin{tabular}{lcccc}
~~~Parameter & Units & Priors & Values\\
\hline
\multicolumn{2}{l}{\bf Stellar Parameters:}&\smallskip\\
~~~~$M_*$\dotfill &Mass (\mst)\dotfill & - &$0.801^{+0.037}_{-0.034}$\\
~~~~$R_*$\dotfill &Radius (\rst)\dotfill & - &$0.768\pm0.011$\\
~~~~$L_*$\dotfill &Luminosity (\lst)\dotfill & - &$0.325\pm0.017$\\
~~~~$\rho_*$\dotfill &Density (cgs)\dotfill & - &$2.50^{+0.14}_{-0.12}$\\
~~~~$\log{g}$\dotfill &Surface gravity (cgs)\dotfill & -  &$4.571^{+0.020}_{-0.019}$\\
~~~~$T_{\rm eff}$\dotfill &Effective Temperature (K)\dotfill & $\mathcal{G}[4850,150]$ &$4973\pm84$\\
~~~~$[{\rm Fe/H}]$\dotfill &Metallicity (dex)\dotfill & $\mathcal{G}[-0.7,0.2]$ &$0.04\pm0.14$\\
~~~~$Age$\dotfill &Age (Gyr)\dotfill & - &$6.4^{+4.5}_{-4.1}$\\
~~~~$A_V$\dotfill &V-band extinction (mag)\dotfill & $\mathcal{U}[0,0.14043]$ &$0.053^{+0.053}_{-0.038}$\\
~~~~$\sigma_{SED}$\dotfill &SED photometry error scaling \dotfill & - &$1.19^{+0.66}_{-0.36}$\\
~~~~$\varpi$\dotfill &Parallax (mas)\dotfill & $\mathcal{G}[2.2378, 0.0166]$ &$2.264\pm0.016$\\
~~~~$d$\dotfill &Distance (pc)\dotfill & - & $441.6\pm3.2$\\
\hline
\multicolumn{1}{l}{\bf Brown Dwarf Parameters:}& \\
~~~~$M_b$\dotfill &Mass (\mj)\dotfill  & - &$80.4^{+2.5}_{-2.3}$\\
~~~~$R_b$\dotfill &Radius (\rj)\dotfill & - & $0.988^{+0.020}_{-0.021}$\\
~~~~$P_{\rm orb}$\dotfill & Period (days)\dotfill & - &$30.36049\pm0.00024$\\
~~~~$T_C$\dotfill &Time of conjunction (\bjdtdb)\dotfill & - &$2459444.083\pm0.030$\\
~~~~$a$\dotfill &Semi-major axis (AU)\dotfill  & - &$0.1824^{+0.0027}_{-0.0026}$\\
~~~~$i$\dotfill &Orbital inclination (Degrees)\dotfill & - &$89.623^{+0.050}_{-0.042}$\\
~~~~$e$\dotfill &Eccentricity \dotfill  & - &$0.2740^{+0.0039}_{-0.0044}$\\
~~~~$ecos{\omega_*}$\dotfill & \dotfill & - &$-0.1359^{+0.0028}_{-0.0030}$\\
~~~~$esin{\omega_*}$\dotfill & \dotfill & - &$-0.2379^{+0.0061}_{-0.0055}$\\
~~~~$\fave$\dotfill &Incident Flux (\fluxcgs)\dotfill & - &$0.01234^{+0.00078}_{-0.00074}$\\
~~~~$T_{eq}$\dotfill &Equilibrium temperature (K)\dotfill  & - &$491.9^{+7.6}_{-7.5}$\\
~~~~$K$\dotfill &RV semi-amplitude ($\rm m\, s^{-1}$)\dotfill  & - &$5943^{+36}_{-38}$\\
~~~~$R_b/R_*$\dotfill &Radius of planet in stellar radii \dotfill  & - &$0.1323^{+0.0012}_{-0.0013}$\\
~~~~$a/R_*$\dotfill &Semi-major axis in stellar radii \dotfill  & - &$51.08^{+0.90}_{-0.85}$\\
~~~~$\delta$\dotfill &Transit depth (fraction)\dotfill  & - &$0.01750^{+0.00033}_{-0.00035}$\\
~~~~$T_{14}$\dotfill &Total transit duration (days)\dotfill & - &$0.2522\pm0.0011$\\
~~~~$\tau$\dotfill &Ingress/egress transit duration (days)\dotfill  & - &$0.0347\pm0.0017$\\
~~~~$b$\dotfill &Transit Impact parameter \dotfill  & - &$0.408^{+0.039}_{-0.048}$\\
~~~~$logg_b$\dotfill &Surface gravity \dotfill & -  &$5.310^{+0.023}_{-0.022}$\\
~~~~$M_b\sin i$\dotfill &Minimum mass (\mj)\dotfill & -  &$80.4^{+2.5}_{-2.3}$\\
~~~~$M_b/M_*$\dotfill &Mass ratio \dotfill & -  &$0.0958\pm0.0016$\\
\hline
\multicolumn{2}{l}{Wavelength Parameters:}&&Kepler\smallskip\\
~~~~$u_{1}$\dotfill &linear limb-darkening coeff \dotfill &-&$0.568\pm0.031$\\
~~~~$u_{2}$\dotfill &quadratic limb-darkening coeff \dotfill &-&$0.144\pm0.053$\\
\multicolumn{2}{l}{Transit Parameters:}&&Kepler UT 2009-05-29\smallskip\\
~~~~$\sigma^{2}$\dotfill &Added Variance \dotfill &-&$0.0000000176^{+0.0000000017}_{-0.0000000016}$\\
~~~~$F_0$\dotfill &Baseline flux \dotfill &-&$1.0000054\pm0.0000042$\\
\multicolumn{2}{l}{Telescope Parameters:}&&SOPHIE\smallskip\\
~~~~$\gamma_{\rm rel}$\dotfill &Relative RV Offset ($\rm m\, s^{-1}$)\dotfill &-&$-72587^{+27}_{-23}$\\
~~~~$\sigma_J$\dotfill &RV Jitter ($\rm m\, s^{-1}$)\dotfill &-&$42^{+31}_{-20}$\\

\hline
\end{tabular}
\end{table*}

\begin{table*}
\centering
\caption[]{MIST median values and 68\% confidence interval for KOI-205, created using {\tt EXOFASTv2} commit number f8f3437. Here, $\mathcal{U}$[a,b] is the uniform prior bounded between $a$ and $b$, and $\mathcal{G}[a,b]$ is a Gaussian prior of mean $a$ and width $b$.} \label{tab:exofast_koi205}

\begin{tabular}{lcccc}
~~~Parameter & Units & Priors & Values\\
\hline
\multicolumn{2}{l}{\bf Stellar Parameters:}&\smallskip\\
~~~~$M_*$\dotfill &Mass (\mst)\dotfill & - &$0.876^{+0.046}_{-0.035}$\\
~~~~$R_*$\dotfill &Radius (\rst)\dotfill & - &$0.904^{+0.022}_{-0.025}$\\
~~~~$L_*$\dotfill &Luminosity (\lst)\dotfill & - &$0.554^{+0.041}_{-0.039}$\\
~~~~$\rho_*$\dotfill &Density (cgs)\dotfill & - &$1.67^{+0.16}_{-0.11}$\\
~~~~$\log{g}$\dotfill &Surface gravity (cgs)\dotfill & -  &$4.467^{+0.030}_{-0.021}$\\
~~~~$T_{\rm eff}$\dotfill &Effective Temperature (K)\dotfill & $\mathcal{G}[5237,100]$ &$5238^{+87}_{-85}$\\
~~~~$[{\rm Fe/H}]$\dotfill &Metallicity (dex)\dotfill & $\mathcal{G}[0.14,0.15]$ &$0.15^{+0.14}_{-0.13}$\\
~~~~$Age$\dotfill &Age (Gyr)\dotfill & - &$10.2^{+2.6}_{-4.3}$\\
~~~~$A_V$\dotfill &V-band extinction (mag)\dotfill & $\mathcal{U}[0,0.51956]$ &$0.15^{+0.20}_{-0.11}$\\
~~~~$\sigma_{SED}$\dotfill &SED photometry error scaling \dotfill & - &$2.62^{+1.8}_{-0.99}$\\
~~~~$\varpi$\dotfill &Parallax (mas)\dotfill & $\mathcal{G}[1.6514,0.0159]$ &$1.674\pm0.016$\\
~~~~$d$\dotfill &Distance (pc)\dotfill & - & $597.4^{+5.7}_{-5.6}$\\
\hline
\multicolumn{1}{l}{\bf Brown Dwarf Parameters:}& \\
~~~~$M_b$\dotfill &Mass (\mj)\dotfill  & - &$39.7^{+2.1}_{-1.8}$\\
~~~~$R_b$\dotfill &Radius (\rj)\dotfill & - & $0.870^{+0.022}_{-0.024}$\\
~~~~$P_{\rm orb}$\dotfill & Period (days)\dotfill & - &$11.72021\pm0.00011$\\
~~~~$T_C$\dotfill &Time of conjunction (\bjdtdb)\dotfill & - &$2455139.25544\pm0.00023$\\
~~~~$a$\dotfill &Semi-major axis (AU)\dotfill  & - &$0.0980^{+0.0017}_{-0.0013}$\\
~~~~$i$\dotfill &Orbital inclination (Degrees)\dotfill & - &$88.198^{+0.10}_{-0.078}$\\
~~~~$e$\dotfill &Eccentricity \dotfill  & - &$0.026^{+0.036}_{-0.019}$\\
~~~~$ecos{\omega_*}$\dotfill & \dotfill & - &$-0.0008^{+0.0090}_{-0.011}$\\
~~~~$esin{\omega_*}$\dotfill & \dotfill & - &$-0.022^{+0.021}_{-0.036}$\\
~~~~$\fave$\dotfill &Incident Flux (\fluxcgs)\dotfill & - &$0.0781^{+0.0058}_{-0.0056}$\\
~~~~$T_{eq}$\dotfill &Equilibrium temperature (K)\dotfill  & - &$766\pm14$\\
~~~~$K$\dotfill &RV semi-amplitude ($\rm m\, s^{-1}$)\dotfill  & - &$3770\pm130$\\
~~~~$R_b/R_*$\dotfill &Radius of planet in stellar radii \dotfill  & - &$0.09876^{+0.00039}_{-0.00019}$\\
~~~~$a/R_*$\dotfill &Semi-major axis in stellar radii \dotfill  & - &$23.29^{+0.71}_{-0.51}$\\
~~~~$\delta$\dotfill &Transit depth (fraction)\dotfill  & - &$0.009753^{+0.000077}_{-0.000038}$\\
~~~~$T_{14}$\dotfill &Total transit duration (days)\dotfill & - &$0.1320\pm0.0012$\\
~~~~$\tau$\dotfill &Ingress/egress transit duration (days)\dotfill  & - &$0.02488^{+0.00091}_{-0.00073}$\\
~~~~$b$\dotfill &Transit Impact parameter \dotfill  & - &$0.7492^{+0.0096}_{-0.0085}$\\
~~~~$logg_b$\dotfill &Surface gravity \dotfill & -  &$5.113^{+0.033}_{-0.026}$\\
~~~~$M_b\sin i$\dotfill &Minimum mass (\mj)\dotfill & -  &$39.6^{+2.1}_{-1.8}$\\
~~~~$M_b/M_*$\dotfill &Mass ratio \dotfill & -  &$0.0432\pm0.0017$\\
\hline
\multicolumn{2}{l}{Wavelength Parameters:}&&Kepler\smallskip\\
~~~~$u_{1}$\dotfill &linear limb-darkening coeff \dotfill &-&$0.531^{+0.045}_{-0.046}$\\
~~~~$u_{2}$\dotfill &quadratic limb-darkening coeff \dotfill &-&$0.162^{+0.048}_{-0.047}$\\
\multicolumn{2}{l}{Transit Parameters:}&&Kepler UT 2009-10-23 \smallskip\\
~~~~$\sigma^{2}$\dotfill &Added Variance \dotfill &-&$0.0000000321\pm0.0000000017$\\
~~~~$F_0$\dotfill &Baseline flux \dotfill &-&$1.0000133\pm0.0000043$\\
\multicolumn{2}{l}{Telescope Parameters:}&&SOPHIE\smallskip\\
~~~~$\gamma_{\rm rel}$\dotfill &Relative RV Offset ($\rm m\, s^{-1}$)\dotfill &-&$14919^{+79}_{-85}$\\
~~~~$\sigma_J$\dotfill &RV Jitter ($\rm m\, s^{-1}$)\dotfill &-&$195^{+200}_{-86}$\\

\hline
\end{tabular}
\end{table*}

\begin{table*}
\centering
\caption[]{MIST median values and 68\% confidence interval for KOI-415, created using {\tt EXOFASTv2} commit number f8f3437. Here, $\mathcal{U}$[a,b] is the uniform prior bounded between $a$ and $b$, and $\mathcal{G}[a,b]$ is a Gaussian prior of mean $a$ and width $b$.} \label{tab:exofast_koi415}

\begin{tabular}{lcccc}
~~~Parameter & Units & Priors & Values\\
\hline
\multicolumn{2}{l}{\bf Stellar Parameters:}&\smallskip\\
~~~~$M_*$\dotfill &Mass (\mst)\dotfill & - &$0.931^{+0.14}_{-0.040}$\\
~~~~$R_*$\dotfill &Radius (\rst)\dotfill & - &$1.355^{+0.039}_{-0.037}$\\
~~~~$L_*$\dotfill &Luminosity (\lst)\dotfill & - &$1.85^{+0.12}_{-0.10}$\\
~~~~$\rho_*$\dotfill &Density (cgs)\dotfill & - &$0.537^{+0.072}_{-0.051}$\\
~~~~$\log{g}$\dotfill &Surface gravity (cgs)\dotfill & -  &$4.148^{+0.055}_{-0.032}$\\
~~~~$T_{\rm eff}$\dotfill &Effective Temperature (K)\dotfill & $\mathcal{G}[5810,100]$ &$5783^{+84}_{-81}$\\
~~~~$[{\rm Fe/H}]$\dotfill &Metallicity (dex)\dotfill & $\mathcal{G}[-0.24,0.11]$ &$-0.24\pm0.10$\\
~~~~$Age$\dotfill &Age (Gyr)\dotfill & - &$11.1^{+1.9}_{-11}$\\
~~~~$A_V$\dotfill &V-band extinction (mag)\dotfill & $\mathcal{U}[0,0.40176]$ &$0.107^{+0.14}_{-0.077}$\\
~~~~$\sigma_{SED}$\dotfill &SED photometry error scaling \dotfill & - &$0.90^{+0.65}_{-0.31}$\\
~~~~$\varpi$\dotfill &Parallax (mas)\dotfill & $\mathcal{G}[1.0312,0.0129]$ &$1.052\pm0.013$\\
~~~~$d$\dotfill &Distance (pc)\dotfill & - & $950^{+12}_{-11}$\\
\hline
\multicolumn{1}{l}{\bf Brown Dwarf Parameters:}& \\
~~~~$M_b$\dotfill &Mass (\mj)\dotfill  & - &$64.8^{+6.2}_{-2.0}$\\
~~~~$R_b$\dotfill &Radius (\rj)\dotfill & - & $0.860^{+0.026}_{-0.024}$\\
~~~~$P_{\rm orb}$\dotfill & Period (days)\dotfill & - &$166.78780\pm0.00034$\\
~~~~$T_C$\dotfill &Time of conjunction (\bjdtdb)\dotfill & - &$2456078.87024\pm0.00089$\\
~~~~$a$\dotfill &Semi-major axis (AU)\dotfill  & - &$0.5915^{+0.028}_{-0.0085}$\\
~~~~$i$\dotfill &Orbital inclination (Degrees)\dotfill & - &$89.05^{+0.13}_{-0.10}$\\
~~~~$e$\dotfill &Eccentricity \dotfill  & - &$0.7003^{+0.0029}_{-0.0031}$\\
~~~~$ecos{\omega_*}$\dotfill & \dotfill & - &$0.4989^{+0.0057}_{-0.0054}$\\
~~~~$esin{\omega_*}$\dotfill & \dotfill & - &$0.4915^{+0.0051}_{-0.0058}$\\
~~~~$\fave$\dotfill &Incident Flux (\fluxcgs)\dotfill & - &$0.00458^{+0.00033}_{-0.00035}$\\
~~~~$T_{eq}$\dotfill &Equilibrium temperature (K)\dotfill  & - &$420.7^{+7.4}_{-8.3}$\\
~~~~$K$\dotfill &RV semi-amplitude ($\rm m\, s^{-1}$)\dotfill  & - &$3363^{+26}_{-25}$\\
~~~~$R_b/R_*$\dotfill &Radius of planet in stellar radii \dotfill  & - &$0.06518^{+0.00029}_{-0.00014}$\\
~~~~$a/R_*$\dotfill &Semi-major axis in stellar radii \dotfill  & - &$94.4^{+4.0}_{-3.1}$\\
~~~~$\delta$\dotfill &Transit depth (fraction)\dotfill  & - &$0.004249^{+0.000038}_{-0.000018}$\\
~~~~$T_{14}$\dotfill &Total transit duration (days)\dotfill & - &$0.2478^{+0.0028}_{-0.0029}$\\
~~~~$\tau$\dotfill &Ingress/egress transit duration (days)\dotfill  & - &$0.0208^{+0.0014}_{-0.0016}$\\
~~~~$b$\dotfill &Transit Impact parameter \dotfill  & - &$0.535^{+0.039}_{-0.054}$\\
~~~~$logg_b$\dotfill &Surface gravity \dotfill & -  &$5.341^{+0.037}_{-0.030}$\\
~~~~$M_b\sin i$\dotfill &Minimum mass (\mj)\dotfill & -  &$64.8^{+6.2}_{-2.0}$\\
~~~~$M_b/M_*$\dotfill &Mass ratio \dotfill & -  &$0.0663^{+0.0012}_{-0.0030}$\\
\hline
\multicolumn{2}{l}{Wavelength Parameters:}&&Kepler\smallskip\\
~~~~$u_{1}$\dotfill &linear limb-darkening coeff \dotfill &-&$0.348\pm0.048$\\
~~~~$u_{2}$\dotfill &quadratic limb-darkening coeff \dotfill &-&$0.246^{+0.047}_{-0.048}$\\
\multicolumn{2}{l}{Transit Parameters:}&&Kepler UT 2009-09-03 \smallskip\\
~~~~$\sigma^{2}$\dotfill &Added Variance \dotfill &-&$0.0000000430^{+0.0000000017}_{-0.0000000016}$\\
~~~~$F_0$\dotfill &Baseline flux \dotfill &-&$1.0000094\pm0.0000042$\\
\multicolumn{2}{l}{Telescope Parameters:}&&SOPHIE\smallskip\\
~~~~$\gamma_{\rm rel}$\dotfill &Relative RV Offset ($\rm m\, s^{-1}$)\dotfill &-&$-1491^{+14}_{-16}$\\
~~~~$\sigma_J$\dotfill &RV Jitter ($\rm m\, s^{-1}$)\dotfill &-&$38^{+22}_{-15}$\\
\hline
\end{tabular}
\end{table*}

\begin{table*}
\centering
\caption[]{MIST median values and 68\% confidence interval for WASP-30, created using {\tt EXOFASTv2} commit number f8f3437. Here, $\mathcal{U}$[a,b] is the uniform prior bounded between $a$ and $b$, and $\mathcal{G}[a,b]$ is a Gaussian prior of mean $a$ and width $b$.} \label{tab:exofast_wasp30}

\begin{tabular}{lcccc}
~~~Parameter & Units & Priors & Values\\
\hline
\multicolumn{2}{l}{\bf Stellar Parameters:}&\smallskip\\
~~~~$M_*$\dotfill &Mass (\mst)\dotfill & - &$1.226^{+0.046}_{-0.034}$\\
~~~~$R_*$\dotfill &Radius (\rst)\dotfill & - &$1.423^{+0.038}_{-0.030}$\\
~~~~$L_*$\dotfill &Luminosity (\lst)\dotfill & - &$2.72^{+0.12}_{-0.11}$\\
~~~~$\rho_*$\dotfill &Density (cgs)\dotfill & - &$0.605^{+0.037}_{-0.047}$\\
~~~~$\log{g}$\dotfill &Surface gravity (cgs)\dotfill & -  &$4.222^{+0.019}_{-0.024}$\\
~~~~$T_{\rm eff}$\dotfill &Effective Temperature (K)\dotfill & $\mathcal{G}[6100,150]$ &$6208^{+78}_{-85}$\\
~~~~$[{\rm Fe/H}]$\dotfill &Metallicity (dex)\dotfill & $\mathcal{G}[-0.03,0.1]$ &$0.032^{+0.088}_{-0.071}$\\
~~~~$Age$\dotfill &Age (Gyr)\dotfill & - &$4.3^{+2.4}_{-1.6}$\\
~~~~$A_V$\dotfill &V-band extinction (mag)\dotfill & $\mathcal{U}[0,0.11532]$ &$0.054^{+0.040}_{-0.037}$\\
~~~~$\sigma_{SED}$\dotfill &SED photometry error scaling \dotfill & - &$1.48^{+0.51}_{-0.33}$\\
~~~~$\varpi$\dotfill &Parallax (mas)\dotfill & $\mathcal{G}[2.7654, 0.0202]$ &$2.809\pm0.020$\\
~~~~$d$\dotfill &Distance (pc)\dotfill & - & $356.1\pm2.5$\\
\hline
\multicolumn{1}{l}{\bf Brown Dwarf Parameters:}& \\
~~~~$M_b$\dotfill &Mass (\mj)\dotfill  & - &$61.7^{+1.5}_{-1.1}$\\
~~~~$R_b$\dotfill &Radius (\rj)\dotfill & - & $0.964^{+0.032}_{-0.028}$\\
~~~~$P_{\rm orb}$\dotfill & Period (days)\dotfill & - &$4.156775^{+0.000034}_{-0.000033}$\\
~~~~$T_C$\dotfill &Time of conjunction (\bjdtdb)\dotfill & - &$2458269.67025\pm0.00027$\\
~~~~$a$\dotfill &Semi-major axis (AU)\dotfill  & - &$0.05501^{+0.00066}_{-0.00050}$\\
~~~~$i$\dotfill &Orbital inclination (Degrees)\dotfill & - &$88.58^{+0.91}_{-0.78}$\\
~~~~$e$\dotfill &Eccentricity \dotfill  & - &$0.0023^{+0.0016}_{-0.0015}$\\
~~~~$ecos{\omega_*}$\dotfill & \dotfill & - &$0.0017\pm0.0015$\\
~~~~$esin{\omega_*}$\dotfill & \dotfill & - &$-0.0000^{+0.0015}_{-0.0016}$\\
~~~~$\fave$\dotfill &Incident Flux (\fluxcgs)\dotfill & - &$1.217^{+0.059}_{-0.052}$\\
~~~~$T_{eq}$\dotfill &Equilibrium temperature (K)\dotfill  & - &$1522^{+18}_{-17}$\\
~~~~$K$\dotfill &RV semi-amplitude ($\rm m\, s^{-1}$)\dotfill  & - &$6600\pm11$\\
~~~~$R_b/R_*$\dotfill &Radius of planet in stellar radii \dotfill  & - &$0.0696\pm0.0011$\\
~~~~$a/R_*$\dotfill &Semi-major axis in stellar radii \dotfill  & - &$8.33^{+0.16}_{-0.22}$\\
~~~~$\delta$\dotfill &Transit depth (fraction)\dotfill  & - &$0.00484\pm0.00016$\\
~~~~$T_{14}$\dotfill &Total transit duration (days)\dotfill & - &$0.1670^{+0.0016}_{-0.0015}$\\
~~~~$\tau$\dotfill &Ingress/egress transit duration (days)\dotfill  & - &$0.01137^{+0.00073}_{-0.00047}$\\
~~~~$b$\dotfill &Transit Impact parameter \dotfill  & - &$0.21^{+0.11}_{-0.13}$\\
~~~~$logg_b$\dotfill &Surface gravity \dotfill & -  &$5.219^{+0.024}_{-0.030}$\\
~~~~$M_b\sin i$\dotfill &Minimum mass (\mj)\dotfill & -  &$61.7^{+1.5}_{-1.1}$\\
~~~~$M_b/M_*$\dotfill &Mass ratio \dotfill & -  &$0.04807^{+0.00047}_{-0.00061}$\\
\hline
\multicolumn{2}{l}{Wavelength Parameters:}&&I\smallskip\\
~~~~$u_{1}$\dotfill &linear limb-darkening coeff \dotfill &-&$0.232^{+0.044}_{-0.045}$\\
~~~~$u_{2}$\dotfill &quadratic limb-darkening coeff \dotfill &-&$0.307^{+0.048}_{-0.049}$\\
\multicolumn{2}{l}{Wavelength Parameters:}&&TESS\smallskip\\
~~~~$u_{1}$\dotfill &linear limb-darkening coeff \dotfill &-&$0.206^{+0.037}_{-0.039}$\\
~~~~$u_{2}$\dotfill &quadratic limb-darkening coeff \dotfill &-&$0.268^{+0.046}_{-0.044}$\\
\multicolumn{2}{l}{Transit Parameters:}&&TRAPPIST UT 2010-10-15 (I)\smallskip\\
~~~~$\sigma^{2}$\dotfill &Added Variance \dotfill &-&$-0.00000084^{+0.00000016}_{-0.00000015}$\\
~~~~$F_0$\dotfill &Baseline flux \dotfill &-&$0.99997\pm0.00013$\\
\multicolumn{2}{l}{Transit Parameters:}&&TESS UT 2018-08-26\smallskip\\
~~~~$\sigma^{2}$\dotfill &Added Variance \dotfill &-&$-0.000000058^{+0.000000050}_{-0.000000049}$\\
~~~~$F_0$\dotfill &Baseline flux \dotfill &-&$1.000115\pm0.000022$\\
\multicolumn{2}{l}{Telescope Parameters:}&&CORALIE\smallskip\\
~~~~$\gamma_{\rm rel}$\dotfill &Relative RV Offset ($\rm m\, s^{-1}$)\dotfill &-&$7928.7^{+9.8}_{-9.7}$\\
~~~~$\sigma_J$\dotfill &RV Jitter ($\rm m\, s^{-1}$)\dotfill &-&$21.9^{+9.0}_{-9.4}$\\
\hline
\end{tabular}
\end{table*}

\begin{table*}
\centering
\caption[]{MIST median values and 68\% confidence interval for WASP-128, created using {\tt EXOFASTv2} commit number f8f3437. Here, $\mathcal{U}$[a,b] is the uniform prior bounded between $a$ and $b$, and $\mathcal{G}[a,b]$ is a Gaussian prior of mean $a$ and width $b$.} \label{tab:exofast_wasp128}

\begin{tabular}{lcccc}
~~~Parameter & Units & Priors & Values\\
\hline
\multicolumn{2}{l}{\bf Stellar Parameters:}&\smallskip\\
~~~~$M_*$\dotfill &Mass (\mst)\dotfill & - &$1.213^{+0.041}_{-0.046}$\\
~~~~$R_*$\dotfill &Radius (\rst)\dotfill & - &$1.179^{+0.017}_{-0.018}$\\
~~~~$L_*$\dotfill &Luminosity (\lst)\dotfill & - &$1.712\pm0.095$\\
~~~~$\rho_*$\dotfill &Density (cgs)\dotfill & - &$1.044^{+0.039}_{-0.043}$\\
~~~~$\log{g}$\dotfill &Surface gravity (cgs)\dotfill & -  &$4.379^{+0.013}_{-0.015}$\\
~~~~$T_{\rm eff}$\dotfill &Effective Temperature (K)\dotfill & $\mathcal{G}[5950,100]$ &$6081^{+76}_{-78}$\\
~~~~$[{\rm Fe/H}]$\dotfill &Metallicity (dex)\dotfill & $\mathcal{G}[0.01,0.12]$ &$0.117^{+0.093}_{-0.082}$\\
~~~~$Age$\dotfill &Age (Gyr)\dotfill & - &$0.95^{+1.0}_{-0.63}$\\
~~~~$A_V$\dotfill &V-band extinction (mag)\dotfill & $\mathcal{U}[0,28706]$ &$0.157^{+0.066}_{-0.072}$\\
~~~~$\sigma_{SED}$\dotfill &SED photometry error scaling \dotfill & - &$0.64^{+0.27}_{-0.17}$\\
~~~~$\varpi$\dotfill &Parallax (mas)\dotfill & $\mathcal{G}[2.3799,0.0125]$ &$2.407\pm0.012$\\
~~~~$d$\dotfill &Distance (pc)\dotfill & - & $415.4\pm2.1$\\
\hline
\multicolumn{1}{l}{\bf Brown Dwarf Parameters:}& \\
~~~~$M_b$\dotfill &Mass (\mj)\dotfill  & - &$39.26^{+0.88}_{-1.00}$\\
~~~~$R_b$\dotfill &Radius (\rj)\dotfill & - & $0.957^{+0.017}_{-0.016}$\\
~~~~$P_{\rm orb}$\dotfill & Period (days)\dotfill & - &$2.2088250^{+0.0000030}_{-0.0000036}$\\
~~~~$T_C$\dotfill &Time of conjunction (\bjdtdb)\dotfill & - &$2459282.92277\pm0.00030$\\
~~~~$a$\dotfill &Semi-major axis (AU)\dotfill  & - &$0.03576^{+0.00040}_{-0.00045}$\\
~~~~$i$\dotfill &Orbital inclination (Degrees)\dotfill & - &$87.75\pm0.44$\\
~~~~$e$\dotfill &Eccentricity \dotfill  & - &$0.0041^{+0.0040}_{-0.0029}$\\
~~~~$ecos{\omega_*}$\dotfill & \dotfill & - &$-0.0010^{+0.0024}_{-0.0044}$\\
~~~~$esin{\omega_*}$\dotfill & \dotfill & - &$-0.0010^{+0.0026}_{-0.0041}$\\
~~~~$\fave$\dotfill &Incident Flux (\fluxcgs)\dotfill & - &$1.825^{+0.095}_{-0.094}$\\
~~~~$T_{eq}$\dotfill &Equilibrium temperature (K)\dotfill  & - &$1684\pm22$\\
~~~~$K$\dotfill &RV semi-amplitude ($\rm m\, s^{-1}$)\dotfill  & - &$5273\pm21$\\
~~~~$R_b/R_*$\dotfill &Radius of planet in stellar radii \dotfill  & - &$0.08341^{+0.00067}_{-0.00069}$\\
~~~~$a/R_*$\dotfill &Semi-major axis in stellar radii \dotfill  & - &$6.523^{+0.080}_{-0.091}$\\
~~~~$\delta$\dotfill &Transit depth (fraction)\dotfill  & - &$0.00696\pm0.00011$\\
~~~~$T_{14}$\dotfill &Total transit duration (days)\dotfill & - &$0.11419^{+0.00079}_{-0.00078}$\\
~~~~$\tau$\dotfill &Ingress/egress transit duration (days)\dotfill  & - &$0.00943^{+0.00029}_{-0.00024}$\\
~~~~$b$\dotfill &Transit Impact parameter \dotfill  & - &$0.257^{+0.046}_{-0.049}$\\
~~~~$logg_b$\dotfill &Surface gravity \dotfill & -  &$5.027^{+0.013}_{-0.015}$\\
~~~~$M_b\sin i$\dotfill &Minimum mass (\mj)\dotfill & -  &$39.22^{+0.89}_{-1.0}$\\
~~~~$M_b/M_*$\dotfill &Mass ratio \dotfill & -  &$0.03089^{+0.00042}_{-0.00037}$\\
\hline
\multicolumn{2}{l}{Wavelength Parameters:}&&TESS\smallskip\\
~~~~$u_{1}$\dotfill &linear limb-darkening coeff \dotfill &-&$0.276^{+0.042}_{-0.041}$\\
~~~~$u_{2}$\dotfill &quadratic limb-darkening coeff \dotfill &-&$0.306\pm0.048$\\
\multicolumn{2}{l}{Transit Parameters:}&&TESS UT 2021-03-08 \smallskip\\
~~~~$\sigma^{2}$\dotfill &Added Variance \dotfill &-&$-0.000000073^{+0.000000036}_{-0.000000035}$\\
~~~~$F_0$\dotfill &Baseline flux \dotfill &-&$1.000048^{+0.000018}_{-0.000019}$\\
\multicolumn{2}{l}{Telescope Parameters:}&&CORALIE\smallskip\\
~~~~$\gamma_{\rm rel}$\dotfill &Relative RV Offset ($\rm m\, s^{-1}$)\dotfill &-&$14785\pm29$\\
~~~~$\sigma_J$\dotfill &RV Jitter ($\rm m\, s^{-1}$)\dotfill &-&$176^{+29}_{-25}$\\
\multicolumn{2}{l}{Telescope Parameters:}&&HARPS\smallskip\\
~~~~$\gamma_{\rm rel}$\dotfill &Relative RV Offset ($\rm m\, s^{-1}$)\dotfill &-&$14720\pm20$\\
~~~~$\sigma_J$\dotfill &RV Jitter ($\rm m\, s^{-1}$)\dotfill &-&$84^{+20}_{-16}$\\

\hline
\end{tabular}
\end{table*}

% Don't change these lines
%\bsp	% typesetting comment
\label{lastpage}

\end{document}